\documentclass[submission,Phys]{SciPost}

\usepackage[utf8]{inputenc} % input encodings (allow UTF-8 input)
\usepackage[T1]{fontenc} 	% selecting font encodings (use 8-bit T1 fonts)
\usepackage[english]{babel} % multilingual support (English language/hyphenation)

\usepackage[bitstream-charter]{mathdesign}
\usepackage{geometry} 		% flexible and complete interface to document dimensions
\usepackage{amsmath} 		% math package (American Mathematical Society)
\usepackage{mathtools} 		% math package (fixes various deficiencies of amsmath)
\usepackage{float} 			% floating objects such as figures and tables
\usepackage{graphicx} 		% enhanced support for graphics
\usepackage{tabularx} 		% tabulars with adjustable-width columns
\usepackage{booktabs} 		% professional-quality tables
\usepackage{color, xcolor} 	% foreground and background colour management
\usepackage{pdfpages} 		% inclusion of external multi-page PDF documents
\usepackage{extarrows} 		% extra arrows beyond those provided in amsmath
\usepackage{multirow} 		% create tabular cells spanning multiple rows
\usepackage{multicol} 		% intermix single and multiple columns
\usepackage[subrefformat=parens]{subcaption}
\usepackage{enumitem} 		% control layout of itemize, enumerate, description
\usepackage{xspace} 		% define commands that appear not to eat spaces
\usepackage{stackrel} 		% enhancement to the \stackrel command
\usepackage{tikz} 			% create PostScript and PDF graphics
\usetikzlibrary{calc}
\usepackage{braket} 		% Dirac bra-ket notation
\usepackage{bm} 			% access bold symbols in maths mode
\usepackage{tensor} 		% typeset tensors (tensor-style super- and subscripts)
\usepackage{slashed} 		% slash through characters (Feynman slash notation)
\usepackage{siunitx} % SI units package (typesetting values with units)
\usepackage{lastpage} 		% reference last page
\usepackage{cite} 			% improved citation handling
\usepackage[normalem]{ulem} % package for underlining
\usepackage{fontawesome} 	% access to web-related icons
\usepackage{tocloft} 		% control over the typography of the TOC, LOF, and LOT
\usepackage{titlesec} 		% interface to sectioning commands (various title styles)
\usepackage{doi} 			% create correct hyperlinks for DOI numbers
\usepackage{hyperref} 		% hypertext links (handel cross-referencing commands)
\usepackage[most]{tcolorbox} 					% coloured and framed text boxes
\usepackage[nameinlink, capitalize]{cleveref} 	% intelligent cross-referencing
\usepackage[nottoc, notlot, notlof]{tocbibind} 	% add references/index/contents to TOC
\usepackage[ruled, vlined]{algorithm2e} 		% floating algorithm environment
\usepackage{makecell}
\usepackage{setspace}

\makeatletter
\def\BState{\State\hskip-\ALG@thistlm}
\makeatother

\makeatletter
\@ifundefined{pdfoutput}{}{\DeclareGraphicsRule{*}{mps}{*}{}}
\makeatother

\makeatletter
\DeclareRobustCommand*{\bfseries}{%
   \not@math@alphabet\bfseries\mathbf
   \fontseries\bfdefault\selectfont
   \boldmath
}
\makeatother

\hypersetup{
	pdftitle={MadNIS},
	pdfauthor={Heimel et al.},
	colorlinks=true, 			% false: boxed links, true: colored links
	linkcolor={red!50!black}, 	% color of internal links (sections, pages, etc.)
	citecolor={blue!50!black}, 	% color of citation links (links to bibliography)
	urlcolor={blue!80!black} 	% color of URL links (external links)
} 
\DeclareSymbolFont{usualmathcal}{OMS}{cmsy}{m}{n}
\DeclareSymbolFontAlphabet{\mathcal}{usualmathcal}

\SetArgSty{textnormal}
\SetKwComment{Comment}{{\small\#}~}{}
\SetCommentSty{mycommfont}

\setitemize{itemsep=0pt, parsep=0pt} 				% adjust itemize environment
\setenumerate{itemsep=0pt, parsep=0pt} 				% adjust enumerate environment
\setitemize{itemsep=2pt,topsep=2pt,parsep=0pt,partopsep=0pt,leftmargin=*}
\setenumerate{itemsep=0pt,topsep=2pt,parsep=0pt,partopsep=0pt,labelindent=3pt,leftmargin=*}
\definecolor{EmeraldGreen}{HTML}{1ea78d}
\definecolor{EnglishRed}{HTML}{b02427}
\hypersetup{colorlinks=true,urlcolor=EmeraldGreen,citecolor=EmeraldGreen,linkcolor=EnglishRed}

\newcommand{\ie}{\text{i.e.}\;}

\newcommand{\gbar}{\overline{G}}

\newcommand{\mwith}{\text{with}}
\newcommand{\mand}{\text{and}}
\newcommand{\mfor}{\text{for}}

\newcommand{\qqquad}{\qquad\quad}

\def\d{\mathrm{d}}

\newcommand\one{\leavevmode\hbox{\small1\normalsize\kern-.33em1}}
\newcommand{\imag}{\mathrm{i}} 				% imaginary unit
\newcommand{\pT}{p_{\mathrm{T}}} 	% transverse momentum
\newcommand{\loss}{\mathcal{L}} 	% loss value

\newcommand{\vegas}{\textsc{Vegas}\xspace}
\newcommand{\madgraph}{\textsc{MadGraph5\_aMC@NLO}\xspace}
\newcommand{\mg}{\textsc{MG5aMC}\xspace}

\newcommand{\tensorflow}{\textsc{TensorFlow}\xspace}

\newcommand{\sherpa}{\textsc{Sherpa}\xspace}

\newcommand{\madnis}{\textsc{MadNIS}\xspace}
\newcommand{\lhapdf}{\textsc{LHAPDF6}\xspace}
\newcommand{\python}{\textsc{Python}\xspace}
\newcommand{\whizard}{\textsc{Whizard}\xspace}

\newcommand{\arXiv}[2][]{%
	\ifthenelse{\equal{#1}{}}%
	{\href{http://arxiv.org/abs/#2}{arXiv:#2}}%
	{\href{http://arxiv.org/abs/#2}{arXiv:#2~[#1]}}}

\newcommand{\gev}{\text{GeV}}

\def\slashchar#1{\setbox0=\hbox{$#1$}           % set a box for #1
   \dimen0=\wd0                                 % and get its size
   \setbox1=\hbox{/} \dimen1=\wd1               % get size of /
   \ifdim\dimen0>\dimen1                        % #1 is bigger
      \rlap{\hbox to \dimen0{\hfil/\hfil}}      % so center / in box
      #1                                        % and print #1
   \else                                        % / is bigger
      \rlap{\hbox to \dimen1{\hfil$#1$\hfil}}   % so center #1
      /                                         % and print /
   \fi}

\newcommand{\tikznode}[2]{%
\ifmmode%
\tikz[remember picture,baseline=(#1.base),inner sep=0pt] \node (#1) {$#2$};%
\else
\tikz[remember picture,baseline=(#1.base),inner sep=0pt] \node (#1) {#2};%
\fi}

\def\mathswitchr#1{\relax\ifmmode{\mathrm{#1}}\else$\mathrm{#1}$\xspace\fi}
\def\mathswitch#1{\relax\ifmmode#1\else$#1$\xspace\fi}

\newcommand{\PW}{\mathswitchr W}

\newcommand{\PZ}{\mathswitchr Z}
\newcommand{\PZp}{\mathswitchr {Z'}}

\newcommand{\Pep}{\mathswitchr {e^+}}
\newcommand{\Pem}{\mathswitchr {e^-}}

\newcommand{\Pp}{\mathswitchr p}

\newcommand{\Pb}{\mathswitchr b}
\newcommand{\Pt}{\mathswitchr t}

\newcommand{\Ptbar}{\mathswitchr{\bar t}}

\newcommand{\jets}{\mathrm{jets}}

\newcommand{\MZ}{\mathswitch {M_\PZ}}
\newcommand{\MZp}{\mathswitch {M_\PZp}}

\graphicspath{{./figs/}}
\begin{document}

%%% Only use in draft mode to mark changes
%\draft

\vspace*{-2.5em}
\hfill{\small IRMP-CP3-22-56, MCNET-22-22, FERMILAB-PUB-22-915-T}
\vspace*{0.5em}

\begin{center}{\Large \textbf{
MadNIS -- Neural Multi-Channel Importance Sampling
}}\end{center}

\begin{center}
Theo Heimel\textsuperscript{1},
Ramon Winterhalder\textsuperscript{2}, \\
Anja Butter\textsuperscript{1,3},
Joshua Isaacson\textsuperscript{4},
Claudius Krause\textsuperscript{1},\\
Fabio Maltoni\textsuperscript{2,5},
Olivier Mattelaer\textsuperscript{2}, and 
Tilman Plehn\textsuperscript{1}
\end{center}

\begin{center}
{\bf 1} Institut für Theoretische Physik, Universität Heidelberg, Germany
\\
{\bf 2} CP3, Universit\'e catholique de Louvain, Louvain-la-Neuve, Belgium
\\
{\bf 3} LPNHE, Sorbonne Universit\'e, Universit\'e Paris Cit\'e, CNRS/IN2P3, Paris, France
\\
{\bf 4} Theoretical Physics Division, Fermi National Accelerator Laboratory, Batavia, IL, USA
\\
{\bf 5} Dipartimento di Fisica e Astronomia, Universit\'a di Bologna, Italy
\\[1em]
ramon.winterhalder@uclouvain.be
\end{center}

%\begin{center}
%\today
%\end{center}

% For convenience during refereeing: line numbers
%\linenumbers
\vspace{-1cm}
\section*{Abstract}
{\bf Theory predictions for the LHC require precise numerical
  phase-space integration and generation of unweighted events. We
  combine machine-learned multi-channel weights with a
  normalizing flow for importance sampling, to improve classical
  methods for numerical integration. We develop an efficient bi-directional
  setup based on an invertible network, combining online and buffered 
  training for potentially expensive integrands. We illustrate our
  method for the Drell-Yan process with an additional narrow
  resonance.}

\vspace{10pt}
\noindent\rule{\textwidth}{1pt}
\tableofcontents\thispagestyle{fancy}
\noindent\rule{\textwidth}{1pt}
%\vspace{10pt}

\clearpage
%%%%%%%%%%%%%%%%%%%%%%%%%%%%%%%%%%%%%%%%%%%%%%%%%%%
\section{Introduction}
\label{sec:intro}

The comparison of data with first-principle predictions defines LHC
physics. Event generators provide and evaluate fundamental theory
predictions as the key part of a comprehensive forward simulation
chain~\cite{Campbell:2022qmc}. Given that event generation is an
inherently numerical task, it can be improved and accelerated by
modern machine learning in, essentially, all
aspects~\cite{Butter:2022rso,Plehn:2022ftl}. In view of the upcoming
HL-LHC, such an improvement in speed and precision is crucial to avoid
a situation where theory predictions limit the entire relevant LHC
program.

Starting with the integration of matrix elements over phase space, we
can use neural networks to replace expensive loop amplitudes with fast
and precise
surrogates~\cite{Bishara:2019iwh,Badger:2020uow,Aylett-Bullock:2021hmo,Maitre:2021uaa,Badger:2022hwf}. The
precise knowledge of the amplitude structure can then be used to
significantly improve the phase-space integration for a given
process~\cite{Danziger:2021eeg}. Generally, it is possible to improve
numerical integration through neural networks by directly learning the
primitive function~\cite{Maitre:2022xle}, or using modified and
enhanced implementations of importance
sampling~\cite{Klimek:2018mza,Chen:2020nfb,Gao:2020vdv,Bothmann:2020ywa,Gao:2020zvv,Winterhalder:2021ngy}.
Technically, this promising approach encodes a change of integration
variables in a normalizing flow~\cite{nflow1} and then uses
\textsl{online training}~\cite{Butter:2022lkf} while generating
weighted phase space configurations, or weighted events.

This rough online training is successful because normalizing flows, or
invertible networks (INNs)~\cite{inn,cinn}, are especially
well-suited, stable, and precise in LHC physics
applications~\cite{Butter:2020tvl}.  This has been shown in many
instances, including event
generation~\cite{Verheyen:2020bjw,Bellagente:2021yyh,Butter:2021csz,Verheyen:2022tov},
detector
simulations~\cite{Krause:2021ilc,Krause:2021wez,Krause:2022jna,Cresswell:2022tof},
unfolding or inverse simulations~\cite{cinn,Bellagente:2020piv},
kinematic reconstruction~\cite{Leigh:2022lpn}, Bayesian
inference~\cite{Bieringer:2020tnw,Bister:2021arb}, or inference using
the matrix element method~\cite{Butter:2022vkj}. On the other hand,
for expensive integrands online training is clearly not optimal,
because it does not make use of all previously generated data at
subsequent stages of the network training.

For a more efficient training we can use the main structural feature
of normalizing flows, their bijective structure best realized in the
fully symmetric INN variant introduced in
Ref.~\cite{coupling1,coupling2, inn}. It allows us to train the same
INN online and on previously generated events in parallel. Such a
\textsl{buffered training} makes optimal use of potentially expensive
integrands, but requires a dedicated loss function and training
strategy, as we will explain in detail.

In multi-purpose LHC event generators like \madgraph~\cite{Alwall:2014hca} (\mg),
\sherpa~\cite{Sherpa:2019gpd} or \whizard~\cite{Kilian:2007gr} importance sampling is combined with
a multi-channel split of the phase space integration. As it is not
guaranteed that an enhanced importance sampling method provides
optimal results when combined with standard multi-channel algorithms,
we complement our flow-based integration with trainable channel
weights. Finally, we introduce a new implementation of rotation layers
in the normalizing flow architecture, to aid our ML-importance
sampling for high-dimensional phase spaces.

In this paper, we present \madnis (\textbf{Mad}graph-ready
\textbf{N}eural Networks for Multi-Channel \- \textbf{I}mportance
\textbf{S}ampling), a comprehensive framework for ML-based phase space
sampling ready to be used in a multi-purpose event generator. In
Sec.~\ref{sec:classic}, we briefly review the basic concepts of
multi-channel integration and importance sampling, before we introduce
our new ML-implementations in Sec.~\ref{sec:madnis}. We illustrate the
ML-channel weights and their interplay with our new bi-directional
training for neural importance sampling in Sec.~\ref{sec:toys}. In
Sec.~\ref{sec:lhc}, we show how our method works for an actual LHC
process, the Drell-Yan process with an additional narrow
\PZp-resonance. In the Appendix, we provide a detailed description of
possible loss functions for our online and buffered training and
potential issues with the implementation of this new training
approach.

%%%%%%%%%%%%%%%%%%%%%%%%%%%%%%%%%%%%%%%%%%%%%%%%%%%
\section{Classic multi-channel integration}
\label{sec:classic}

The main structure of LHC phase space generators and integrators is
the combination of importance sampling and multi-channel
factorization~\cite{KLEISS1994141}. The reason is that even advanced
sampling methods are not powerful enough to probe all phase space
features with the required precision, and that we know the leading
features from the construction of the helicity amplitudes based on
Feynman diagrams. Before we introduce a network-based implementation,
we briefly review the standard methods.

%%%%%%%%%%%%%%%%%%%%%%%%%%%%%%%%%%%%%%%%%%%%%%%%%%%
\subsection{Multi-channel decomposition}
\label{sec:classic_mc}

A generic integral of a function $f \sim \vert\mathcal{M}\vert^2$ over
the $d$-dimensional phase space $x \in \Phi\subseteq\mathbb{R}^d$ can
be represented by
\begin{align}
    I[f] = \int_\Phi\d^d x\,f(x) \; .
    \label{eq:psinteg}
\end{align}
The standard multi-channel method~\cite{KLEISS1994141, Weinzierl:2000wd}, which is also
followed by \sherpa, starts by introducing
several mappings $G_i:\Phi\to U_i=[0,1]^d$ denoted as $x \to y = G_i(x)$, of
the phase-space variables to obtain individual densities
\begin{align}
   g_i(x)=\left\vert\frac{\partial G_i(x)}{\partial x}\right\vert
   \qquad \mwith \qquad \int \d x\,g_i(x)=1
   \quad \mfor \quad i=1,\dots,m\;,
   \label{eq:channel_densities}
\end{align}
where $m$ is the total number of channels. Typically, the mappings
$G_i(x)$ are initially fixed and based on prior physics knowledge,
like the structure of the underlying Feynman diagrams. In practice,
current event generators like \mg, \sherpa or \whizard do not solely rely on physics-inspired mappings $G_i(x)$, but also combine it with an adaptive \vegas algorithm~\cite{vegas1,vegas2, Ohl:1998jn, Brass:2018xbv, Lepage:2020tgj}. 
Ignoring this for now, the different channels
can still be optimized with respect to some channel weights $\alpha_i$
by combining the individual channel densities into a total density
\begin{align}
    g(x)=\sum^m_i \alpha_i g_i(x)
    \qquad \mwith \qquad \sum^m_i \alpha_i =1
    \quad \mand \quad \alpha_i\ge 0\;,
    \label{eq:total_density}
\end{align}
which also renders $g(x)$ normalized. With this, Eq.\eqref{eq:psinteg}
becomes
\begin{align}
    I[f] 
    &=\sum^m_i \int_{\Phi}\d^d x\,\alpha_i\,g_i(x)\,\frac{f(x)}{g(x)}=\sum^m_i\int_{U_i}\d^d y\,\alpha_i\,\left.\frac{f(x)}{g(x)}\right\vert_{x=\gbar_i(y)}\;,
    \label{eq:multi-channel-standard}
\end{align}
Where $\gbar$ denotes the inverse transformation to $G$. The
optimization finds the set of global $\alpha_i$ that minimizes the
total variance~\cite{KLEISS1994141, Weinzierl:2000wd}.

The single-diagram-enhanced method in \mg~\cite{Maltoni:2002qb,
  Mattelaer:2021xdr} defines local, phase-space dependent, channel
weights $\alpha_i(x)$ as
\begin{align}
  f(x)=\sum^m_i \alpha_i(x) f(x)\,
  \qquad \mwith \qquad \sum^m_i \alpha_i(x) =1
  \quad \mand \quad \alpha_i(x)\ge 0\;.
\label{eq:norm_alpha}
\end{align}
Inserting this into Eq.\eqref{eq:psinteg}, we can decompose and parameterize the phase-space integral as
\begin{align}
  I[f]
  = \sum^m_i\int_\Phi\d^d x\; \alpha_i(x)f(x)
  = \sum_i^m\int_{U_i}\d^d y\; \left.\alpha_i(x)\,\frac{f(x)}{g_i(x)}\right\vert_{x=\gbar_i(y)}\;.
  \label{eq:multi-channel-mg}
\end{align}
Once an appropriate decomposition in terms of $\alpha_i(x)$ is found,
the channel weights are fixed and not further optimized.  The
difference between Eq.\eqref{eq:multi-channel-standard} and
Eq.\eqref{eq:multi-channel-mg} can be understood just as different
channel splittings. If we define the local weights as
\begin{align}
  \alpha_i(x)=\alpha_i\,\frac{g_i(x)}{g(x)}\; ,
  \label{eq:multi-channel-unified}
\end{align}
the two approaches coincide. For more details about the differences
of both multi-channel strategies when used in practice, we refer to Ref.~\cite{Maltoni:2002qb}.

%%%%%%%%%%%%%%%%%%%%%%%%%%%%%%%%%%%%%%%%%%%%%%%%%%%%
\subsubsection*{Single diagram enhancement}

While for a generic integral, finding suitable weights $\alpha_i(x)$
might be unfeasible, \mg introduces two different sets of
$\alpha_i(x)$ for phase-space integration. In the first
basis~\cite{Maltoni:2002qb}, we can parameterize the integral as
\begin{align}
  I[|\mathcal{M}|^2]
%  = \int_\Phi\d^d x\,|\mathcal{M}(x)|^2
%  &= \sum_i \int_\Phi\d^d x\,|\mathcal{M}_i(x)|^2 \frac{|\mathcal{M}(x)|^2}{\sum_j |\mathcal{M}_j(x)|^2},
  = \sum^m_i \int_\Phi\d^d x\; \alpha_i(x)\, |\mathcal{M}(x)|^2 
  \quad \text{with} \quad
  \alpha_i(x)= \frac{|\mathcal{M}_i(x)|^2}{\sum_j |\mathcal{M}_j(x)|^2}\; ,
  \label{eq:sde}
\end{align}
where $i$ indicates individual Feynman diagrams. This choice of
$\alpha_i$ is motivated by the classical limit without interference,
\begin{align}
  I[|\mathcal{M}|^2]
  =  \sum^{m}_i \int_\Phi\d^d x\; |\mathcal{M}_i(x)|^2 \; \frac{|\mathcal{M}(x)|^2}{\sum_j |\mathcal{M}_j(x)|^2} \approx \sum^m_i \int_\Phi\d^d x\, |\mathcal{M}_i(x)|^2 \times 1 \; .
  \label{eq:sde-limit}
\end{align}
In this limit each channel is behaving as a squared diagram, its
features are easily identifiable, and importance sampling is easy to
implement. In general, the number of channels $m$ are completely
arbitrary and will often be less than the number of Feynman diagrams
$M$, \ie $m\le M$.

An alternative choice of channel weights in
\mg~\cite{Mattelaer:2021xdr} replaces the $|\mathcal{M}_i|^2$ by the
product of all propagator denominators appearing in a given diagram
and normalizes them as needed,
\begin{align}
  \alpha_i(x) = \frac{\bar \alpha_i(x)}{\sum_j \bar \alpha_j(x)} 
  \quad \text{with} \quad
  \bar \alpha_i(x) =  \prod_{k \in \text{prop}} \frac{1}{|p_k(x)^2-m_k^2 -\imag m_k\Gamma_k|^2} \; .
  \label{eq:sdenew-wgt} 
\end{align}
While this works extremely well for VBF-like or multi-jet processes,
this does not seem to be a good choice for $\PW/\PZ + \jets$ or $\Pt
\Ptbar + \jets$ production~\cite{Mattelaer:2021xdr}.

%%%%%%%%%%%%%%%%%%%%%%%%%%%%%%%%%%%%%%%%%%%%%%%%%%%%%%%%%%%%%%
\subsection{Monte-Carlo error}
\label{sec:classic_err}

To efficiently calculate an integral, we rely on a smart choice for the
variable transformation $y=G(x)$ introduced in
Eq.\eqref{eq:multi-channel-standard},
\begin{align}
  I[f] = \int_\Phi \d^d x\,f(x)
  =\int_{U}\d^d y \,\left. \frac{f(x)}{g(x)} \right\vert_{x=\gbar(y)}
  \qquad \text{with} \qquad
  g(x)=\left\vert\frac{\partial G(x)}{\partial x}\right\vert \;,
  \label{eq:is_def}
\end{align}
which can be any combination of analytic
remappings~\cite{Weinzierl:2000wd}, a \vegas-like numerical
remapping~\cite{Press:1989vk,vegas1,vegas2,Ohl:1998jn, Brass:2018xbv, Lepage:2020tgj}, or a
normalizing
flow~\cite{Bothmann:2020ywa,Gao:2020zvv,Gao:2020vdv,Winterhalder:2021ngy}.
To construct an optimal variable transformation we need a figure of
merit for the phase space integration. While the integral is unchanged
under the above reparametrization, the variance $\sigma^2$ of the new
integrand is given by
\begin{align}
  \sigma^2\equiv\sigma^2\left[\frac{f}{g}\right] = \int\d^d x\,\left(\frac{f(x)}{g(x)}-I[f] \right)^2\;,
  \label{eq:def_sigma}
\end{align}
and becomes minimal for a perfect mapping with $g(x) = f(x)/I[f]$. In
practice, we evaluate the Monte Carlo estimate of our integral with
discrete sampled points,
\begin{align}
  I[f]
  = \int_\Phi \d^d x \, g(x) \; \frac{f(x)}{g(x)} 
  = \left\langle\frac{f(x)}{g(x)} \right\rangle_{x\sim g(x)} %\notag \\
  \approx \frac{1}{N}\sum_{j=1}^{N}\left.\frac{f(x_j)}{g(x_j)}\right\vert_{
    x_j=\gbar(y_j)} \; .
%  \equiv E_N \; .
\end{align}
In this case the error of the Monte Carlo estimate is itself estimated
through the variance defined in
Eq.\eqref{eq:def_sigma}~\cite{Weinzierl:2000wd},
\begin{align}
  \Delta_{N}^2=\frac{\sigma^2}{N}
%   &= \frac{1}{N} \, \frac{1}{N-1}\sum_{j=1}^{N}\left.\left(\frac{f(x_j)}{g(x_j)}-E_N\right)^2\right\vert_{x_j=\gbar(y_j)}\notag\\
  &=\frac{1}{N-1}\left[
    \left\langle \frac{f(x)^2}{g(x)^2} \right\rangle_{x\sim g(x)}
      - \left\langle \frac{f(x)}{g(x)} \right\rangle_{x\sim g(x)}^2 \right] \; .
  \label{eq:var_all}
\end{align}
Note the correction factor $N/(N-1)$ to obtain the unbiased result.

Next, we split the integral into independent channels, as defined in
Eq.\eqref{eq:multi-channel-mg}.  The Monte Carlo estimate of the
integral is given by the sum of the individual estimates
\begin{align}
  I[f]
%  \approx E_N
  \approx \sum_i \left\langle \alpha_i(x)\frac{f(x)}{g_i(x)} \right\rangle_{x\sim g_i(x)} \; ,
  \label{eq:int_multi_mc}
\end{align}
where the individual channels are evaluated using $N_i$ points and
$\sum_i N_i = N$. The error on the total integral is given by the
uncorrelated combination of the channel-wise errors,
\begin{align}
  \Delta^2_{N} &= \sum_i \Delta^2_{N_i, i}=\sum_i \frac{\sigma^2_i}{N_i} \notag \\
  \text{with} \quad 
  \sigma^2_{i}
  %&= \sum_i \sigma^2_{E_{N_i}}
  %=\sum_i \frac{\sigma^2_i}{N_i}\\
  &= \frac{N_i}{N_i-1} \left[
    \left\langle \alpha_i(x)^2\frac{f(x)^2}{g_i(x)^2} \right\rangle_{x\sim g_i(x)} 
  - \left\langle \alpha_i(x)\frac{f(x)}{g_i(x)} \right\rangle_{x\sim g_i(x)}^2
  \right] \; .
  \label{eq:var_channel}
\end{align}
As known from stratified sampling~\cite{Press:1989vk}, the optimal
number of points per channel, defined by the minimized combined error
is a function of the standard deviations $\sigma_i$
\begin{align}
  N_i
 = N \frac{\sigma_{i}}{\sum_k \sigma_{k}} \; .
 \label{eq:stratified-sampling}
\end{align}
In practice, the $\sigma_i$ are calculated during training, and the
numbers of points $N_i$ used for the numerical integration are
subsequently updated.

%%%%%%%%%%%%%%%%%%%%%%%%%%%%%%%%%%%%%%%%%%%%%%%%%%%
\section{\textsc{MadNIS}}
\label{sec:madnis}

While the state-of-the-art event generators work sufficiently well for
simple processes, they require significant computing time for complex
LHC processes. Consequently, there have been
attempts~\cite{Klimek:2018mza,Gao:2020zvv,Gao:2020vdv,Bothmann:2020ywa,Chen:2020nfb}
to replace \vegas~\cite{vegas1,vegas2, Lepage:2020tgj} with a neural
network equivalent. We add several new components to improve the
precision of the network-based integrator and sampler.

%%%%%%%%%%%%%%%%%%%%%%%%%%%%%%%%%%%%%%%%%%%%%%%%%%%
\subsection{Neural multi-channel weights}
\label{sec:madnis_nms}

First, \madnis replaces the local multi-channel weights from
Sec.\eqref{sec:classic_mc} with trainable channel-weight networks (CWnets),
\begin{align}
  \alpha_i(x) \to \alpha_i(x|\theta)\; .
  \label{eq:trained_weights}
\end{align}
In analogy to classification networks, we encode the normalization of
Eq.\eqref{eq:norm_alpha} into the network architecture. Two possible
methods are
\begin{align}
  \bar{\alpha}_i(x|\theta) = \frac{\exp\alpha_i(x|\theta)}{\sum_j \exp\alpha_j(x|\theta)} \in [0,1]
  \qquad \text{or} \qquad 
  \tilde{\alpha}_i(x|\theta) = \frac{\alpha_i(x|\theta)}{\sum_j \alpha_j(x|\theta)} \in\mathbb{R} \;.
  \label{eq:normalization}
\end{align}
Note that the second normalization also allows for negative channel weights for a generic and unconstrained network output $\alpha_i(x|\theta)$. While this is mathematically allowed and satisfies the requirements in Eq.~\eqref{eq:channel_densities}, these channel weights lose their interpretation as probabilities.
Our tests, however, indicate that the first version, corresponding to a softmax
activation, is more stable during training.We can improve the
training by using physics knowledge. For instance, we can learn a
correction to a prior weight $\alpha^*_i$ given by \mg,
\begin{align}
  \alpha_i(x|\theta) = \log \alpha^*_i(x) + \theta_i\cdot\Delta_i(x|\theta) \; .
\end{align}
This specific form gives the normalized weight
\begin{align}
  \bar{\alpha}_i(x|\theta) = \frac{\alpha^*_i(x)\cdot \exp\left[\theta_i\cdot\Delta_i(x|\theta)\right]}{\sum_j \alpha^*_j(x)\cdot \exp\left[\theta_j\cdot\Delta_j(x|\theta)\right]}
  \qquad \text{with} \qquad \sum_i\alpha^*_i(x)=1\; .
\end{align}
In addition, we can provide the neural network with derived quantities
such as invariant masses alongside the event representation $x$.

%%%%%%%%%%%%%%%%%%%%%%%%%%%%%%%%%%%%%%%%%%%%%%%%%%%
\subsection{Neural importance sampling}
\label{sec:madnis_nis}

\begin{figure}[b!]
  \centering
  \includegraphics[width=0.75\textwidth]{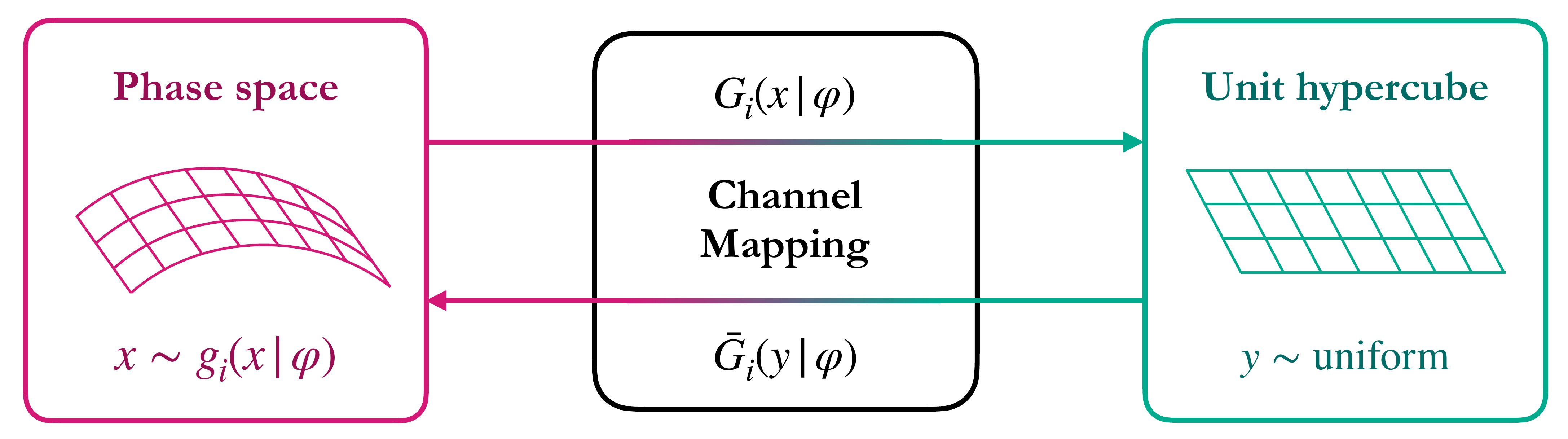}
  \caption{Structure of the INN channel mappings. The latent space
    $y\sim\text{uniform}$ is mapped onto the phase space $x\sim
    g_i(x|\varphi)$ for each channel $i$.}
  \label{fig:cond_flow}
\end{figure}

Second, \madnis augments the physics-inspired phase space mappings with
an INN~\cite{inn}
\begin{align}
%  G_i(x) \to g_i(x|\varphi)
%  \qquad \text{or} \qquad
  y = G_i(x) \to G_i(x|\varphi)
  \qquad \text{and} \qquad
  x = \gbar_i(y|\varphi)
  \; .
  \label{eq:trained_mappings}
\end{align} 
This replaces the classic importance sampling density $g_i(x)$ with a
network-based variable transformation $g_i(x|\varphi)$ in
Eqs.\eqref{eq:multi-channel-mg} and~\eqref{eq:int_multi_mc}
\begin{align}
  I[f]
  %  = \sum_i\int_\Phi\d^d x\; \alpha_i(x)\,f(x)&
  = \sum_i\int_{U_i}\d^d y\,\left.\alpha_i(x)\,\frac{f(x)}{g_i(x|\varphi)} \right\vert_{x=\gbar_i(y|\varphi)} %\notag\\
  \text{with} \quad g_i(x|\varphi)&=\left\vert\frac{\partial G_i(x|\varphi)}{\partial x}\right\vert \; , 
  \label{eq:def_is}
\end{align}
where we assume the latent distribution in $y$ to be uniform.  The
INN-encoded phase space mapping is trained to provide a surrogate
density
\begin{align}
  g_i(x|\varphi)
  \approx f_i(x)  
  =\alpha_i(x)f(x)\; ,
  \label{eq:q_x}
\end{align}
The INN variant of a normalizing flow, illustrated in
Fig.~\ref{fig:cond_flow}, ensures that the training and the evaluation
of the network are symmetric and equally fast in both directions. We
will make use of this structural advantage in our training setup.

%------------------------------------------------------
%------------------------------------------------------

To clearly separate the discussion of the neural importance sampling
from the channel weights defined in Eq.\eqref{eq:trained_weights},
$\alpha_i(x|\theta)$, we denote its network weights as $\varphi$.  In
principle, the bijective mapping $G_i(x|\varphi)$ can be any
combination of a fixed physics-inspired mapping and a normalizing flow.

%------------------------------------------------------
\begin{figure}[t]
  \centering
  \includegraphics[page=1, width=0.98\textwidth]{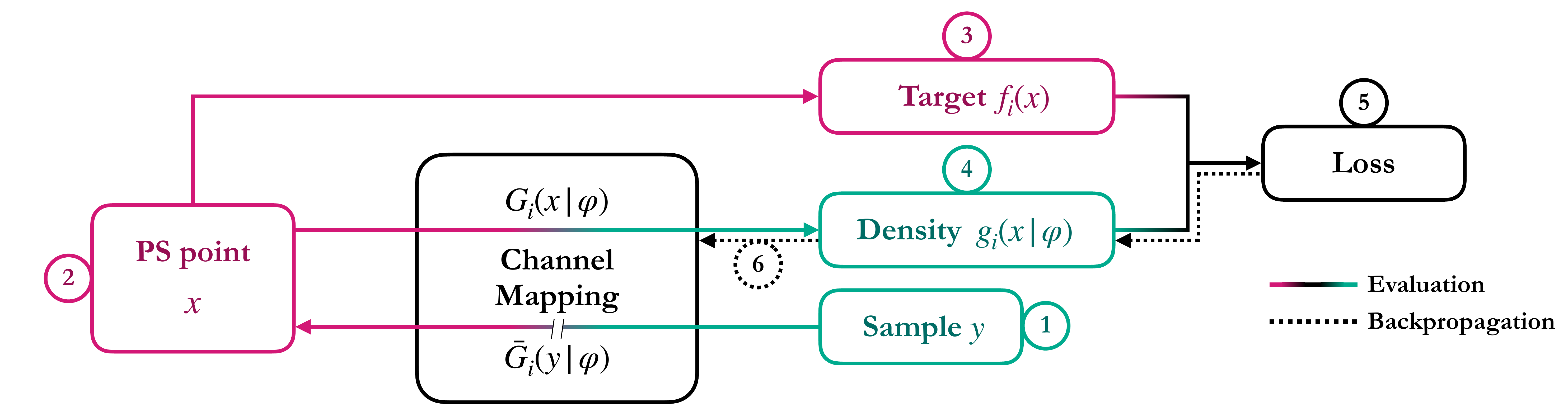}
  \caption{Workflow of the online training of the INN. The
    discontinuous line from (1) to (2) indicates that it only allows
    forward sampling but no gradient backpropagation.}
  \label{fig:gen_training}
\end{figure}
%------------------------------------------------------

Normalizing flows are already used to improve numerical integration
over phase space~\cite{Bothmann:2020ywa} or the Feynman parameters in
loop integrations~\cite{Winterhalder:2021ngy}. The standard i-flow
algorithm~\cite{Gao:2020vdv,Gao:2020zvv} for importance sampling is
\begin{enumerate}
    \item Draw samples from the latent space $y\sim\text{uniform}$;
    \item Transform them into phase-space points $x = \gbar(y|\varphi)$, without gradient calculation;
    \item Evaluate the integrand or target distribution $f(x)$;
    \item Pass the network in the other direction, $y = G(x|\varphi)$, to evaluate the density $g(x|\varphi)$;
    \item Compute divergence-based loss between $f(x)$ and $g(x|\varphi)$;
    \item Compute gradients of the loss and optimize the network.
\end{enumerate}
We illustrate the algorithm in Fig.~\ref{fig:gen_training}. The
additional pass in step 4 is important to evaluate $g(x|\varphi)$ as a
proper function of $x$ and obtain the correct gradients for training,
as explained in the Appendix. Note that the two passes in step 2 and 4
are inverse to each other. We refer to this approach as online
training, because the training data $x$ is continuously generated and
immediately used once for training. It implies that a potentially
expensive integrand $f(x)$ has to be evaluated for every event used to
train the network, which makes it inefficient. One way to alleviate
this problem is to buffer already generated samples and use them for a
limited number of training passes~\cite{Butter:2022lkf}.

%%%%%%%%%%%%%%%%%%%%%%%%%%%%%%%%%%%%%%%%%%%%%%%%%%%
\subsection{Buffered training}
\label{sec:madnis_buff}

An alternative training method for the phase-space mapping would be
traditional sample-based training, where the same samples can be used
every epoch. Pure sample-based training only requires one pass through
the INN, but it is not a sensible choice for neural importance
sampling, because all training data needs to be available from the
beginning. Instead, we iterate between online training, where samples are generated and directly used for training, and buffered training on previously generated events. Because memory constraints inhibit storing all generated phase-space points, we only save a fraction of events in a buffer which is replaced during the next online training phase. 

Before looking into the training algorithm in detail, we need to define a
common loss function for online and the buffered training, so the combination
converges towards a common minimum. The buffered loss has to account
for the fact that training happens after sampling, so the network
weights will change in between. The sampling probability
$q_i(x|\hat{\varphi})$ is different from the density $g_i(x|\varphi)$
at the time of training, even though the two might be related as
\begin{align}
  g_i(x|\varphi)
  \xrightarrow{\varphi\to\hat{\varphi}}
  q_i(x|\hat{\varphi})\;.
\end{align}
Consequently, the buffered form of a KL-loss has to be modified
according to
\begin{align}
  \loss \to \loss \times
%  _{\text{gen},i}\to\mathcal{L}_{\text{2-stage},i}=w(x|i,\varphi)\cdot\mathcal{L}_{\text{gen},i}
%    \quad \text{with} \quad
  %    w(x|i,\varphi) =
  \frac{g_i(x|\varphi)}{q_i(x|\hat{\varphi})}\; ,
\end{align}
which is a generalization of the weighted log-likelihood loss in
Ref.~\cite{Verheyen:2020bjw}.  This means we have to buffer $x$,
$f_i(x)$, and the sampling density $q_i(x|\hat{\varphi})$ to be able
to evaluate the loss.  More details about the corresponding losses can
be found in the Appendix.

%------------------------------------------------------
\begin{figure}[t]
    \centering
    \includegraphics[page=2, width=0.98\textwidth]{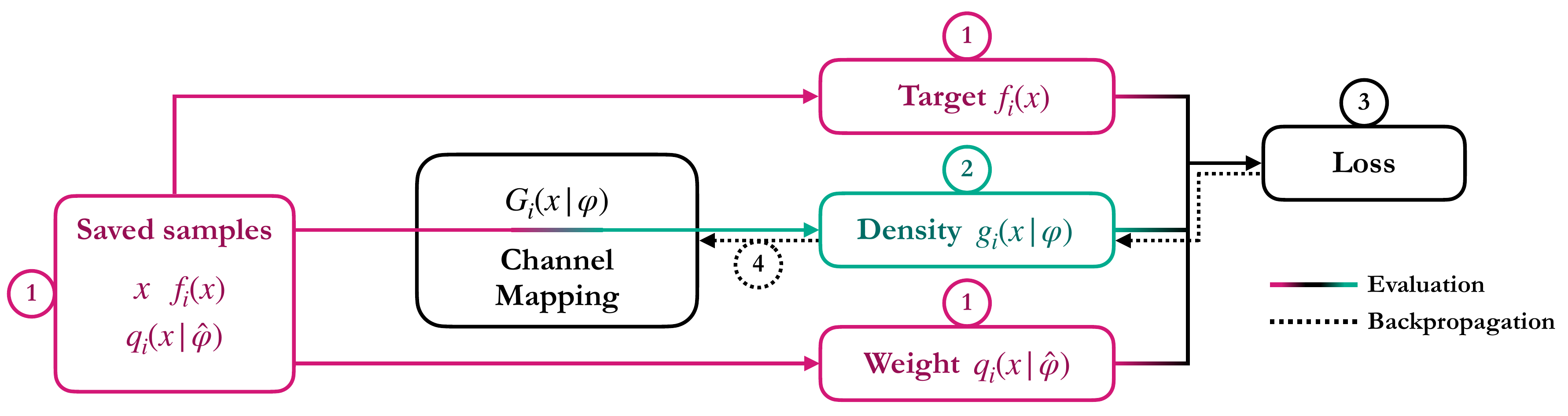}
    \caption{Workflow of the buffered training of the INN.}
    \label{fig:sample_training}
\end{figure}
%------------------------------------------------------

In Fig.~\ref{fig:sample_training}, we illustrate the workflow of the buffered based training:
\begin{enumerate}
  \item Start with a buffered phase space point $x$ with $f(x)$,
    and $q_i(x|\hat{\varphi})$;
  \item Pass it through the INN and compute the density $g_i(x|\varphi)$;
  \item Compute the weighted loss from $g_i(x|\varphi)$ and $f_i(x)$,
    using $q_i(x|\hat{\varphi})$;
  \item Compute gradients and optimize the network.
\end{enumerate}
This training can be combined with the online training introduced in
Sec.~\ref{sec:madnis_nis}, and the balance of the two training
strategies can be adjusted depending on how computationally expensive
the integrand evaluation is.

%%%%%%%%%%%%%%%%%%%%%%%%%%%%%%%%%%%%%%%%%%%%%%%%%%%
\subsubsection*{Training time statistics}

To illustrate the trade-off between training time and weight updates,
we consider a training taking the time $T$, split into buffered
$(T_\text{buff}=T\cdot (1-r_\text{@}))$ and online $(T_\text{@}=T\cdot
r_\text{@})$ training. Let $t_\text{buff}$ and $t_\text{@}$ be the
time for a weight update in the buffered training and online training
(excluding the integrand evaluation), respectively. Note that
$t_\text{@}$ requires an additional sampling without gradient updates,
as explained in the Appendix. We find that
$t_\text{@}/t_\text{buff}\approx 1.33$.

\begin{figure}[t]
  \includegraphics[width=0.495\textwidth]{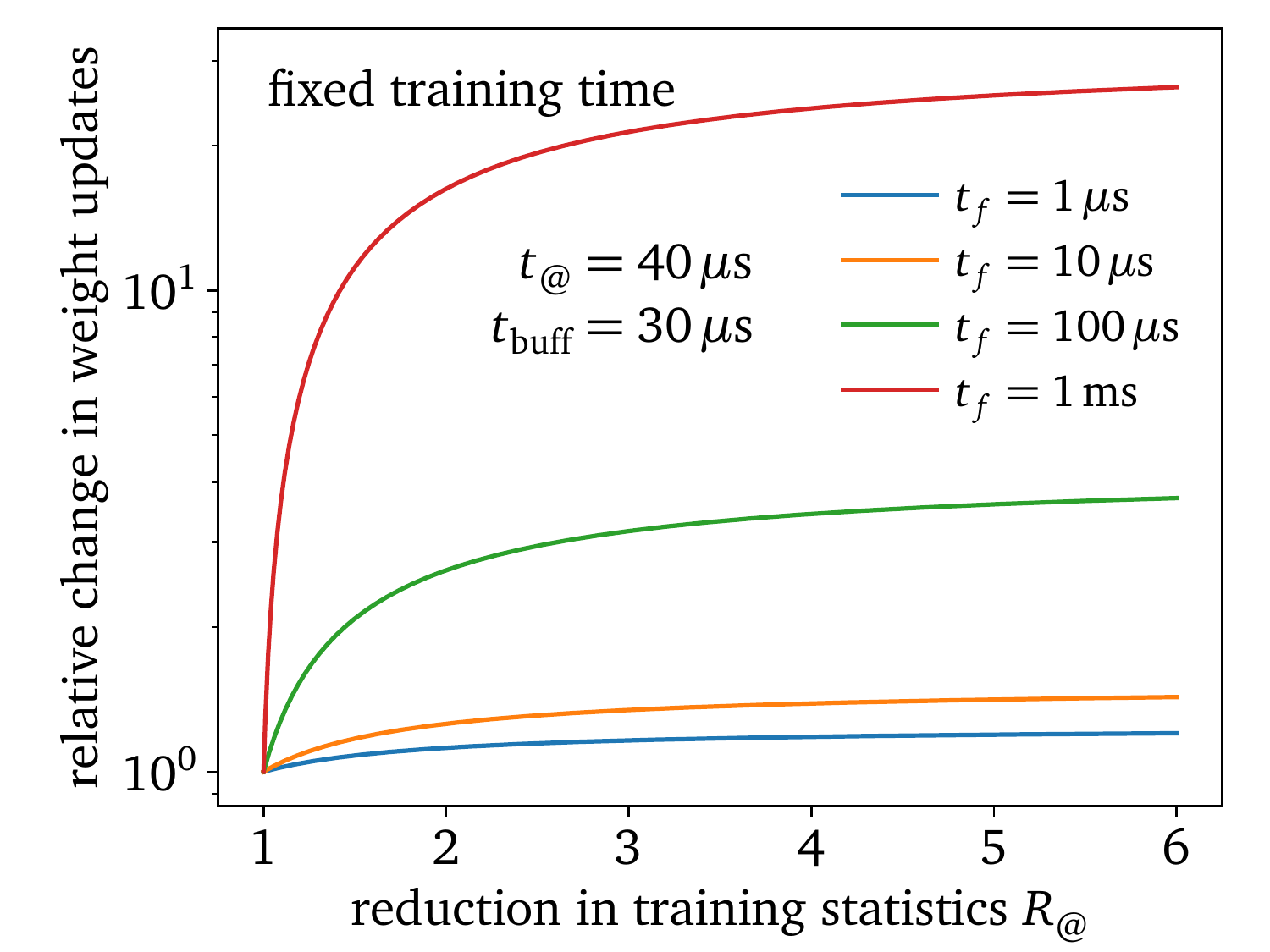}
  \includegraphics[width=0.495\textwidth]{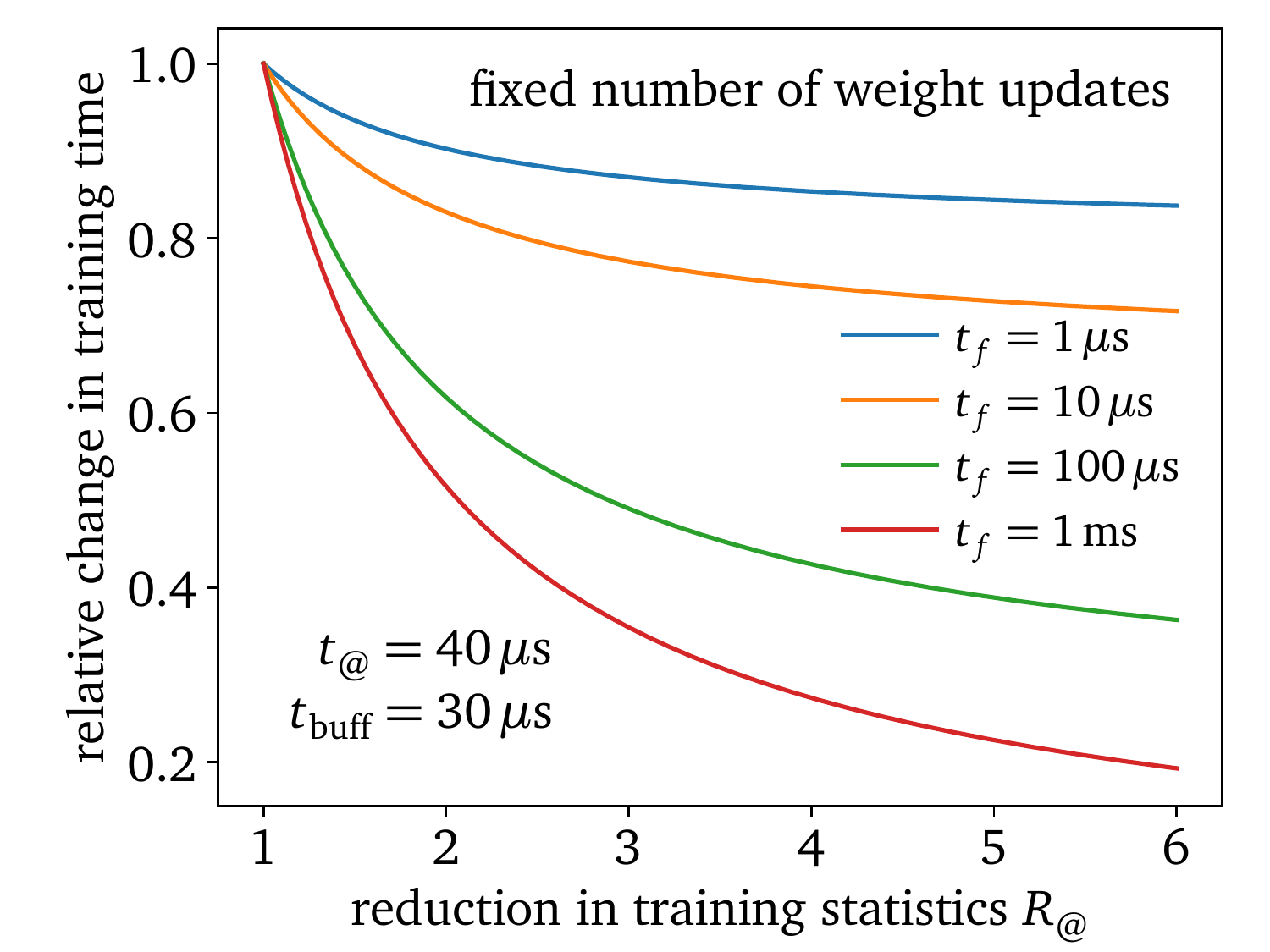}
  \caption{Hypothetical change in weight updates (left panel) and
    training time (right panel) as a function of the
    reduction in training statistics $R_\text{@}$ for integrands with different
    computational costs.}
  \label{fig:two-stage-statistics}
\end{figure}

If $t_f$ is the time it takes to evaluate the integrand, the time for
a weight update in online training will be $t_\text{@} + t_f$,
compared to $t_\text{buff}$ for the buffered training. The number of
weight updates is divided between the training modes,
\begin{align}
  n
  = n_\text{buff} + n_\text{@} 
  = \frac{T(1-r_\text{@})}{t_\text{buff}} + \frac{Tr_\text{@}}{t_\text{@}+t_f} \; ,
\end{align}
As a baseline we can look at the number of weight updates
$n_\text{base} = T/(t_\text{@}+t_f)$ for pure online training, giving
a increase factor in weight updates of
\begin{align} \label{eq:train_red}
  \frac{n}{n_\text{base}}
  = \left(1 - \frac{1}{R_\text{@}}\right) \frac{t_\text{@} + t_f}{t_\text{buff}} + \frac{1}{R_\text{@}}
  \qquad \mwith \qquad R_\text{@}=\frac{n_\text{base}}{n_\text{@}}=\frac{1}{r_\text{@}}\;,
\end{align}
in terms of the reduction factor in training statistics $R_\text{@}$
which coincides with the inverse of the relative training time
$r_\text{@}$. The left panel of Fig.~\ref{fig:two-stage-statistics}
shows the increase factor in weight updates for integrands with
different computational cost $t_f$; $t_\text{@}$ and $t_\text{buff}$
are extracted from a test run on a CPU. In a similar fashion, we can
also fix the number of weight updates $n$ and instead compare the
reduction factor in training time $T/T_\text{base}$ depending on
$R_\text{@}$, which is shown in the right panel of
Fig.~\ref{fig:two-stage-statistics}.

%------------------------------------------------------
%------------------------------------------------------

%%%%%%%%%%%%%%%%%%%%%%%%%%%%%%%%%%%%%%%%%%%%%%%%%%%
\subsubsection*{Variance-weighted training}

Stratified sampling~\cite{Press:1989vk} 
minimize the variances discussed in Sec.~\ref{sec:classic_err}, but it can
also improve the network training. We use the variance
from Eq.\eqref{eq:stratified-sampling} to sample more events in poor
channels and weight them accordingly in the loss function.  This
forces the network to focus on improving these channels, which should
ultimately lead to a better convergence of the network.  Such a
variance-weighted training can be easily combined with both, online
and buffered training.  To stabilize the training when using
variance-weighted channel sampling, we fix a small fraction of events
to be uniformly distributed across all channels.  This guarantees that
no channel is empty during training, which would otherwise lead to an
error.  This is in contrast to integration and pure sampling, where
the algorithm is encouraged to ignore channels with vanishing
contributions.

%%%%%%%%%%%%%%%%%%%%%%%%%%%%%%%%%%%%%%%%%%%%%%%%%%%
\subsection{Trainable rotations}

The INN employed in our study is based on a bipartite
architecture~\cite{coupling1, coupling2} and requires permutations in
the order of the coordinates between the coupling blocks to learn all
correlations.  The simplest implementation is an exchange of the
bipartite sets~\cite{coupling1, coupling2}. It ensures that
correlations between the variables can be learned stacking a few
coupling blocks. Shuffling the elements of the two sets with each
other is more efficient, but comes with a small probability that some
elements are never modified.
%In practice, it is primarily used with the exchange permutation.
Another solution is a deterministic set of permutations based on a
logarithmic decomposition of the integral
dimension~\cite{Gao:2020vdv}.  It ensures that every pair of elements
appears in different bipartite sets at least once. This relates the
number of required coupling layers to the dimensionality of the
integrand, and is particularly efficient for integrals of dimension
$d=2^k$.

For an integration over $\mathbb{R}^d$ we can generalize these
permutations to rotations described by $SO(d)$. Introduced in the
context of image generation, a randomly initialized but fixed $SO(d)$
matrix (soft permutation~\cite{cinn}) allows for mixing of color
channel information~\cite{glow,cinn}. A trainable implementation~\cite{glow} first adjusts all $d^2$ parameters and then projects the
trained matrix back onto $SO(d)$. This implementation as a independent
$d\times d$ matrix with a subsequent projection is not efficient.

%%%%%%%%%%%%%%%%%%%%%%%%%%%%%%%%%%%%%%%%%%%%%%%%%%%
\subsubsection*{Generalized Euler angles}

We construct a trainable soft permutation that only optimizes the
relevant degrees of freedom.  The elements of $SO(d)$ are described by
a $d (d-1) / 2$-dimensional Lie algebra and can be parametrized by
$D=d (d-1) / 2$ real parameters, interpreted as angles.  The common
parametrization of rotations in $\mathbb{R}^3$ are the Euler
angles~\cite{EulerAngles}. They can be generalized to
$\mathbb{R}^d$~\cite{doi:10.1063/1.1666011}. To efficiently construct
our rotation matrix $R$, we start with an orthonormal basis
$\vec{a}_i$, connected to the standard basis $\vec{e}_i$ by
\begin{align}
  \vec{a}_k=\sum_{i=1}^d \vec{e}_{i} R_{ik}
  \quad \leftrightarrow \quad
  \vec{e}_i= \sum_{i=1}^d R_{ki} \vec{a}_i\; .
  \label{eq:rotation.def}
\end{align}
To properly construct the corresponding rotation matrix we proceed
iteratively:
\begin{enumerate}
\item Define one direction with the unit-vector $\vec{a}_d$ in terms of
  $d-1$ angles $\vartheta^{(d)}_i$;
\item Construct an orthonormal basis $\{\vec{b}^{(d)}_i\}$, which
  contains $\vec{a}_d$ as last basis vector;
\item Fix next direction
  %within
  %$\{\vec{b}^{(d)}_1,\dots,\vec{b}^{(d)}_{d-1}\}$
  %by defining a
  %unit-vector
  $\vec{a}_{d-1}$ in terms of $d-2$ angles
  $\vartheta^{(d-1)}_i$ and construct new basis
  $\{\vec{b}^{(d-1)}_i\}$;
\item Iterate until the basis $\{\vec{a}_i\}$ determines $R$.
\end{enumerate}
For the three steps of this algorithms we provide the details below.

%%%%%%%%%%%%%%%%%%%%%%%%%%%%%%%%%%%%%%%%%%%%%%%%%%%
\paragraph{1. Definition of unit-vector $\vec{a}_d$}

We start by defining the unit-vector $\vec{a}_d$ in terms of $d-1$ angles $\vartheta^{(d)}_i$  or $d$-dimensional spherical coordinates,
\begin{align}
  \vec{a}_d =& \sin{\vartheta^{(d)}_{1}}\vec{e}_1 + \notag \\
    \phantom{=}& \cos{\vartheta^{(d)}_1} \sin{\vartheta^{(d)}_2}\vec{e}_2+ \notag \\
    \phantom{=}&\hspace{1cm}\vdots \hspace{2cm}\ddots \notag \\
    \phantom{=}&\cos{\vartheta^{(d)}_1}\dots\, \cos{\vartheta^{(d)}_{d-2}}\sin{\vartheta^{(d)}_{d-1}}\vec{e}_{d-1}+
    \label{eq:vec.a.d} \\    
    \phantom{=}&\cos{\vartheta^{(d)}_1}\dots\, \cos{\vartheta^{(d)}_{d-2}}\cos{\vartheta^{(d)}_{d-1}}\vec{e}_d
    = \sum_{i=1}^d\,\vec{e}_i \frac{\sin{\vartheta^{(d)}_{i}}}{\cos{\vartheta^{(d)}_{i}}}\prod_{j=1}^i \cos{\vartheta^{(d)}_{j}} \qquad \text{with} \quad \sin{\vartheta^{(d)}_d}=1\; .  \notag 
\end{align}
While
%$\sin{\vartheta^{(d)}_k}$ can be positive or negative, the
$\cos{\vartheta^{(d)}_k}$ are assumed to be positive,
$\cos{\vartheta^{(d)}_{d-1}}$ can have either sign. 
%as $\sin{\vartheta^{(d)}_d}=1$ is fixed.
%This construction
%determines the domain of the angles to
%%
%\begin{align}
%-\frac{\pi}{2} &\leq \vartheta^{(d)}_k \leq \frac{\pi}{2} \quad \text{for} \quad k=1, 2, \dots, d-2\,,     \notag\\
%-\pi &\leq \vartheta^{(d)}_{d-1} < \pi\;.
%\label{eq:angle.domain}
%\end{align}
%%
%------------------------------------------------------
\begin{figure}[t]
  \centering
  \includegraphics[width=0.40\textwidth]{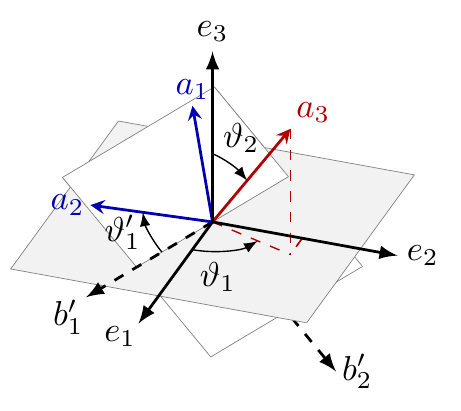}
  \caption{Exemplary rotation of the standard 3-dimensional basis
    $\vec{e}_i$ into another orthonormal basis $\vec{a}_i$
    parametrized by three Euler angles $\vartheta_i$. The different
    colors of the new basis $\vec{a}_i$ indicate there iterative
    construction.}
  \label{fig:euler_angle}
\end{figure}
%------------------------------------------------------

%%%%%%%%%%%%%%%%%%%%%%%%%%%%%%%%%%%%%%%%%%%%%%%%%%%
\paragraph{2. Orthonormal basis $\vec{b}_i^{(d)}$}

To construct an orthonormal basis which contains $\vec{a}_d$ as one of
its basis vectors, we can define the new basis 
\begin{align}
\vec{b}^{(d)}_k&=\left(\prod_{j=1}^{k-1}\cos{\vartheta^{(d)}_j}\right)^{-1}\frac{\partial\vec{a}_d}{\partial \vartheta^{(d)}_k}
\qquad (k=1,...,d-1) 
\qquad \text{and} \qquad 
\vec{b}^{(d)}_d=\vec{a}_d\;.
\label{eq:basis.vectors}
\end{align}
This definition fulfills the orthogonality and normalization
condition~\cite{doi:10.1063/1.1666011},
$\vec{b}^{(d)}_i\cdot\vec{b}^{(d)}_k=\delta_{ik}$.  The rotation into
this basis is given by
\begin{align}
\vec{b}^{(d)}_k =  \sum_i \vec{e}_i A^{(d)}_{ik} 
\qquad \text{with} \qquad 
A^{(d)} = \left(
\begin{array}{ccc|c}
     \tikznode{a}{}&  &  \text{IV} & \\
     \tikznode{c}{}&  \text{I} & & \text{II} \\
    \text{III} &  & \tikznode{d}{}& \tikznode{b}{}\\
\end{array}
\right)
    \label{eq:angle.to.matrix}
\begin{tikzpicture}[remember picture, overlay,shorten >=1pt,shorten <=1pt]
  \draw ($(a)+(0em, 1em)$) -- ($(b)+(-1em, 1em)$);
  \draw ($(c)+(-0.5em, 1em)$) -- ($(d)+(1em, -0.5em)$);
\end{tikzpicture} \; .
\end{align}
In the regions we have
\begin{align}
  &\text{I} \quad
  &A^{(d)}_{ii} &= \cos{\vartheta^{(d)}_i} \qquad
  &\text{for \quad} i &= 1~...~d-1 \notag \\
  &\text{II}
  &A^{(d)}_{id} &=\frac{\sin{\vartheta^{(d)}_i}}{\cos{\vartheta^{(d)}_i}}\prod_{j=1}^i \cos{\vartheta^{(d)}_j}
  &\text{for \quad} i &= 1~...~d \notag \\
  &\text{III}
  &A^{(d)}_{ik} &= -\frac{\sin{\vartheta^{(d)}_i}\sin{\vartheta^{(d)}_k}}{\cos{\vartheta^{(d)}_i}\cos{\vartheta^{(d)}_k}}\prod_{j=k}^i\cos{\vartheta^{(d)}_j}
  &\text{for \quad} i &> k  \notag \\
  &\text{IV}
  &A^{(d)}_{ik} &= 0  &\text{for \quad} i &< k < d\;. 
\label{eq:matrix.regions}
\end{align}
For $\vartheta^{(d)}_1 = \dots=\vartheta^{(d)}_{d-1}=0$ this gives
$A^{(d)}_{ik} = \delta^{(d)}_{ik}$, so the transformation is
continuously connected to the identity and $\det A^{(d)} = 1$.

%%%%%%%%%%%%%%%%%%%%%%%%%%%%%%%%%%%%%%%%%%%%%%%%%%%
\paragraph{3. Subsequent basis vectors $\vec{b}^{(l)}_i$}

Next, we consider $\vec{a}_{d-1}$ in
$\{\vec{b}^{(d)}_1,\dots,\vec{b}^{(d)}_{d-1}\}$. As in
Eq.\eqref{eq:vec.a.d} we define this vector in terms of 
a new set of $d-2$ angles
$\vartheta^{(d-1)}_i$ and construct an orthonormal
basis $b^{(d-1)}_i$ which contains $\vec{a}_{d-1}$.
%Similarly, we can proceed for the vectors $\vec{a}_{d-2},\dots,\vec{a}_{1}$.
Similarly, we can proceed for the remaining vectors $\vec{a}_{d-2},\dots,\vec{a}_{2}$.
A general step $l$ in this iterative basis transformation leads from a basis
\begin{align}
\vec{b}^{(l+1)}_1,\dots,\vec{b}^{(l+1)}_{l}
\quad \text{and} \quad \vec{b}^{(l+1)}_{l+1}=\vec{a}_{l+1},\dots,\vec{b}^{(l+1)}_{d}= \vec{a}_{d}
\end{align}
to the basis
\begin{align}
\vec{b}^{(l)}_1,\dots,\vec{b}^{(l)}_{l-1}
\quad \text{and} \quad
\vec{b}^{(l)}_{l}=\vec{a}_{l},\dots,\vec{b}^{(l)}_{d}=\vec{a}_{d}\;,
\end{align}
where we have defined $\vec{a}_{l}$ by
\begin{align}
    \vec{a}_{l}=\sum_{i=1}^{l}\,\vec{b}^{(l+1)}_i \frac{\sin{\vartheta^{(l)}_{i}}}{\cos{\vartheta^{(l)}_{i}}}\prod_{j=1}^i \cos{\vartheta^{(l)}_{j}} \qquad \text{with} \quad \sin{\vartheta^{(l)}_{l}}=1\;.
\end{align}
The corresponding transformation into this basis is defined by
\begin{align}
\vec{b}^{(l)}_k = \sum_i \vec{b}^{(l+1)}_i B^{(l)}_{ik}\;, 
\end{align}
where $B^{(l)}$ is the matrix
\begin{align}
B^{(l)} = \left( \begin{array}{c|c}
    A^{(l)} &  0\\
    \hline
     0& \mathbb{1}^{(d-l)}
\end{array}\right)\;,
\label{eq:basis.iterative}
\end{align}
and where $A^{(l)}$ is defined following
Eq.\eqref{eq:angle.to.matrix}--\eqref{eq:matrix.regions} with angles
$\vartheta^{(l)}_i$.

%%%%%%%%%%%%%%%%%%%%%%%%%%%%%%%%%%%%%%%%%%%%%%%%%%%
\paragraph{4. Iteration}

At the end of the procedure, we have an orthonormal basis defined by
$d(d-1)/2$ angles which yields the desired $d$-dimensional rotation
matrix
\begin{align}
R = B^{(d)}B^{(d-1)}\dots B^{(3)}B^{(2)}\;,
\label{eq:final.rotation}
\end{align}
as introduced in Eq.~\eqref{eq:rotation.def}. An illustration of this
procedure in 3 dimensions is shown in Fig.~\ref{fig:euler_angle}:
First the new basis vector $\vec{a}_3$ is defined by rotations with
angles $\vartheta_1$ and $\vartheta_2$. Using
Eq.\eqref{eq:basis.vectors} we can construct the new basis
$\vec{b}'_1, \vec{b}'_2,\vec{a}_3$. Afterwards, we define the vector
$\vec{a_2}$ in this basis by a rotation with angle $\vartheta_1'$
which also fixes the last basis vector $\vec{b}''_1=\vec{a_1}$ and
determines the procedure.  We implement these angles $\vartheta^{l}_i$
as trainable parameters.

%%%%%%%%%%%%%%%%%%%%%%%%%%%%%%%%%%%%%%%%%%%%%%%%%%%
\section{Toy examples}
\label{sec:toys}

To check and benchmark the various ideas presented in
Sec.~\ref{sec:madnis} we first consider two parametric toy models, a
1-dimensional camel back, and a 2-dimensional crossed ring. The camel back
allows us to illustrate how to train channel weights to optimize a
simple bi-modal integration. The crossed ring we use to illustrate how
learnable local channel weights can be combined with an INN-importance
sampling successfully. A discussion of the trainable rotations and the
mixed online and buffered training will only become relevant for the
LHC example in Sec.~\ref{sec:lhc}.

%%%%%%%%%%%%%%%%%%%%%%%%%%%%%%%%%%%%%%%%%%%%%%%%%%%
\subsection{One-dimensional camel back}
\label{sec:toy_camel}

Our first toy example just illustrates how the neural integrator
learns channel weights for pre-defined channels. We define a
normalized 1-dimensional camel back or Gaussian mixture,
\begin{align}
  f_\text{GM}(x)&=
  \frac{a_1}{\sqrt{2\pi}\sigma_1}\,\exp \left[-\frac{(x-\mu_1)^2}{2\sigma_1^2}\right]
  \; + \; 
  \frac{1-a_1}{\sqrt{2\pi}\sigma_2}\,\exp \left[-\frac{(x-\mu_2)^2}{2\sigma_2^2}\right] \notag \\
  &\text{with} \qquad
  \mu_1=2 \qquad \sigma_1=0.5 \qquad 
  \mu_2=5 \qquad \sigma_2=0.1 \qquad
  a_1=0.35 \; .
  \label{eq:def_gm}
\end{align}
If we want to describe each of the hardly overlapping Gaussians by an
integration channel we need reasonable mappings which should not be
identical to the Gaussian integrand.  We choose a Cauchy or Breit-Wigner
mapping~\cite{Plehn:2015dqa}
\begin{align}
  x &= \gbar_i(y)= \mu_i + \sqrt{2}\sigma_i \,\tan\!\left[\pi \left( y-{\frac {1}{2}} \right)\right] \notag \\
  g_i(x) &= \frac{1}{\pi}
  \frac{\sqrt{2} \sigma_i}{(x-\mu_i)^2+2\sigma^2_i} \; .
%    \quad \text{with} \quad \left\vert\frac{\partial G_i(x)}{x}\right\vert=g_i(x)\,,
  \label{eq:bw_mapping1}
\end{align}
With these definitions the widths of the Gaussian and the Breit-Wigner
functions are roughly the same.
The multi-channel form of Eq.\eqref{eq:multi-channel-mg} using a known mapping is
\begin{align}
    I[f_\text{GM}] = \int_{-\infty}^{\infty}\d x\,f_\text{GM}(x)
    &=\sum_{i=1}^2 \int_{-\infty}^{\infty}\d x\,\alpha_i(x|\theta)\,f_\text{GM}(x) \notag \\
    &=\sum_{i=1}^2 \int_0^1 \d y \,\left.\alpha_i(x|\theta)\,\frac{f_\text{GM}(x)}{g_i(x)}\right\vert_{x=\gbar_i(y)}\,.
    \label{eq:multi-channel-1d}
\end{align}
%

%------------------------------------------------------
\begin{figure}[t]
  \includegraphics[width=0.495\textwidth]{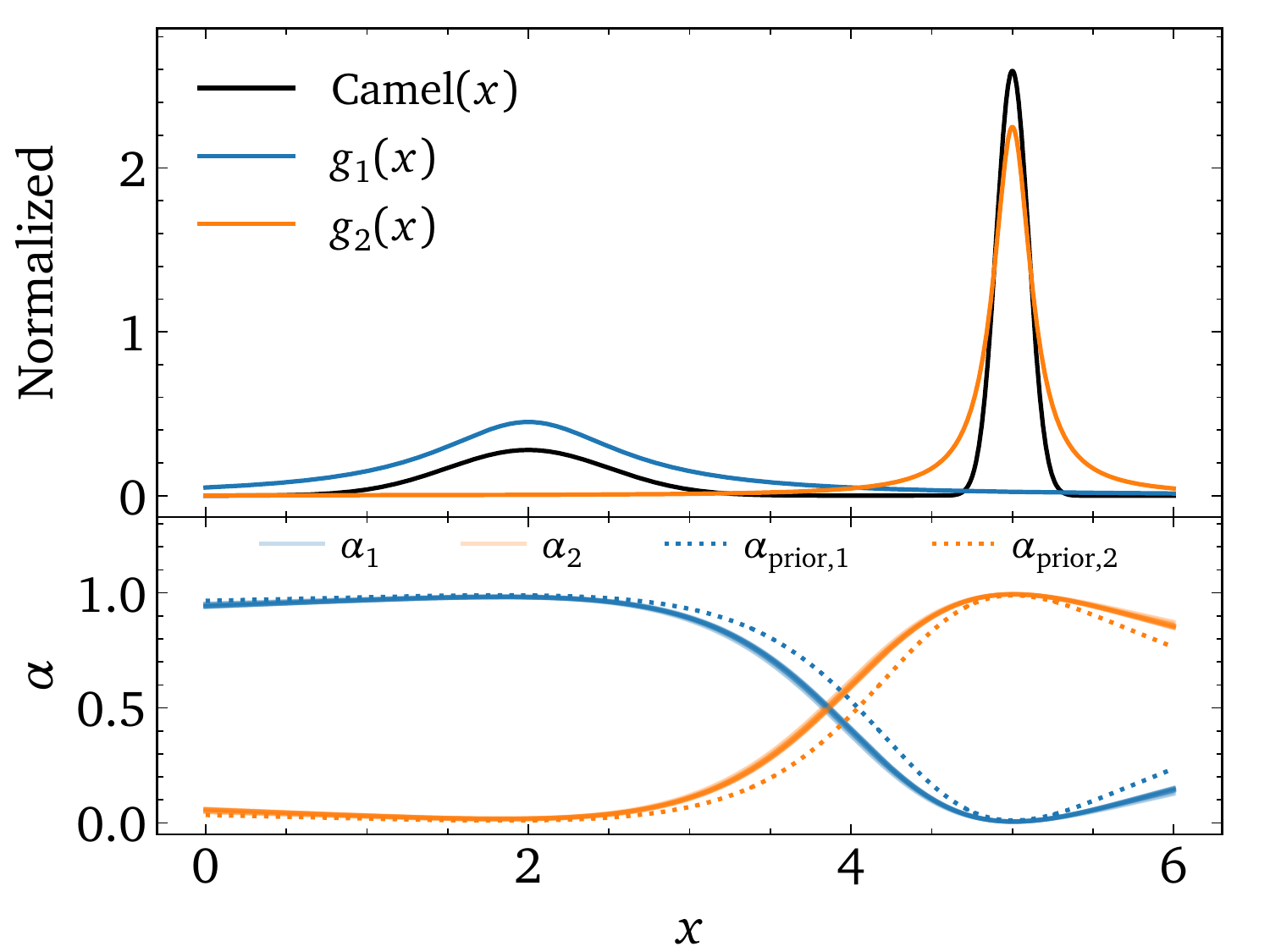}
  \includegraphics[width=0.495\textwidth]{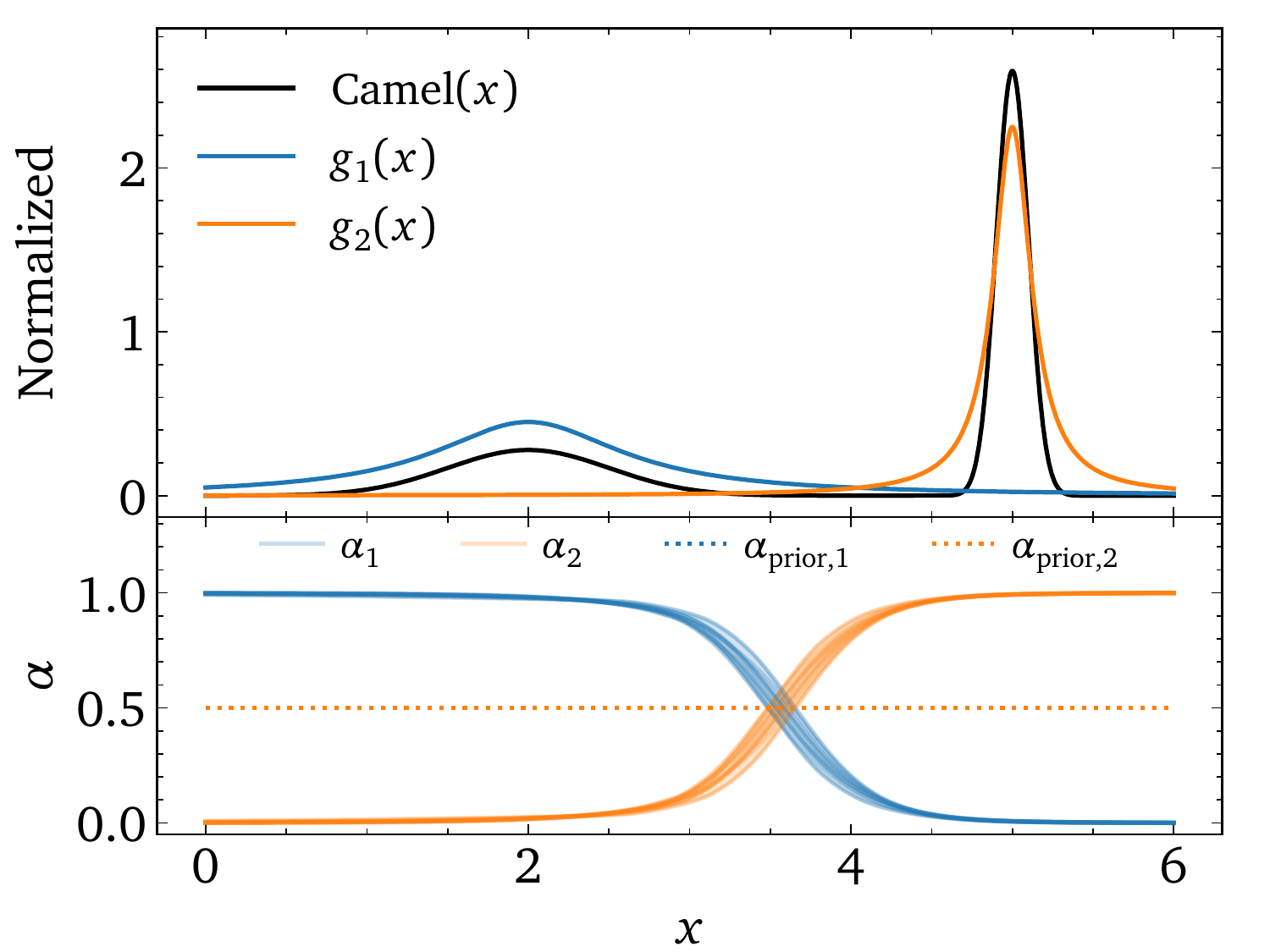}
  \caption{Learned weights for the camel back function for ten different trainings. We train NN-weights starting from a near-optimal (left) or flat (right) prior. The prior weights are illustrated as dotted lines.}
  \label{fig:alphas_camel}
\end{figure}
%------------------------------------------------------

As mentioned above, the camel back toy model only serves as an
illustration that a simple regression network can learn the channel
weights $\alpha_i(x|\theta)$, as described in
Sec.~\ref{sec:madnis_nms}. We provide the hyperparameters for this
simple network to the left in Tab.~\ref{tab:camel}. The only
noteworthy setting is that the loss function of the network is defined
as the variance of the integral given in
Eq.\eqref{eq:var_channel}. The amount of training data is
comparably large, to give the network a chance to learn the channel
weights with enough precision and to allow for a test of the stability
using an ensemble of networks.

To the right in Tab.~\ref{tab:camel} we compare the error on the
integral just using uniform, constant weights $\alpha_i$, the (nearly)
optimal choice $\alpha_i(x) = g_i(x)/\sum_i g_i(x)$, and local channel
weights $\alpha_i(x|\theta)$ optimizing the actual variance. We see
that the optimal and the trained weights provide the same results,
significantly improving over the naive choice.

%------------------------------------------------------
\begin{table}[b!]
\centering
\begin{small} \begin{tabular}{cc}
\begin{tabular}[t]{l|l}
\toprule
Parameter & Value \\
\midrule
Loss function & variance \\
Learning rate & 0.001 \\
LR schedule & inverse time decay \\
Decay rate & 0.01 \\
Batch size & 128 \\
Epochs & 20 \\
Batches per Epoch & 100 \\
Number of layers & 3 \\
Hidden nodes & 16 \\
Activation function & leaky ReLU \\
\bottomrule
\end{tabular} 
\hspace*{1cm}
&\begin{tabular}[t]{ll|l}
\toprule
Function & $\alpha_i(x)$ & Rel.~Error [\%]  \\
\midrule
Camel back      & Uniform         & $2.553 \pm 0.017$\\
                & Optimal         & $0.769 \pm 0.006$\\
                & NN (flat prior) & $0.770 \pm 0.005$  \\
                & NN (opt.~prior) & $0.767 \pm 0.006$\\
\midrule
Cut camel back  & Uniform         & $3.412 \pm 0.048$\\
                & Optimal         & $1.031 \pm 0.006$\\
                & NN (flat prior) & $1.032 \pm 0.017$\\
                & NN (opt.~prior) & $1.030 \pm 0.009$\\
\midrule
\multicolumn{3}{c}{Based on  $10^4$ events}
\end{tabular}
\end{tabular} \end{small}
\caption{Left: hyperparameters of the multi-channel weight network for
  the 1-dimensional camel back. Right: relative errors of the camel
  back integrals using the trained channel weights (means and standard
  deviations from ten runs).}
\label{tab:camel}
\end{table}
%------------------------------------------------------

In Fig.~\ref{fig:alphas_camel} we show the target function from
Eq.\eqref{eq:def_gm}, the two pre-defined channels $g_i(x)$, and, in
the lower panel, the learned channel weights $\alpha_i(x|\theta)$ and
their prior or starting points. For the left and right panels network
training starts from the near optimal $\alpha_i(x) \propto g_i(x)$ or
a flat prior $\alpha_i(x)=$~const. While the first version converges
on the same network weights for ten different trainings, the harder
task leads to a small variation in the training outcome. Nevertheless,
the two learned channel weights are essentially identical, with the
exception of slight deviations in the exponentially suppressed tails
of the two Gaussians. From Tab.~\ref{tab:camel} we know that these
deviations do not have any impact on the evaluation of the integral.

%%%%%%%%%%%%%%%%%%%%%%%%%%%%%%%%%%%%%%%%%%%%%%%%%%%
\subsubsection*{Camel back with cut}

%------------------------------------------------------
\begin{figure}[t]
  \includegraphics[width=0.495\textwidth]{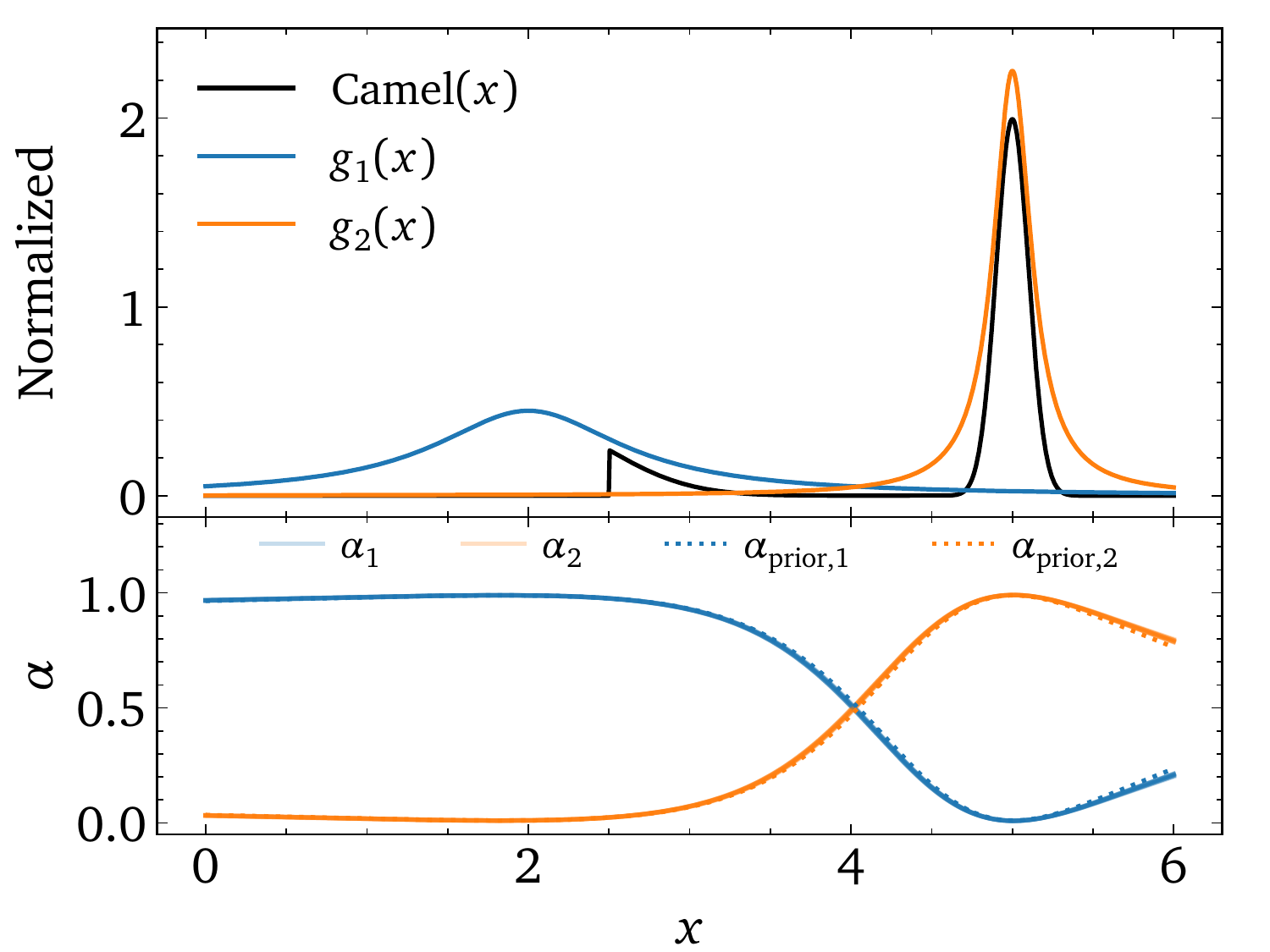}
  \includegraphics[width=0.495\textwidth]{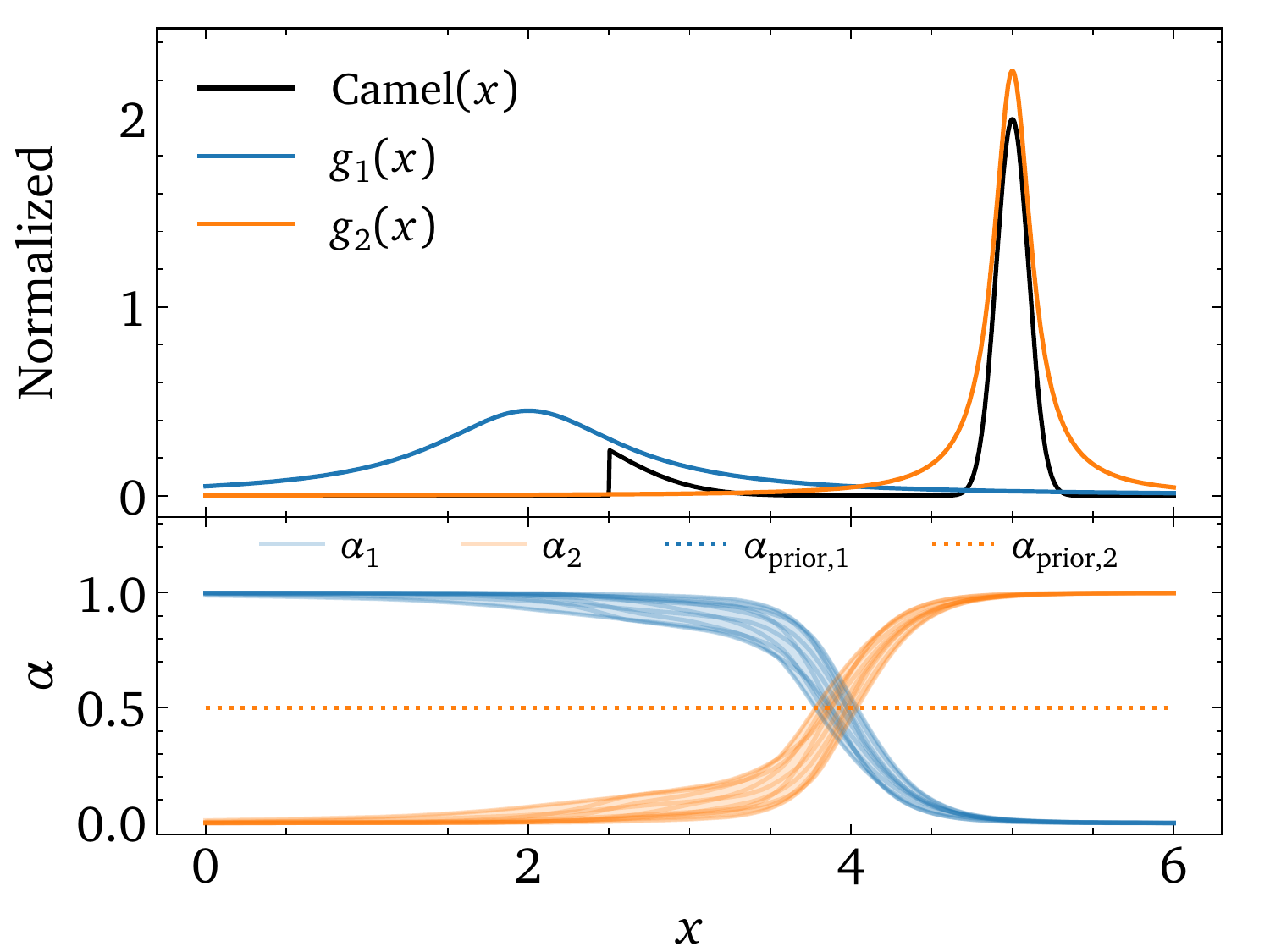}
  \caption{Learned weights for the cut camel back function for 
  ten different trainings. We train NN-weights starting from a near-optimal (left) or flat (right) prior. The prior weights are illustrated as dotted lines.}
  \label{fig:alphas_camelcut}
\end{figure}
%------------------------------------------------------

For the camel-back function in Eq.\eqref{eq:def_gm} our well-suited
choice of channels $g_i(x)$ in Eq.\eqref{eq:bw_mapping1} guarantees
that the learned channel weights converge to a reasonable and stable
solution. An obvious question is what happens with the trained weights
$\alpha_i(x|\theta)$ if the channels $g_i(x)$ are not perfect.  To
investigate the effect of a non-perfect shape of the channels on the
integration we consider a camel back with a cut in the left Gaussian
of Eq.\eqref{eq:def_gm},
\begin{align}
    f_\text{GM}(x) \; \to \; 
    \begin{cases}
      f_\text{GM}(x) & x\ge\mu_1+\sigma_1\\
      0 & x<\mu_1+\sigma_1
\end{cases}\; ,
\end{align}
where $\mu_1+\sigma_1=2.5$.  In Tab.~\ref{tab:camel} we see that for
all methods the integration becomes slightly harder and less
numerically reliable. The level of improvement for the network weights
remains the same as for the perfect camel back, confirming the power
of our NN-channel weights. Finally, in Fig.~\ref{fig:alphas_camelcut}
we also see that the modification of the integrand does not affect a
properly initialized training, but leads to a slightly larger spread
when we train the network from scratch. Such a behavior is expected
for any complication of the network task.

%%%%%%%%%%%%%%%%%%%%%%%%%%%%%%%%%%%%%%%%%%%%%%%%%%%
\subsection{Two-dimensional crossed ring}
\label{sec:toy_ring}

To show how the trained channel weights from Sec.~\ref{sec:madnis_nms}
and the neural importance sampling from Sec.~\ref{sec:madnis_buff}
work in combination, we choose a moderately challenging 2-dimensional toy model. 
It combines a closed Gaussian ring and a diagonal Gaussian line
\begin{align}
  f_\text{no-parking}(x)
  &= \frac{1}{2} \left[ f_\text{ring}(x) + f_\text{line}(x) \right] \notag \\
  f_\text{line}(x)
  &= N_1 \exp \left[-\frac{(\tilde{x}_1-\mu_1)^2}{2\sigma_1^2} \right]
    \;
    \exp \left[-\frac{(\tilde{x}_2-\mu_2)^2}{2\sigma_2^2} \right] \notag \\
  f_\text{ring}(x)
  &= N_2 \,\exp \left[-\frac{\left( \sqrt{x_1^2+x_2^2}-r_0 \right)^2}{2\sigma_0^2}\right] \notag \\
  \text{with}\quad
  r_0 &= 1 \qquad
  \sigma_0 = 0.05 \qquad
  \mu_1 = 0 \qquad 
  \sigma_1 = 3 \qquad
  \mu_2 = 0 \qquad 
  \sigma_2 = 0.05 \; ,
\end{align}
where $N_0$ and $N_1$ are chosen such that $f_\text{ring}(x)$ and
$f_\text{line}$ are both normalized to unity and $\tilde{x}_{1,2} =
(x_1 \mp x_2)/\sqrt{2}$.

%------------------------------------------------------
\begin{table}[b!]
\centering
\begin{small} \begin{tabular}{cc}
\begin{tabular}[t]{l|l}
\toprule
Parameter & Value \\
\midrule
Loss function & variance \\
Learning rate & 0.0005 (0.001) \\
LR schedule & inverse time decay \\
Decay rate & 0.02 \\
Batch size & 1024 \\
Epochs & 100 \\
Batches per Epoch & 500 \\
Coupling blocks & affine \\
Permutations & soft \\
Blocks & 6 \\
Subnet hidden nodes & 32 (16) \\
Subnet layers & 3 (2) \\
CWnet layers & 2 \\
CWnet hidden nodes & 16 \\
Activation function & leaky ReLU \\
\bottomrule
\end{tabular} 
\hspace*{1cm}
\begin{tabular}[t]{ll|l}
\toprule
Fig. & Analytic Mappings & Rel.~Error [\%] \\
\midrule
\ref{fig:ring-flat} & flat & $1.17 \pm 0.13$ \\
\ref{fig:ring-flat} & flat, flat & $0.71 \pm 0.15$ \\
\ref{fig:ring-flat} & flat, flat, flat & $0.50 \pm 0.15$ \\
\ref{fig:ring-rf} & ring, flat & $0.30 \pm 0.11$ \\
& ring, line & $0.14 \pm 0.06$ \\
& ring, line, flat & $0.29 \pm 0.14$ \\
\midrule
\multicolumn{3}{c}{Based on  $10^4$ events}
\end{tabular} 
\end{tabular} \end{small}
\caption{Left: hyperparameters of the INN and the channel weight
  network (CWnet) for the crossed ring. The numbers in parentheses
  indicate that a different setting was used for a ring
  mapping. Right: Relative integration errors for different numbers of
  channels and variations of analytic mappings. We show the means and
  standard deviations for ten independent trainings.}
\label{tab:rings-channels}
\end{table}
%------------------------------------------------------

%%%%%%%%%%%%%%%%%%%%%%%%%%%%%%%%%%%%%%%%%%%%%%%%%%%
\subsubsection*{Channel-mappings}

%------------------------------------------------------
\begin{figure}[b!]
  \centering
  \begin{small} \begin{tabular}{ccccc}
    \rotatebox[origin=c]{90}{1 channel}&
    \raisebox{-0.5\height}{\includegraphics[width=0.23\textwidth]{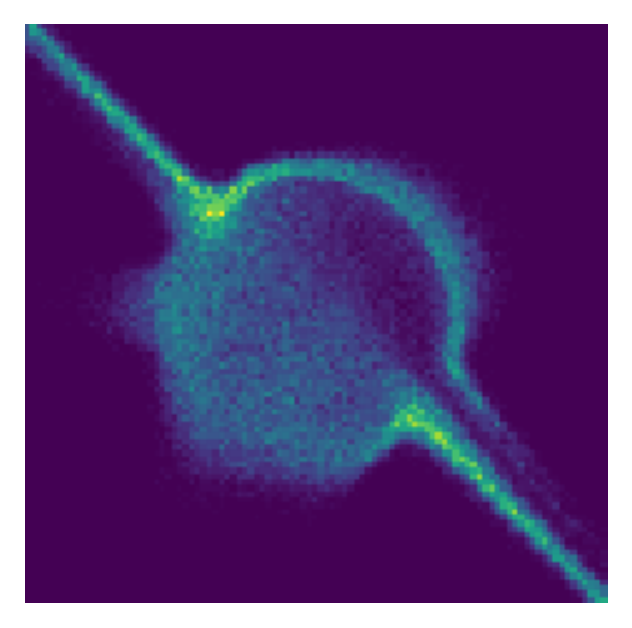}} &
    \raisebox{-0.5\height}{\includegraphics[width=0.23\textwidth]{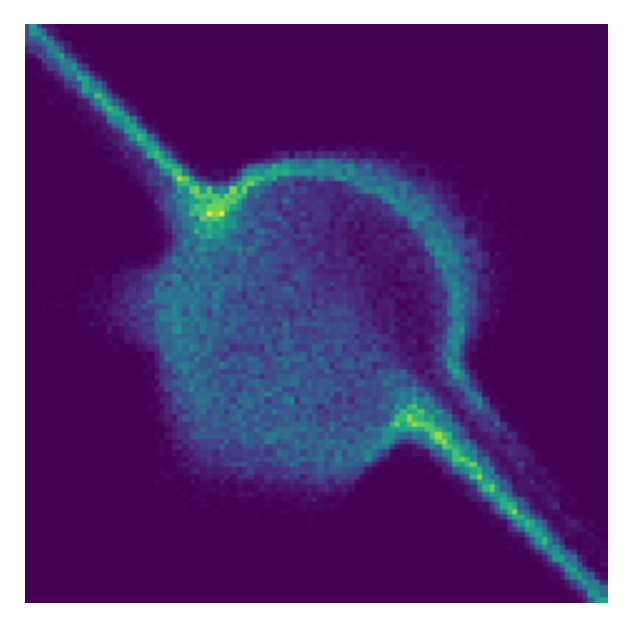}} & & \\
    \rotatebox[origin=c]{90}{2 channels}&
    \raisebox{-0.5\height}{\includegraphics[width=0.23\textwidth]{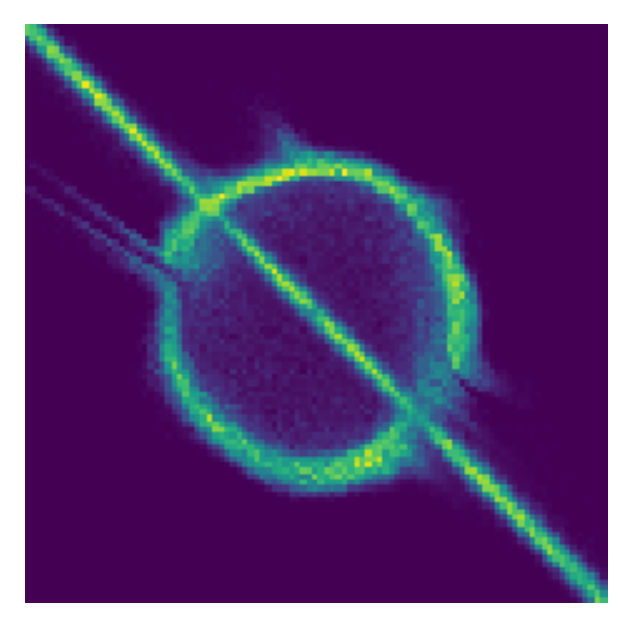}} &
    \raisebox{-0.5\height}{\includegraphics[width=0.23\textwidth]{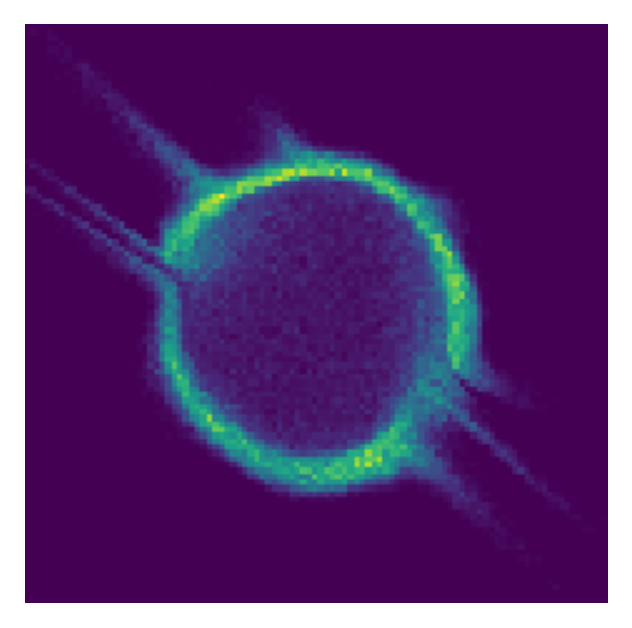}} &
    \raisebox{-0.5\height}{\includegraphics[width=0.23\textwidth]{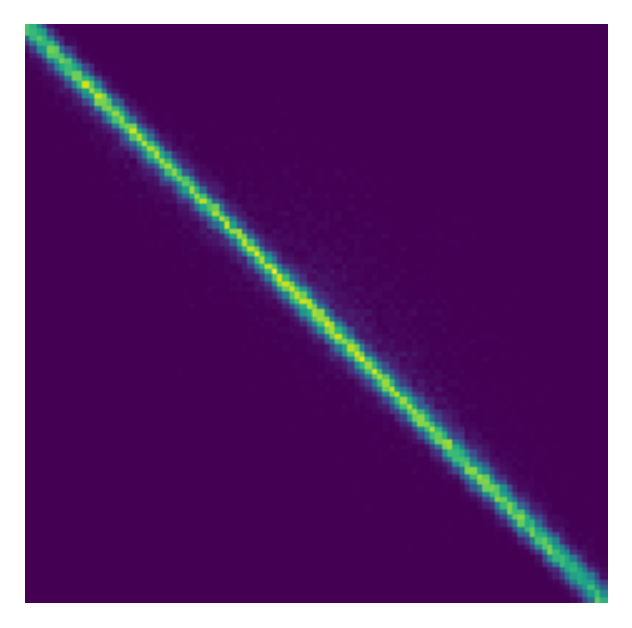}} & \\
    \rotatebox[origin=c]{90}{3 channels}&
    \raisebox{-0.5\height}{\includegraphics[width=0.23\textwidth]{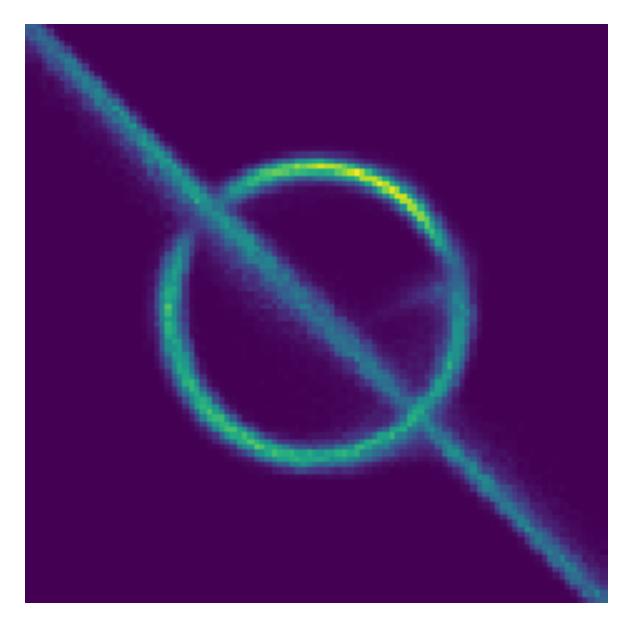}} &
    \raisebox{-0.5\height}{\includegraphics[width=0.23\textwidth]{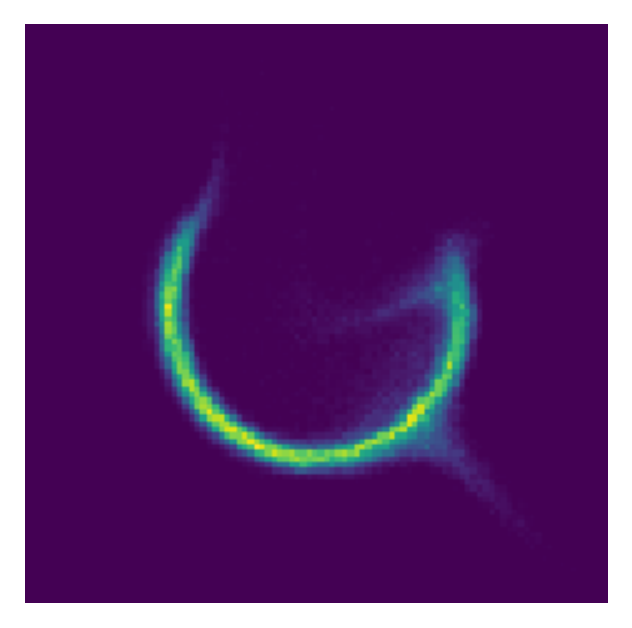}} &
    \raisebox{-0.5\height}{\includegraphics[width=0.23\textwidth]{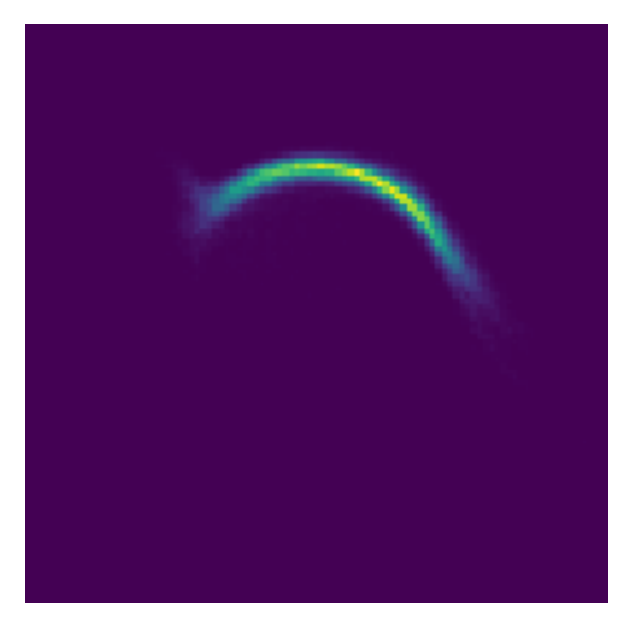}} &
    \raisebox{-0.5\height}{\includegraphics[width=0.23\textwidth]{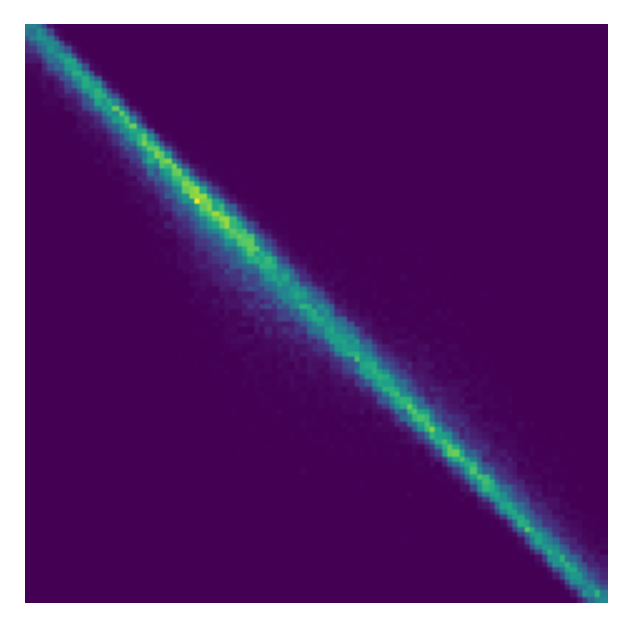}} \\
    & combined & channel 1 & channel 2 & channel 3
  \end{tabular} \end{small}
  \caption{Combined and channel-wise (the latter not weighted by channel weights) distributions learned by a one-,
    two- and three-channel integrator with flat mappings and a mode-specific prior. 
    Note that the splitting in the three-channel case is not unique and learned differently by the network for each run.}
  \label{fig:ring-flat}
\end{figure}
%------------------------------------------------------

To see how much additional analytic mappings help, we construct two
channels for the line and the ring contributing to our integral. We
start with the mapping for the Gaussian line which first aligns the
line with the $x_1$-axis by performing a first change of variables
$x\to y=G_{1}(x)$ as
\begin{align}
%   y_{1,2} = \frac{x_1 \pm x_2}{\sqrt{2}}
%   \qquad \Leftrightarrow \qquad 
  x_{1,2} = \frac{y_2 \pm y_1}{\sqrt{2}}
  \qquad \text{with} \qquad g_{1}(x)=\left|\frac{\partial G_{1}(x)}{\partial x}\right|=1\;.
\end{align}
As for the camel back, Eq.\eqref{eq:bw_mapping1}, we approximate the
Gaussian peak through a Breit-Wigner distribution using the variable
transformation $y \to z=G_2(y)$,
\begin{align}
  y_{1,2}&=\mu_{1,2} + \gamma_{1,2} \tan\left[ \pi\left(z_{1,2}-\frac{1}{2}\right)\right] 
  \qquad \text{with} \qquad 
  g_2(y)= \frac{1}{\pi^2} \prod_{j=1}^2
  \frac{\gamma_j}{\gamma_j^2+ (y_j-\mu_j)^2}\;.
\end{align}
The combined channel density is then $g_\text{line}(x)= 1 \times
g_2(G_1(x))$. The Gaussian ring requires a mapping $x\to (r,\theta)
=G_3(x)$ into polar coordinates
\begin{align}
  x_1 = r \cos \theta
  \qquad \text{and} \qquad 
  x_2 = r \sin \theta \; . %\qquad \text{with} \qquad g_3(x)=r\;.
\end{align}
Its Jacobian is $g_3(x) = r$. Again, we approximate the radial peak by
a Breit-Wigner through the variable transformation $(r,\theta) \to
z=G_4(r,\theta)$,
\begin{align}
  r&=r_0+\gamma_0\tan\left[\pi\left(\omega_0z_1-C_0\right)\right]\notag \\
  \theta&=2\pi z_2
  \qquad \text{with} \qquad 
  g_4(r)=\frac{1}{2\pi} \;
  \frac{1}{\omega_0\pi} \frac{\gamma_0}{\gamma_0^2 + (r-r_0)^2} \; ,
\end{align}
where $\pi C_0=\arctan ( r_0/\gamma_0 )$ and $\omega_0 = (1+2C_0)/2$ ensures $r>0$ and thus
$g_\text{ring}(x)= r\,g_4(G_3(x))$.
 We either augment or replace these mappings with a neural channel
mapping $G_i(x|\varphi)$. To challenge our INN when paired with the
above mappings we pick wide channel widths,
\begin{align}
    \gamma_{0,1,2} =\sqrt{40}\,\sigma_{0,1,2} \; . 
%    \qqquad \gamma_1=\sqrt{40}\,\sigma_1
%    \qqquad \gamma_2=\sqrt{40}\,\sigma_2\;.
\end{align}

%%%%%%%%%%%%%%%%%%%%%%%%%%%%%%%%%%%%%%%%%%%%%%%%%%%
\subsubsection*{Results}

%------------------------------------------------------
\begin{figure}[t]
  \includegraphics[width=0.495\textwidth]{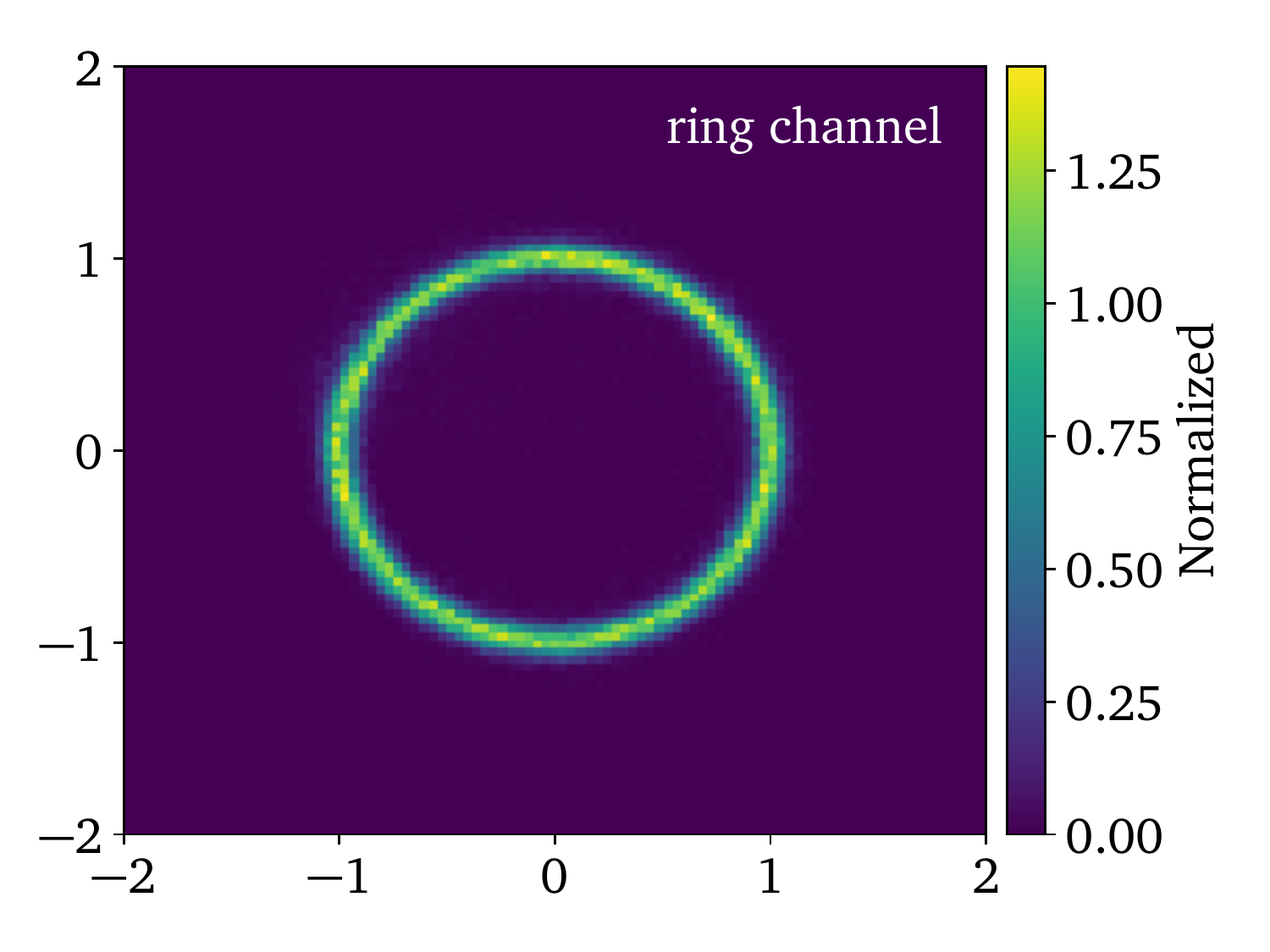}
  \includegraphics[width=0.495\textwidth]{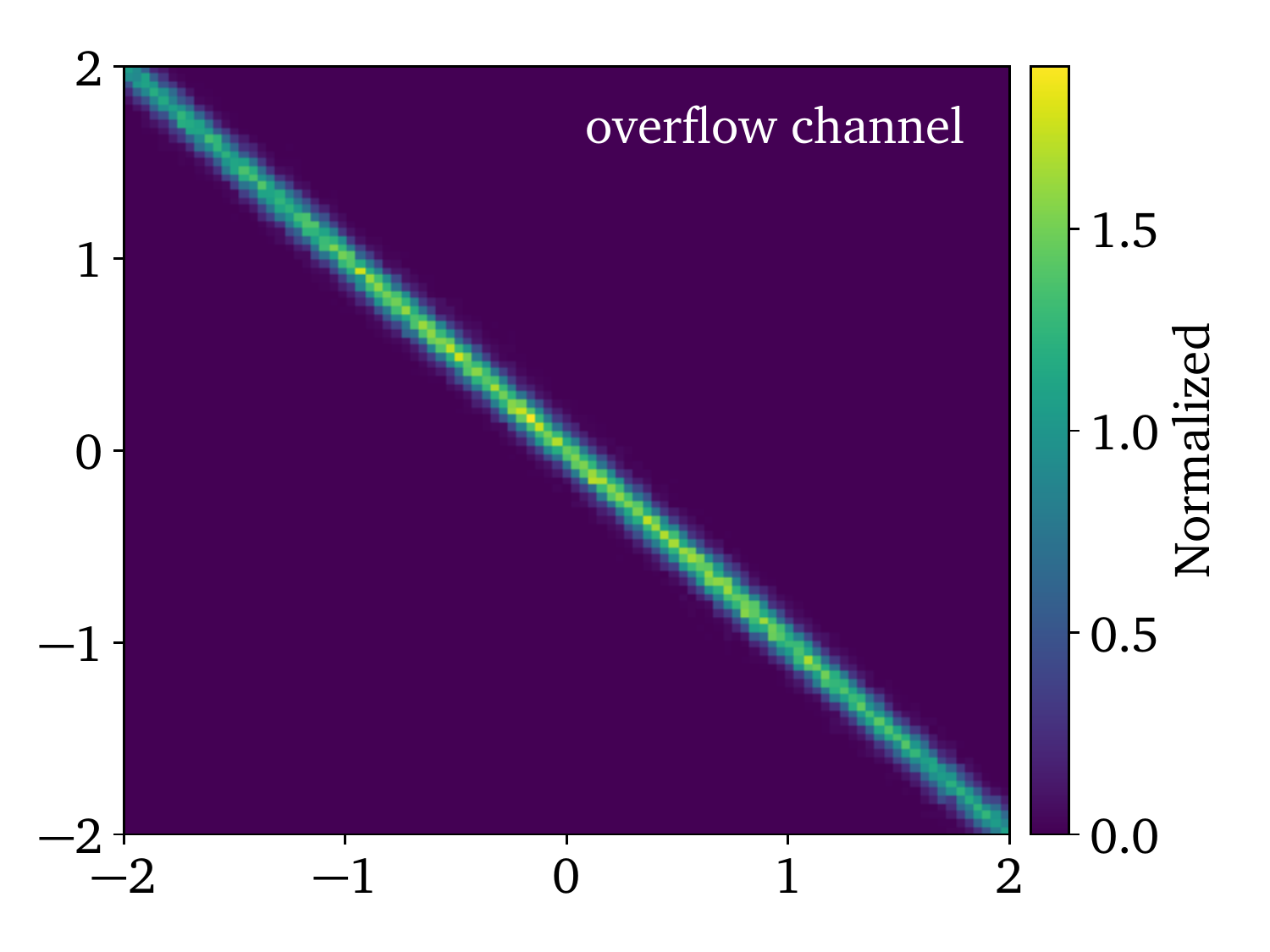}\\
  \includegraphics[width=0.495\textwidth]{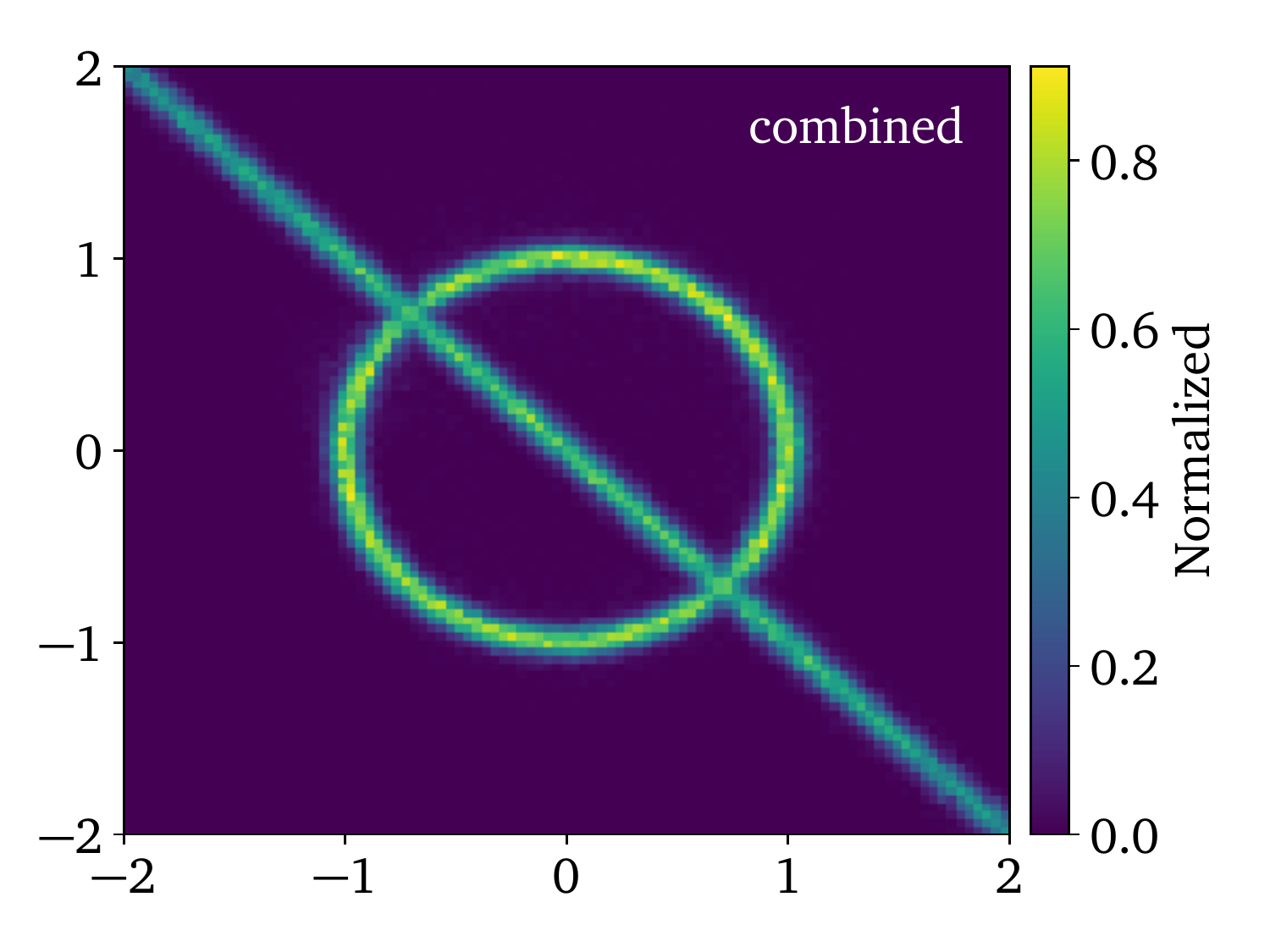}
  \includegraphics[width=0.495\textwidth]{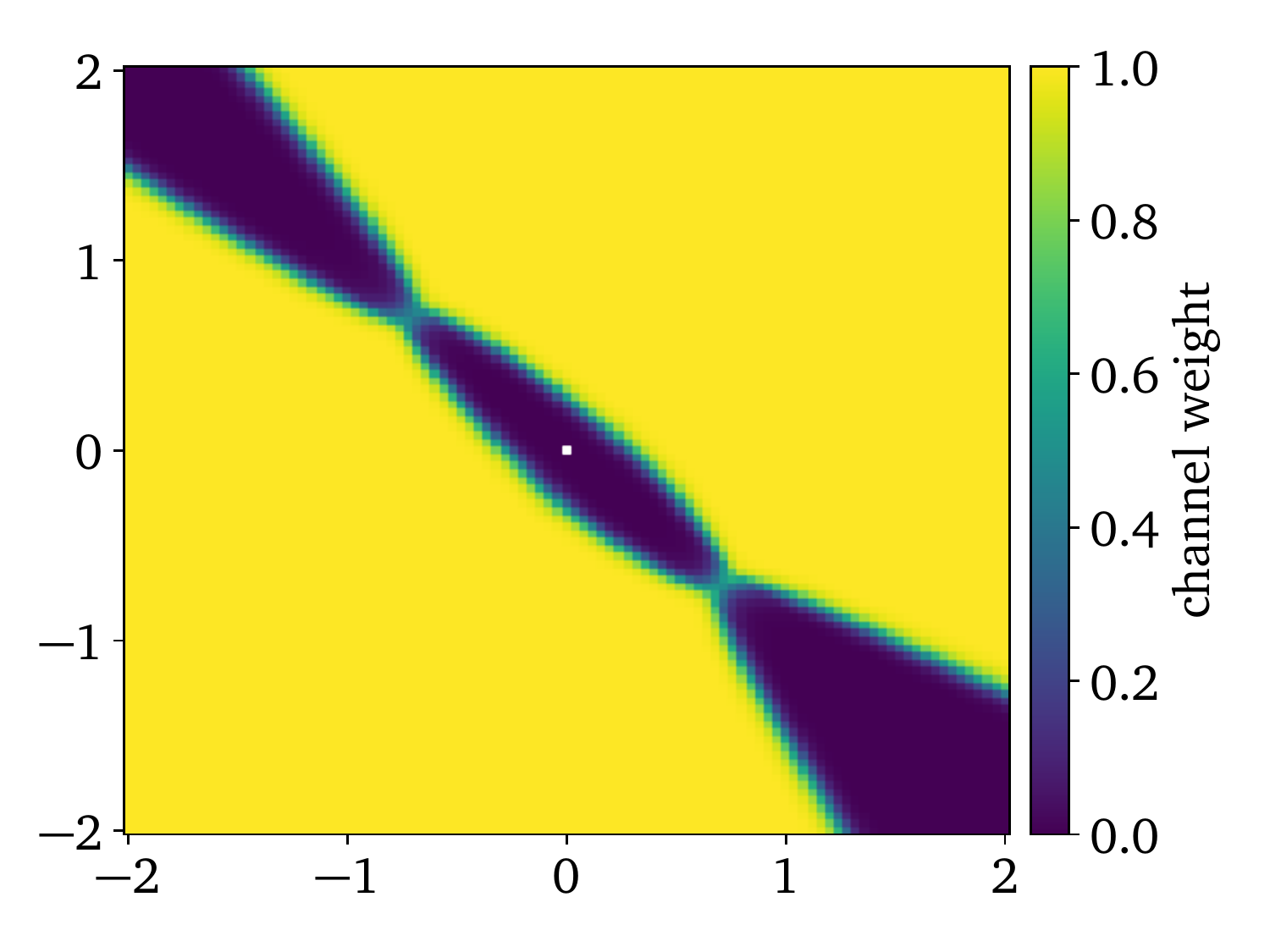}
  \caption{Distribution learned by a 2-channel integrator with a ring
    mapping and a flat mapping and flat prior. Upper: individual
    channels, not weighted by channel weights; Lower: combined
    distribution and channel weight of the ring channel.}
  \label{fig:ring-rf}
\end{figure}
%------------------------------------------------------

As a first check of our combined training of the channel weight and
channel mapping networks, we use the network setup 
given in Tab.~\ref{tab:rings-channels} with one, two or three channels
with flat analytic mappings. Because expressive spline-based coupling
blocks can learn topologically challenging distributions without a
need for multiple channels, we use simple affine coupling
blocks. Ideally, our network should automatically define channels
removing any topological problems for each individual channel. While
this sometimes converges to a reasonable result starting from a flat
prior, we found that a mode-specific weight prior led to much more
stable results. In detail, we used a prior that encourages one or two
channels to focus on the ring and the other to focus on the
line. Still, there is a large variation between the results of
different trainings.  Some examples for the total and channel-wise
distributions for different numbers of channels are shown in
Fig.~\ref{fig:ring-flat}.  The relative uncertainties for different
numbers of channels are given in Tab.~\ref{tab:rings-channels}. The
performance improves significantly after adding more
channels. However, these results are highly sensitive to the choice of
the hyperparameters. This suggests that an unsupervised approach to
channel partitioning, while theoretically possible, might not be
optimal in practice.

Next, we can start with the analytic mapping of the ring and combine
it with a flat mapping. Because the ring mapping greatly simplifies
the training, we can reduce the number of INN parameters. Results for this combination of learned channels
and channel weights is shown in Fig.~\ref{fig:ring-rf}. In the upper
panels, we see that the flat channel learns the line without any
connection to the pre-defined ring.  The integration uncertainties are
given in Tab.~\ref{tab:rings-channels}.  They show that we can define
overflow channels to extract features that are not captured by
pre-defined mappings. The combined distribution in the lower panel
closely matches the truth. The channel weights exhibit a clean cut
between the two channels with a weight close to $0.5$ in the two
points where the ring and the line cross. In addition, we show the
relative uncertainties for a two-channel integrator with a ring and
line mapping and a three-channel integrator with a ring, line and flat
mapping in Tab.~\ref{tab:rings-channels}. It can be seen that using a line mapping instead of a flat mapping further improves the performance. Adding an additional flat mapping as an overflow channel is not beneficial since two channels are already sufficient to map out all the features and it just increases the complexity of the training. In all three
cases, the relative uncertainties improve compared to the trainings
with flat mappings only.

%%%%%%%%%%%%%%%%%%%%%%%%%%%%%%%%%%%%%%%%%%%%%%%%%%%
\section{Drell-Yan plus \texorpdfstring{\PZp}{Z'} at the LHC}
\label{sec:lhc}

After showing how to improve the integration of one- and
two-dimensional toy examples, we now use \madnis for an actual LHC
process. To keep things simple, while still challenging all components
of our framework, we consider the Drell-Yan process with an additional
\PZp-resonance,
\begin{align}
     \Pp \Pp \to \gamma, \PZ^*, {\PZp}^* \to \Pep \Pem \; ,
\label{eq:drell-yan}
\end{align}
assuming 
\begin{align}
  \MZp  = 400.0~\gev
  \qqquad
  \Gamma_\PZp = 0.5~\gev \; ,
\label{eq:bsm-params}
\end{align}
for 13~TeV center-of-mass energy. We use the leading-order NNPDF4.0
PDF set~\cite{NNPDF:2021njg} with a fixed factorization scale
$\mu_F=M_\PZ$ and $\alpha_s(\MZ) = 0.118$. In the four-flavor scheme
we neglect \Pb quarks in the initial state. The $\PZ$-parameters are
$\MZ = 91.19$~GeV and $\Gamma_\PZ = 2.44$~GeV.  We define the fiducial
phase space by requiring only
\begin{align}
m_{\Pep\Pem} > 15~\gev\; .
\end{align}
%

%----------------------------------------------------------
\begin{figure}[b!]
  \centering
  \includegraphics[width=0.33\textwidth]{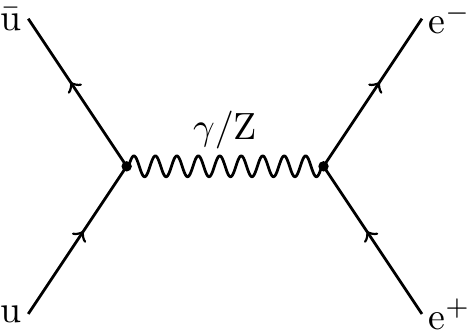}
  \hspace*{0.05\textwidth}
  \includegraphics[width=0.33\textwidth]{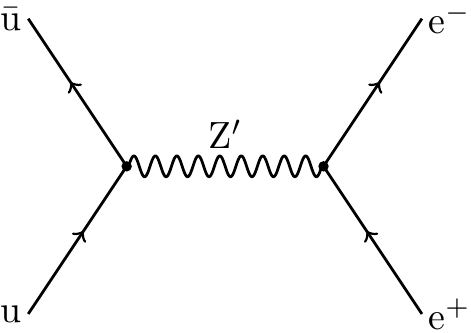}
  \caption{Example LO Feynman diagrams contributing to the \PZp-extended
    Drell-Yan process $\Pp \Pp \to \Pep \Pem$ for one partonic
    channel.}
   \label{fig:feynman_dy}
\end{figure}
%-----------------------------------------------------------

%%%%%%%%%%%%%%%%%%%%%%%%%%%%%%%%%%%%%%%%%%%%%%%%%%%
\subsubsection*{Implementation details}
\label{sec:dy_implementation}

To maintain full control, we implement the \madnis components directly
in \tensorflow, including the matrix element and the phase-space
mappings. The calculation of a hadronic scattering cross section
requires many ingredients which need to combined efficiently to
achieve precise numerical results. In detail, we implement
\begin{enumerate}
\item the full squared spin-color averaged/summed LO amplitude
  \begin{align}
    \langle\vert\mathcal{M}\vert^2\rangle
    = \frac{1}{4N_\text{c}}\sum_{\text{spins}}\vert \mathcal{M}_\gamma + \mathcal{M}_\PZ + \mathcal{M}_\PZp \vert^2\;,
  \end{align}
  with $N_\text{c}=3$.  As the amplitude is implemented in
  \tensorflow, we can evaluate it in a vectorized form on a CPU and
  GPU and have access to its gradient, an option we do not use in this study, but plan to use in the future.
    
\item the hadronic cross section as a convolution of the partonic
  cross section with the PDFs,
  \begin{align}
    \sigma_{\Pp \Pp}=\sum_{a,b}
    \int_0^1 \d x_1 \d x_2 \; f_a(x_1) f_b(x_2) \; \hat{\sigma}_{ab}(x_1x_2s).
    \end{align}
We use \lhapdf~\cite{Buckley:2014ana} and implement our own \python
interface to efficiently evaluate large event batches.

\item a multi-channel integration, where we define suitable mappings
  associated with the different Feynman diagrams.
\end{enumerate}
The hadronic phase space is expressed in terms of $\{x_1, x_2, \cos
\theta, \phi\}$. The sampling requires a mapping from the unit
hypercube $U=[0,1]^4$ to the two-particle phase space. We implement
this mapping sequentially as
\begin{alignat}{2}
    &G_1:  \qquad  \{y_1, y_2, y_3, y_4\} &&\to \{s, y_2, y_3, y_4\} \notag\\
    &G_2: \qquad  \{s, y_2, y_3, y_4\}  &&\to \{x_1, x_2, \cos\theta,\phi\}\;,
\end{alignat}
where the first step takes into account the propagator structure, so
the substitution $y_1 \to s$ maps out the two mass peaks or the photon propagator. 
For a resonance with mass $M$ and width $\Gamma$, the standard mapping 
is again the Breit-Wigner mapping of Eq.\eqref{eq:bw_mapping1}~\cite{lusifer,Plehn:2015dqa}
\begin{align}
    s(y_1) &= M^2 + M \Gamma \tan \Big[ \omega_\text{min} + (\omega_\text{max} - \omega_\text{min}) y_1 \Big] \notag \\
    g_1(s)&= \frac{1}{\omega_\text{max} - \omega_\text{min}}
    \; \frac{M \Gamma}{(s-M^2)^2 + M^2\Gamma^2} \; .
\end{align}
where the limits $s=s_\text{min}\dots s_\text{max} = 4 E_\text{beam}^2$ translate into
\begin{align}
    \omega_\text{min,max} = \arctan \frac{s_\text{min,max}^2 - M^2}{M\Gamma} \; .
\end{align}
For the massless photon we instead use the mapping
\begin{alignat}{2}
    s(y_1) &= \Big[ y_1 s_\text{max}^{1-\nu} + (1-y_1) s_\text{min}^{1-\nu} \Big]^{1/(1-\nu)}\notag \\ 
    g_1(s) &= \frac{1-\nu}{s^\nu \left( s_\text{max}^{1-\nu} - s_\text{min}^{1-\nu} \right)} \; .
\label{eq:lusifer_map_massles_nu}
\end{alignat}
The hyperparameter $\nu\ne1$ can be tuned, but we stick to the naive
assumption $\nu=2$. In the second step, we map to $\{x_1, x_2,
\cos\theta,\phi\}$ using
\begin{alignat}{3}
    x_1 &= \left(\frac{s}{s_\text{max}}\right)^{y_2} \qquad
    &x_2 &= \left(\frac{s}{s_\text{max}}\right)^{1-y_2} && \notag\\
    \cos \theta &= 2 y_3 - 1 \qquad
    &\phi &= 2 \pi y_4 - \pi \qquad \text{with} \quad 
    g_2 =- \frac{s_\text{max}}{4\pi\log (x_1 x_2) } \; .
\end{alignat}

We test our numerical setup by computing the fiducial cross section 
and comparing the result to the standard \mg prediction of 
$\sigma = (4349.7\pm 0.32)$~pb to a relative deviation of $10^{-5}$. 
%We apologize to Stefan Dittmaier for this rather modest level of numerical precision.

%---------------------------------------------
\begin{figure}[t]
    \includegraphics[width=0.495\textwidth]{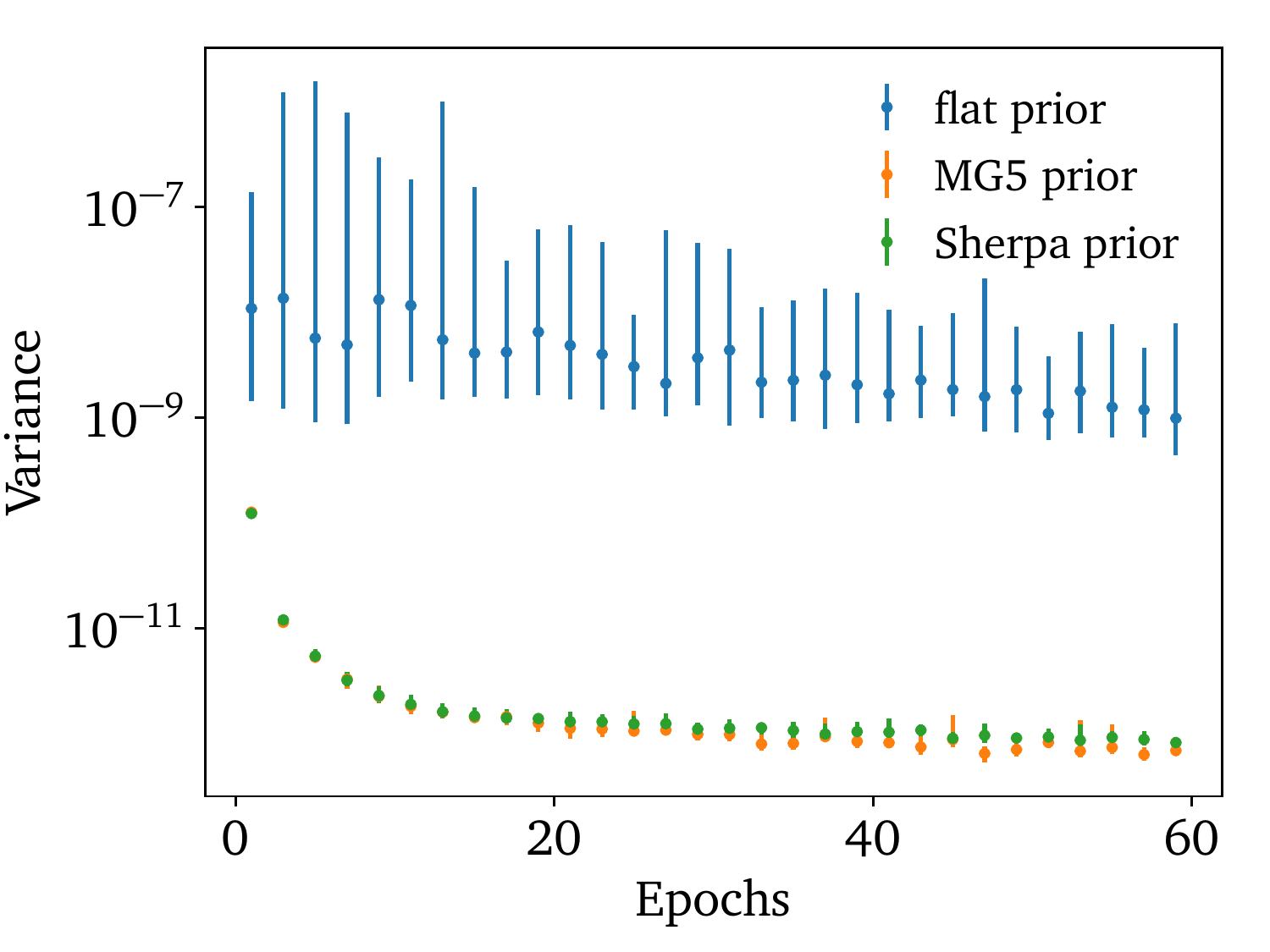}
    \includegraphics[width=0.495\textwidth]{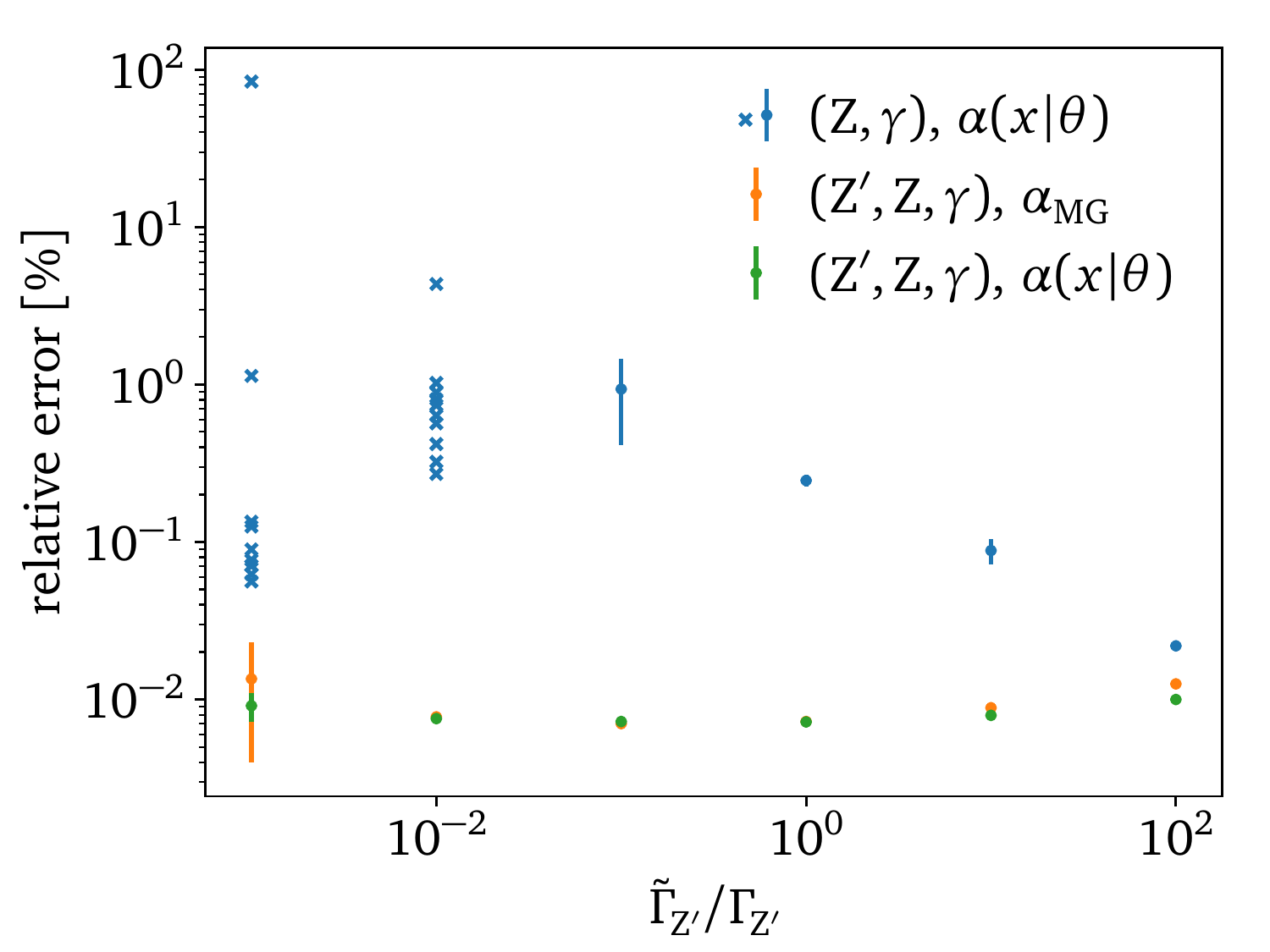}
    \caption{Left: mean and spread ($5\%$ to $95\%$ percentile) of 25
      evaluations of the variance for three priors of the network
      weights $\alpha$.  Right: integration error as a function of
      $\Gamma_\PZp$ for two and three channels, with and without
      trained channel weights. We give means and standard deviations
      for ten runs, or the individual results in case of large
      variation. For very narrow peaks, the two-channel integrator
      misses the \PZp peak entirely.}
    \label{fig:dy_zprime_variance_width}
\end{figure}
%---------------------------------------------

%----------------------------------------------------------
\begin{table}[b!]
    \centering
    \begin{small} \begin{tabular}[t]{ll|ll}
    \toprule
    Parameter & Value & Parameter & Value \\
    \midrule
    Loss function & variance 
    & Coupling blocks & rational-quadratic splines \\
    Learning rate & 0.001 
    & Permutations & exchange \\
    LR schedule & inverse time decay 
    & Blocks & 6 \\
    Decay rate & 0.01 
    & Subnet hidden nodes & 16 \\
    Batch size & 10000 
    & Subnet layers & 2 \\
    Epochs & 60 
    & CWnet layers & 2 \\
    Batches per epoch & 50 
    & CWnet hidden nodes & 16 \\
    && Activation function & leaky ReLU \\
    \bottomrule
    \end{tabular} \end{small}
    \caption{Hyperparameters of the INN and the channel weight network
      (CWnet) for the integration of the Drell-Yan + $\PZp$ cross
      section.}
    \label{tab:dy-hyper}
\end{table}
%------------------------------------------------------

%%%%%%%%%%%%%%%%%%%%%%%%%%%%%%%%%%%%%%%%%%%%%%%%%%%
\subsubsection*{Choice of mappings and priors}

While for the simple parametric toy models affine~\cite{coupling1,
  coupling2} coupling blocks were sufficient when combined with a
multi-channel strategy, the rich phase-space structure in the
\PZp-extended Drell-Yan process benefits from rational-quadratic
spline blocks~\cite{durkan2019neural}. Another advantage of spline
blocks is that they are naturally defined on a compact domain which
makes them especially well-suited for mappings between
unit-hypercubes. The other network parameters for this process are
given in Tab.~\ref{tab:dy-hyper}.

For the toy models we have seen that the choice of mappings and priors
is key to a precise integration. This is especially true once we need
to cover two narrow peaks in $M_{\Pep\Pem}$.
We confirm this using our network trained with a flat prior, the \sherpa-like
prior in Eq.\eqref{eq:multi-channel-unified}, and the \mg-like prior
in Eq.\eqref{eq:sde}. After every second epoch, we extract the
variance of the integrand from 25 batches of generated samples. The
mean and spread of these variances are shown in the left panel of
Fig.~\ref{fig:dy_zprime_variance_width}. For both non-flat priors, the
variance is stable and converges in the course of the training. In
contrast, the flat prior leads to a much larger and unstable
variance. Compared to the physics-informed priors the convergence is
extremely slow. We follow the standard setup of LHC
event generators and include the available physics information through
the \mg-like prior of Eq.\eqref{eq:sde}. 

Second, a powerful physics-informed mapping becomes increasingly
important for integrands with narrower features. To this end, we vary
the $\PZp$-width over several orders of magnitude around the central
value given in Eq.\eqref{eq:bsm-params},
\begin{align}
  \tilde{\Gamma}_\PZp = \Gamma_\PZp \times \{10^{-3}, 10^{-2}, 10^{-1}, 10^{0}, 10^{1}, 10^{2}\}
  \; ,
\end{align}
while keeping the $\PZ$ -width constant.  In the right panel of
Fig.~\ref{fig:dy_zprime_variance_width}, we first compare a two-channel
integrator with mappings tailored for the \PZ and photon diagrams with
a three-channel integrator with an additional mapping for the \PZp.
For the three-channel setup, we either fix the channel weights to the
\mg prior or train them from this prior. For all three scenarios we
give the relative error of the phase-space integral. While the error
remains small for the three-channel integrator, even for very narrow
decay widths, the integration rapidly degrades for two channels only.
For the two narrowest \PZp-peaks we see a large spread in the variance
combined with an overconfident error estimate, indicating that the
sampling misses the peak altogether.  For the three-channel setup the
trainable channel weights lead to a small improvement over the fixed
channel weights, mostly for large $\tilde{\Gamma}_\PZp$. This reflects
the fact that for negligible interferences the \mg
choice of channel weights is essentially optimal.

\begin{figure}[t]
    \includegraphics[page=3, width=0.495\textwidth]{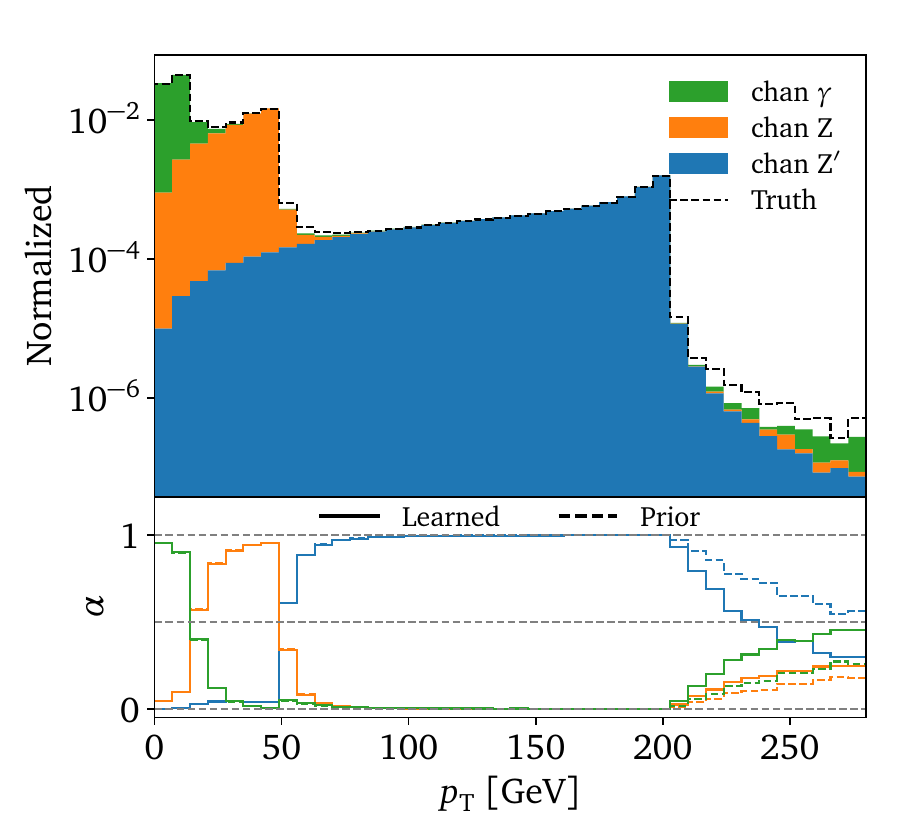}
    \includegraphics[page=9, width=0.495\textwidth]{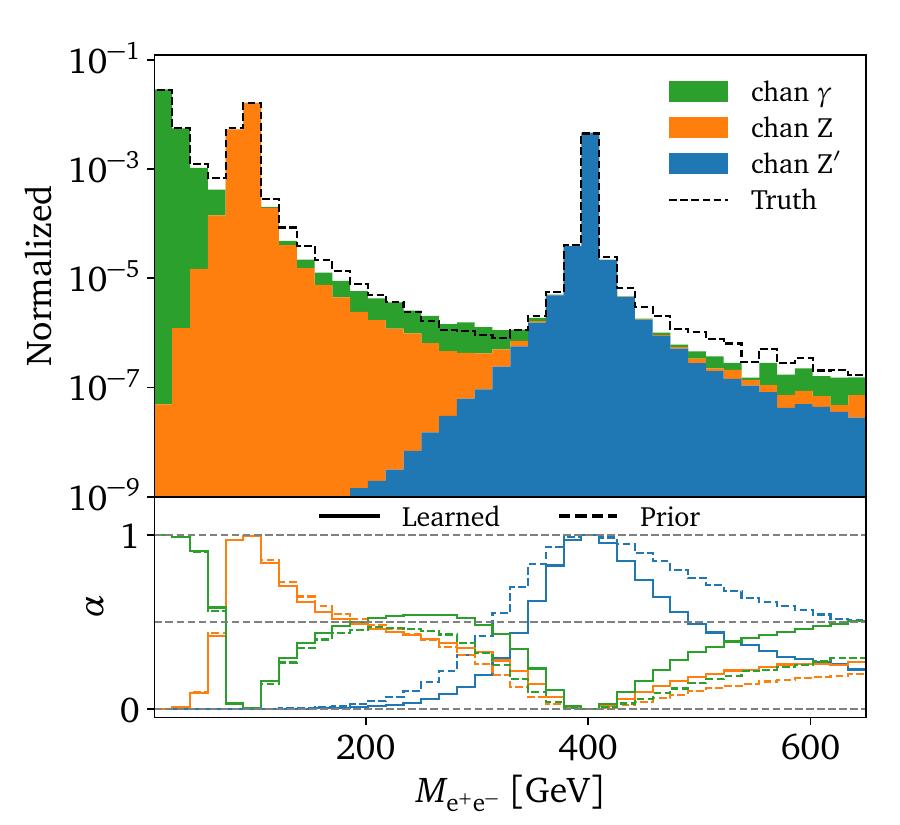}
    \caption{Learned $\pT$ and $M_{\Pep\Pem}$ distributions for the
      \PZp-extended Drell-Yan process. In the lower panels we show the
      learned channel weights.}
    \label{fig:dy_zprime}
\end{figure}

In Fig.~\ref{fig:dy_zprime}, we look at the phase-space coverage for
the distinctive $\pT$ and $M_{\Pep\Pem}$ distributions. We show the
learned local channel weights for a three-channel integrator starting
from the \mg prior. In agreement with the above result the channel
weight network mostly learns small corrections to the prior. Each
channel dominates an $M_{\Pep\Pem}$ region and the combined
distributions are in good agreement with the truth. This means each
channel focuses on a single task, as defined by the initialization,
rather than learning the full distribution.

%---------------------------------------------
%---------------------------------------------

%%%%%%%%%%%%%%%%%%%%%%%%%%%%%%%%%%%%%%%%%%%%%%%%%%%
\subsubsection*{Buffered training}

%---------------------------------------------
\begin{figure}[t]
    \includegraphics[width=0.495\textwidth]{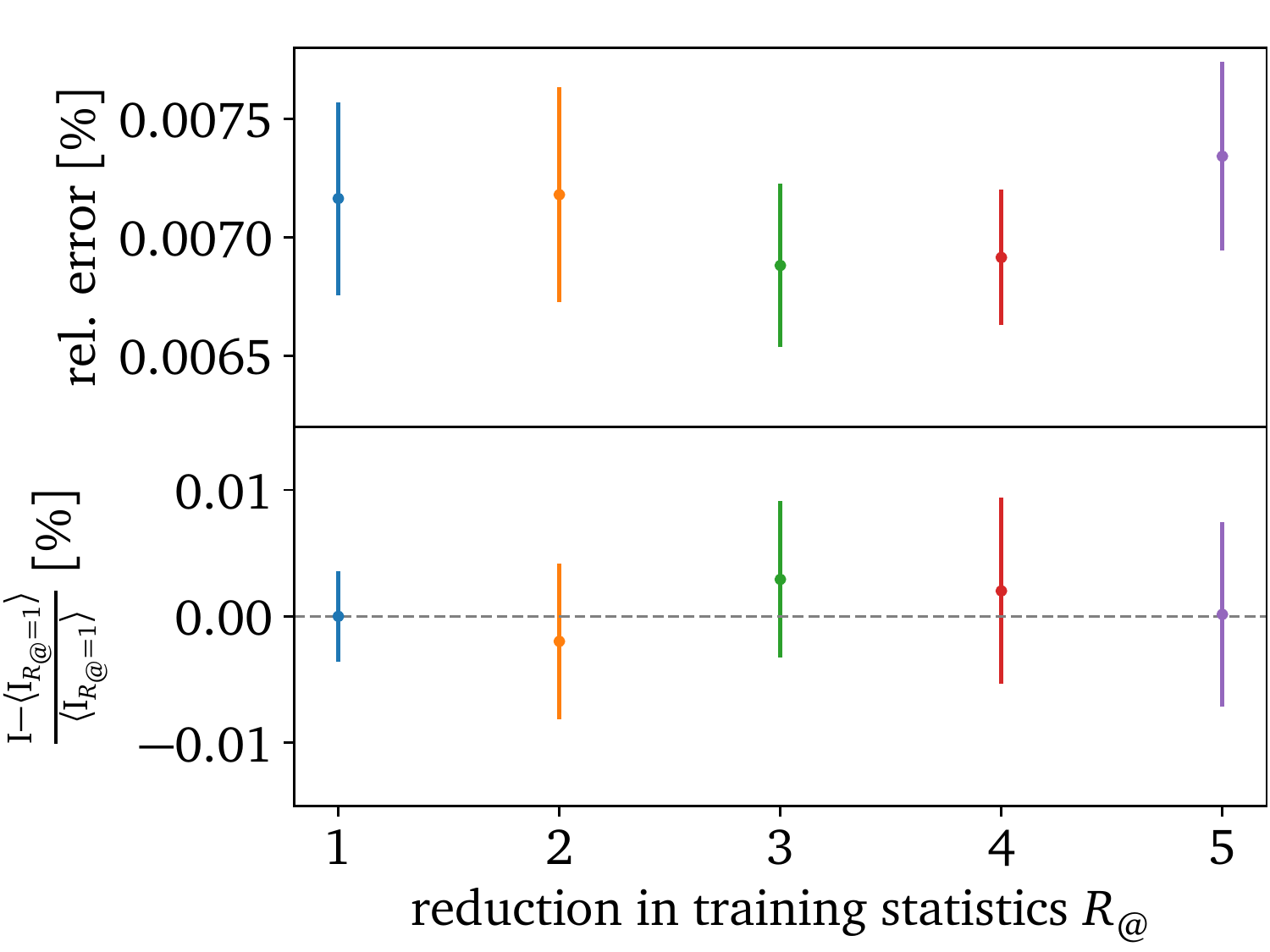}
    \includegraphics[page=3, width=0.495\textwidth]{5_buffered_weights} \\
    \includegraphics[page=2, width=0.495\textwidth]{5_buffered_weights}
    \includegraphics[page=1, width=0.495\textwidth]{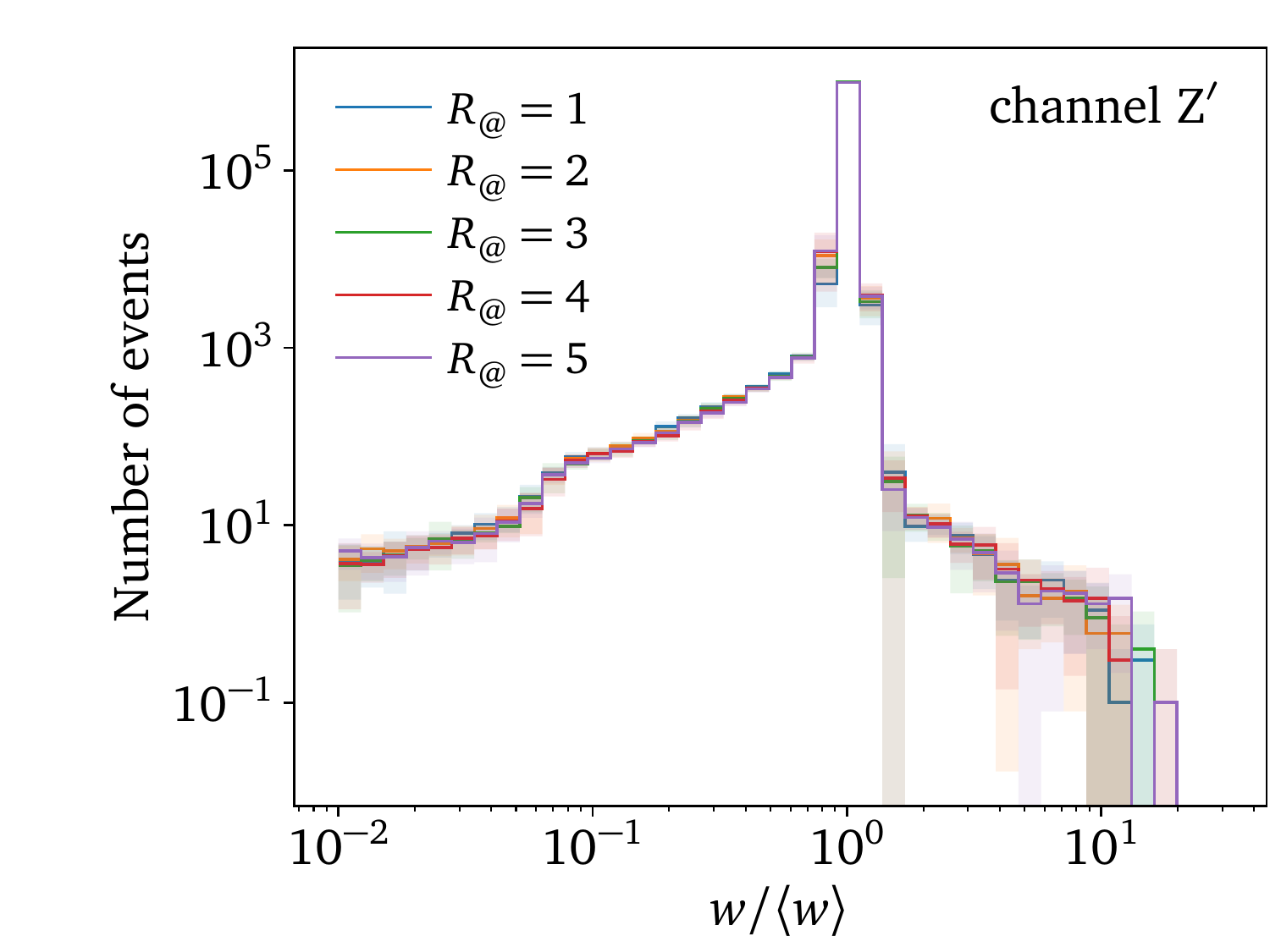}
    \caption{Relative integration error (from $10^6$ events),
      relative deviation from the mean $R_\text{@} = 1$ result,
      and weight distributions for different reduction factors $R_\text{@}$ in
      training statistics for the \PZp-extended Drell-Yan process. The
      points/lines and error bars/bands show means and standard
      deviations over ten runs.}
    \label{fig:dy_zprime_buffered}
\end{figure}
%---------------------------------------------

Even though the integrand for our modified Drell-Yan process is
computationally cheap, we can still use it as a test case for our new
buffered training. Specifically, we first train the network online for
one epoch and save all samples generated during that epoch. Then, we
train the network for $k_\text{buff}$ epochs on the saved samples,
shuffling them every time. After that, we discard the saved
samples.  We find that this training schedule works well for our
application, but it can be easily adapted for other application. For
example, we can save samples from more than one online training epoch.

To benchmark the buffered training, we continue to train the network
for 60~epochs, but replace some of the online training epochs with
training on samples following the above schedule. The training cycle
is then repeated $60 / (k_\text{buff}+1)$ times, and the relative
reduction in the training statistics defined in
Eq.\eqref{eq:train_red} is
\begin{align}
  R_\text{@} = k_\text{buff} + 1
  \qquad \text{with} \qquad
  k_\text{buff} = 0, 1, 2, 3, 4 \; .
\label{eq:def_r_at}
\end{align}
For each value of $k_\text{buff}$ we run our integrator ten times. The
relative integration error, the relative deviation from the mean
$R_\text{@} = 1$ result, and the weight
distributions for the three different channels are shown in
Fig.~\ref{fig:dy_zprime_buffered}. Even for a reduction of the
training statistics by a factor five the performance of the integrator
--- in terms of the relative error and the weight distribution ---
matches the pure online training. Even in this simple case, where the
evaluation time for the integrand is negligible, the training time can
be reduced by around $20\%$ because of the lower number of INN
evaluations.

%%%%%%%%%%%%%%%%%%%%%%%%%%%%%%%%%%%%%%%%%%%%%%%%%%%
%\subsubsection*{Trainable rotations}

%------------------------------------------------------
%\begin{table}[t]
%\centering
%\begin{small} \begin{tabular}[t]{l|l}
%\toprule
% Permutation & Rel.~Error [\%] \\
%\midrule
% exchange & $0.0072 \pm 0.0004$ \\
% random & $0.0085 \pm 0.0011$ \\
% logarithmic & $0.0078 \pm 0.0002$ \\
% soft & $0.0227 \pm 0.0022$ \\
% trainable soft & $0.0106 \pm 0.0014$ \\
%\midrule
%\multicolumn{2}{c}{Based on  $10^6$ events}
%\end{tabular} \end{small}
%\caption{Relative integration errors for different choices of
%  permutations between the normalizing flow blocks. We show the means
%  and standard deviations for ten independent trainings.}
%\label{tab:rot}
%\end{table}
%%------------------------------------------------------

%Table~\ref{tab:rot} shows how different choices for the permutation 
%layer affect the relative error of the integral estimate. The
%trainable soft permutations perform much better than the fixed soft
%permutations. This result shows the advantage of
%trainable angles over fixed angles. However, for this low-dimensional
%%%problem, soft permutations perform slightly worse than simple exchange
%permutations. The reason for this is twofold. First, the features the
%flow has to learn are almost perfectly aligned with the axis of the
%chosen parametrization, and thus no rotation is necessary. Second, spline blocks require us to nest the soft permutations 
%between a pair of logit and sigmoid functions to perform the 
%rotations on the valid domain, leading to potentially 
%slower convergence.

As a side remark, we have tested how different choices for the permutation 
layer affect the integration. While the
trainable soft permutations perform much better than the fixed soft
permutations, soft permutations perform slightly worse than simple exchange
permutations for this low-dimensional
problem. The reason for this is that the features the
flow has to learn are almost perfectly aligned with the axis of the
chosen parametrization without any rotation, and that spline blocks 
require us to nest the soft permutations 
between logit and sigmoid functions, which leads to potentially 
slower convergence.

%---------------------------------------------
\begin{figure}[t]
    \centering
    \includegraphics[width=0.55\textwidth]{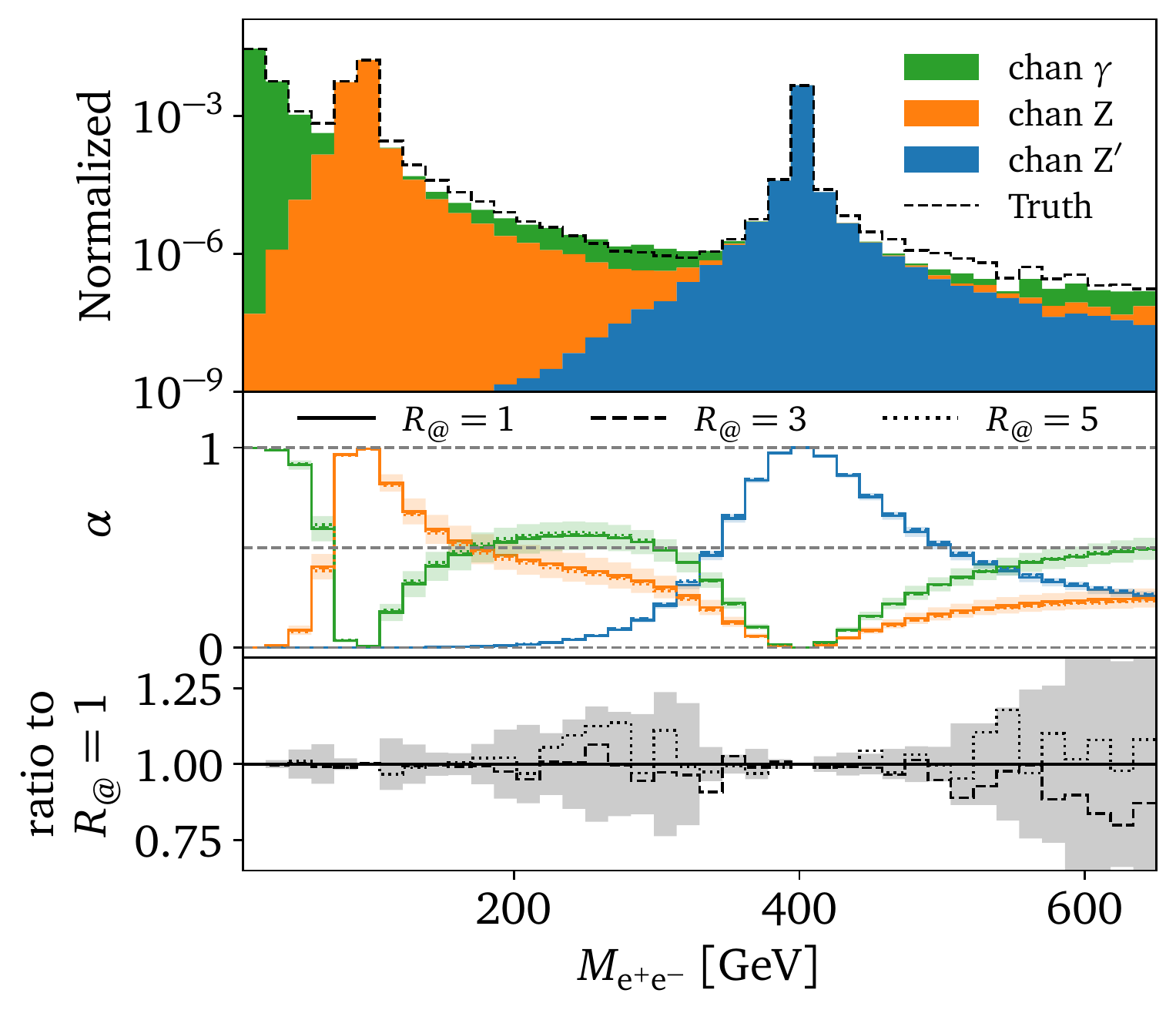}
    \caption{Learned $M_{\Pep\Pem}$ distributions for the
      \PZp-extended Drell-Yan process. The upper panel is the same as
      in Fig.~\ref{fig:dy_zprime}, the middle panel shows the learned
      channel weights, and the lower panel shows the ratio of the
      combined distribution to pure online training for 
      reduction factors $R_\text{@}$ in training
      statistics, see Eq.\eqref{eq:def_r_at}. The lines in the
      lower two panels are obtained by averaging over ten
      independent trainings. The error envelopes are only shown
      for $R_\text{@} = 1$.}
    \label{fig:dy_money}
\end{figure}
%---------------------------------------------

%%%%%%%%%%%%%%%%%%%%%%%%%%%%%%%%%%%%%%%%%%%%%%%%%%%
\section{Outlook}
\label{sec:conclusion}

We introduced the new, comprehensive \madnis approach to importance
sampling and multi-channel integration. The bijective variable
transformations behind importance sampling suggest using normalizing
flows, in our case an INN which is equally fast in both
directions. For LHC event generators, this ML-integrator needs to be
embedded in a common framework with multi-channel integration. We have
shown how to efficiently combine normalizing flows with a
multi-channel strategy by defining local and trainable multi-channel
weights. Finally, we developed trainable rotations as a general
permutation layer between the INN coupling blocks. They will become
beneficial for high-dimensional phase spaces.

For simple parametric examples, we have seen that it is possible to
learn optimal channel weights, including a combination with
normalizing flows. Moreover, we have shown that it is possible to
define single or multiple overflow channels and leave it to the
networks to split the complicated topological structure into
easy-to-learn substructures. More realistically, we have shown that
our framework works for the \PZp-extended Drell-Yan process, which
includes many challenges of a generic LHC process while still
having a low-dimensional phase space. In particular, it requires 
a combination of the normalizing flow with a physics-informed
mapping to achieve a precise integration at low
computational cost.

A bottleneck for current LHC predictions is increasingly expensive
evaluations of the matrix element. To alleviate this problem, we
combine expensive online training with buffered sample training. In
Fig.~\ref{fig:dy_money}, we illustrate the performance of the \madnis
methodology, including an effective reduction in training statistics
by using buffered training in addition to the standard online
training. For our LHC example, our new training
scheme can reduce the number of calls to the matrix element by a
factor of five without losing precision in the integration.

%%%%%%%%%%%%%%%%%%%%%%%%%%%%%%%%%%%%%%%%%%%%%%%%%%%
\section*{Acknowledgements}

The authors would like to express special thanks to the
Mainz Institute for Theoretical Physics (MITP) of the Cluster of
Excellence PRISMA+ (Project ID 39083149), for its hospitality and
support. OM, FM and RW acknowledge support by FRS-FNRS (Belgian
National Scientific Research Fund) IISN projects 4.4503.16. CK was
supported by DOE grant DOE-SC0010008. AB, CK, and TP would like to thank
the Baden-W\"urttemberg-Stiftung for funding through the program
\textit{Internationale Spitzenforschung}, project
\textsl{Uncertainties --- Teaching AI its Limits}
(BWST\_IF2020-010). AB would like to acknowledge support by the BMBF for the AI junior group 01IS22079. TH is supported by the DFG Research Training Group 
GK-1940, \textsl{Particle Physics Beyond the Standard Model}. 
TH's contribution to this project was made possible by funding from the Carl-Zeiss-Stiftung.
Computational resources have been provided by the
supercomputing facilities of the Université catholique de Louvain
(CISM/UCL) and the Consortium des Équipements de Calcul Intensif en
Fédération Wallonie Bruxelles (CÉCI) funded by the Fond de la
Recherche Scientifique de Belgique (F.R.S.-FNRS) under convention
2.5020.11 and by the Walloon Region. 
The work of JI was supported by the Fermi National Accelerator Laboratory 
(Fermilab), a U.S. Department of Energy, Office of Science, HEP User Facility.
Fermilab is managed by Fermi Research Alliance, LLC (FRA), acting under Contract
No. DE--AC02--07CH11359.  This project was supported by the Deutsche
Forschungsgemeinschaft (DFG, German Research Foundation) under grant
396021762 -- TRR~257 \textsl{Particle Physics Phenomenology after the
Higgs Discovery} and through Germany’s Excellence Strategy EXC 2181/1 - 390900948 
\textsl{The Heidelberg STRUCTURES Excellence Cluster}.

\clearpage
\appendix
%%%%%%%%%%%%%%%%%%%%%%%%%%%%%%%%%%%%%%%%%%%%%%%%%%%
\section{Buffered losses and training}
\label{sec:2stageloss}

Splitting the integral of Eq.\eqref{eq:psinteg} using the trained weights defined in Eq.\eqref{eq:trained_weights} we
first define normalized channel-wise probability distributions as
\begin{align}
  I[f]
  = \sum_i I_i(\theta)
  &= \sum_i \int_\Phi\d^d x\,\alpha_i(x|\theta)\,f(x) \notag \\
  \Rightarrow \qquad
  p_i(x|\theta) &=\frac{\alpha_i(x|\theta)\,f(x)}{I_i(\theta)} \; ,
\end{align}
where the channel-wise integrals change during training. The goal is
to approximate these probability distributions with a network function in terms of the weights $\varphi$,
\begin{align}
  p_i(x|\theta) \approx g_i(x|\varphi) \; .
\end{align}
The implicit dependence of $g_i(x|\varphi)$ on $\theta$ enters through this training objective. To quantify the agreement between the two functions we can use a range of divergences $D$, all summed over the channels,
\begin{align}
  \loss = \sum_i a_i \; D_i[p_i;g_i] \; ,
\end{align}
with arbitrary weights $a_i$.  For a combined training of the channel
weights $(\theta)$ and the importance sampling $(\varphi)$ we have to
be careful when updating the losses based on these divergences.
  
%%%%%%%%%%%%%%%%%%%%%%%%%%%%%%%%%%%%%%%%%%%%%%%%%%%
\subsection*{Neyman $\chi_N^2$ divergence}

The first divergence we can use to define our loss is the
Neyman-$\chi_N^2$ divergence
\begin{align}
\begin{split}
D_{\chi_N^2,i}
&=\int_\Phi\d^d x\,\frac{[p_i(x|\theta)-g_i(x|\varphi)]^2}{g_i(x|\varphi)}\\
&=\int_\Phi\d^d x\;\frac{p_i(x|\theta)^2}{g_i(x|\varphi)} - 2 \underbrace{\int_\Phi\d^d x\;p_i(x|\theta)}_{=1} + \underbrace{\int_\Phi\d^d x\;g_i(x|\varphi)}_{=1}\;.
\end{split}
\label{eq:neyman_loss_def}
\end{align}
To minimize $D_{\chi^2,i}$, we need its gradient with respect to $\varphi$ and $\theta$.
\begin{align}
\begin{split}
  \nabla_\varphi D_{\chi_N^2,i}
&=\int_\Phi\d^d x\,p_i(x|\theta)^2 \; \nabla_\varphi\frac{1}{g_i(x|\varphi)}
=-\int_\Phi\d^d x\,\frac{p_i(x|\theta)^2}{g_i(x|\varphi)} \; \nabla_\varphi \log g_i(x|\varphi) \\
&= \left\langle -\frac{p_i(x|\theta)^2}{q_i(x|\hat{\varphi})g_i(x|\varphi)}\; \nabla_\varphi \log g_i(x|\varphi) \right\rangle_{x\sim q_i(x|\hat{\varphi})} 
\\
\nabla_\theta D_{\chi_N^2,i}
&=2\int_\Phi\d^d x\,\frac{p_i(x|\theta)}{g_i(x|\varphi)} \; \nabla_\theta p_i(x|\theta)
=2\int_\Phi\d^d x\,\frac{p_i(x|\theta)^2}{g_i(x|\varphi)} \; \nabla_\theta \log p_i(x|\theta) \\
&= 2 \left\langle \frac{p_i(x|\theta)^2}{q_i(x|\hat{\varphi})g_i(x|\varphi)}\; \nabla_\theta \log p_i(x|\theta) \right\rangle_{x\sim q_i(x|\hat{\varphi})}
\; .
\end{split}
\end{align}
Note that we evaluate the integrals by sampling from a proposal
function $x\sim q_i(x|\hat{\varphi})$, which can be either a totally
independent function that is easy to sample from and the dependence of
$\varphi$ drops out, i.e.\ $q_i(x|\hat{\varphi})=q_i(x)$, or it is
directly linked to the importance weight
$q_i(x|\hat{\varphi})=g_i(x|\hat{\varphi})$ possibly depending on
different network weights $\hat{\varphi}\ne\varphi$ which is relevant
for the buffered training as described in
Sec.~\ref{sec:madnis_buff}. The loss functions are then given by
\begin{align}
\begin{split}
  \loss^\text{int}_{\chi_N^2}
  &=- \sum_i a_i \; \left\langle \frac{{\color{red!80!black}p_i(x|\theta)}^2}{\color{red!80!black}q_i(x|\hat{\varphi})g_i(x|\varphi)}\,\log g_i(x|\varphi) \right\rangle_{x\sim q_i(x|\hat{\varphi})} \\
  \loss^\text{weights}_{\chi_N^2}
  &=2 \sum_i a_i \; \left\langle \frac{{\color{red!80!black}p_i(x|\theta)}^2}{\color{red!80!black}q_i(x|\hat{\varphi})g_i(x|\varphi)}\,\log p_i(x|\theta) \right\rangle_{x\sim q_i(x|\hat{\varphi})} \;,
\end{split}
\label{eq:neyman_losses}
\end{align}
where the red expressions have to be evaluated without gradient
calculation. Note that $p_i(x|\theta)$ indirectly also depends on
$\hat{\varphi}$ as the samples are drawn from $x\sim
q_i(x|\hat{\varphi})$. However, we do not need a gradient calculation
for $p_i$.

%%%%%%%%%%%%%%%%%%%%%%%%%%%%%%%%%%%%%%%%%%%%%%%%%%%
\subsection*{Variance loss}

Alternatively, we can minimize the variance of the normalized
functions $p_i(x|\theta)/g_i(x|\varphi)$,
\begin{align}
\begin{split}
  %\mathbb{V}\left(\frac{p_i(x|\theta)}{q_i(x|\varphi)}\right)\equiv
  \mathbb{V}_i
  &= \left\langle \frac{p_i(x|\theta)^2}{g_i(x|\varphi)^2} \right\rangle_{x\sim g_i(x|\varphi)}
  - \left\langle \frac{p_i(x|\theta)}{g_i(x|\varphi)} \right\rangle_{x\sim g_i(x|\varphi)}^2\\
  &=\int_\Phi\d^d x\,\frac{p_i(x|\theta)^2}{g_i(x|\varphi)} - \underbrace{\left(\int_\Phi\d^d x\,p_i(x)\right)^2}_{=1}\;.
\end{split}
\label{eq:var_loss_def}
\end{align}
This is the same expression as $D_{\chi_N^2}$ in
Eq.\eqref{eq:neyman_loss_def}, so the losses are given by
Eq.\eqref{eq:neyman_losses}.  Note that we can write
Eq.\eqref{eq:var_loss_def} into a MC estimate using the sampling
$x\sim q_i(x|\hat{\varphi})$,
\begin{align}
  \mathbb{V}_i
  &= \left\langle \frac{p_i(x|\theta)^2}{g_i(x|\varphi) q_i(x|\hat{\varphi})} \right\rangle_{x\sim q_i(x|\hat{\varphi})}
  - \left\langle \frac{p_i(x|\theta)}{q_i(x|\hat{\varphi})} \right\rangle_{x\sim q_i(x|\hat{\varphi})}^2\; .
\end{align}
%

%%%%%%%%%%%%%%%%%%%%%%%%%%%%%%%%%%%%%%%%%%%%%%%%%%%
\subsection*{Pearson $\chi_P^2$ divergence}

A similar choice is the Pearson-$\chi_P^2$ divergence,
\begin{align}
\begin{split}
  %D_{\chi_P^2}(p_i|\theta||q_i|\varphi)\equiv
  D_{\chi_P^2,i}
&=\int_\Phi\d^d x\,\frac{(g_i(x|\varphi)-p_i(x|\theta))^2}{p_i(x|\theta)}\\
&=\int_\Phi\d^d x\;\frac{g_i(x|\varphi)^2}{p_i(x|\theta)} - 2 \underbrace{\int_\Phi\d^d x\;g_i(x|\varphi)}_{=1} + \underbrace{\int_\Phi\d^d x\;p_i(x|\theta)}_{=1}\;.
\end{split}
\label{eq:pearson_loss_def}
\end{align}
To minimize $D_{\chi_P^2,i}$ we need the two gradients
\begin{align}
\begin{split}
\nabla_\varphi D_{\chi_P^2,i}
%&=\nabla_\varphi\int_\Phi\d^d x\,\frac{q_i(x|\varphi)^2}{p_i(x)}
&=2\int_\Phi\d^d x\,\frac{g_i(x|\varphi)}{p_i(x|\theta)}\nabla_\varphi g_i(x|\varphi)
=2\int_\Phi\d^d x\,\frac{g_i(x|\varphi)^2}{p_i(x|\theta)}\nabla_\varphi \log g_i(x|\varphi)\\
&= 2 \left\langle \left(\frac{g_i(x|\varphi)^2}{p_i(x|\theta)q_i(x|\hat{\varphi})}\right)\nabla_\varphi \log g_i(x|\varphi) \right\rangle_{x\sim q_i(x|\hat{\varphi})}
\\
\nabla_\theta D_{\chi_P^2,i}
%&=\nabla_\theta\int_\Phi\d^d x\,\frac{q_i(x|\varphi)^2}{p_i(x|\theta)}
&=\int_\Phi\d^d x\,g_i(x|\varphi)^2\nabla_\theta\frac{1}{p_i(x|\theta)}
=-\int_\Phi\d^d x\,\frac{g_i(x|\varphi)^2}{p_i(x|\theta)} \; \nabla_\theta \log p_i(x|\theta)\\
&= - \left\langle \left(\frac{g_i(x|\varphi)^2}{p_i(x|\theta)q_i(x|\hat{\varphi})}\right) \; \nabla_\varphi \log p_i(x|\theta) \right\rangle_{x\sim q_i(x|\hat{\varphi})} \; .
\end{split}
\end{align}
The corresponding losses can be written as
\begin{align}
\begin{split}
  \loss^\text{int}_{\chi_P^2}
  &=2 \sum_i a_i  \; \left\langle \left(\frac{{\color{red!80!black}g_i(x|\varphi)}^2}{\color{red!80!black}p_i(x|\theta)q_i(x|\hat{\varphi})}\right)\log q_i(x|\varphi) \right\rangle_{x\sim q_i(x|\hat{\varphi})} \\
  \loss^\text{weights}_{\chi_P^2}
  &=- \sum_i a_i \; \left\langle \left(\frac{{\color{red!80!black}g_i(x|\varphi)}^2}{\color{red!80!black}p_i(x|\theta)q_i(x|\hat{\varphi})}\right)\log p_i(x|\theta) \right\rangle_{x\sim q_i(x|\hat{\varphi})} \; ,
\end{split}
\end{align}
where, again, the red expressions have to be evaluated without gradient calculation. 

%%%%%%%%%%%%%%%%%%%%%%%%%%%%%%%%%%%%%%%%%%%%%%%%%%%
\subsection*{KL-divergence}

As a fourth option, we can use the KL-divergence to train the network,
\begin{align}
\begin{split}
  %D_\text{KL}(p_i|\theta||q_i|\varphi)\equiv
  D_{\text{KL},i}&=\int_\Phi\d^d x\;p_i(x|\theta)\log\frac{p_i(x|\theta)}{g_i(x|\varphi)}\\
  &=\int_\Phi\d^d x\;p_i(x|\theta)\log p_i(x|\theta) - \int_\Phi\d^d x\;p_i(x|\theta)\log g_i(x|\varphi) \;.
\end{split}
\label{eq:kl_loss_def}
\end{align}
To minimize $D_{\text{KL},i}$ with respect to $\varphi$ we only need to consider the second term, which is the cross entropy,
\begin{align}
\nabla_\varphi D_{\text{KL},i}
=-\int_\Phi\d^d x\;p_i(x|\theta) \; \nabla_\varphi\log q_i(x|\varphi)
= - \left\langle \frac{p_i(x|\theta)}{q_i(x|\hat{\varphi})}\nabla_\varphi \log g_i(x|\varphi) \right\rangle_{x\sim q_i(x|\hat{\varphi})} \;.
\end{align}
To train the channel weight we evaluate
\begin{align}
\begin{split}
\nabla_\theta D_{\text{KL},i}
%&=\nabla_\theta\int_\Phi\d^d x\;p_i(x|\theta)\log p_i(x|\theta) - \nabla_\theta\int_\Phi\d^d x\;p_i(x|\theta)\log q_i(x|\varphi)\\
&=\int_\Phi\d^d x\; \nabla_\theta p_i(x|\theta) \; \log p_i(x|\theta)
+\int_\Phi\d^d x\; p_i(x|\theta) \; \nabla_\theta \log p_i(x|\theta) \\
&\phantom{=}-\int_\Phi\d^d x\;\nabla_\theta p_i(x|\theta)\log g_i(x|\varphi)\\
&=\int_\Phi\d^d x\;p_i(x|\theta)\left(1+\log\frac{p_i(x|\theta)}{g_i(x|\varphi)}\right)\nabla_\theta\log p_i(x|\theta)\\
&=\ \left\langle \frac{p_i(x|\theta)}{q_i(x|\hat{\varphi})}\left(1+\log\frac{p_i(x|\theta)}{g_i(x|\varphi)}\right)\; \nabla_\theta \log p_i(x|\theta) \right\rangle_{x\sim q_i(x|\hat{\varphi})}\;.
\end{split}
\end{align}
The two loss functions are then
\begin{align}
\begin{split}
  \loss^\text{int}_\text{KL}
  &= - \sum_i a_i \; \left\langle \frac{\color{red!80!black}p_i(x|\theta)}{\color{red!80!black}q_i(x|\hat{\varphi})}\log g_i(x|\varphi)\right\rangle_{x\sim q_i(x|\hat{\varphi})}  \\
  \loss^\text{weights}_\text{KL}
  &=\sum_i a_i \; \left\langle\frac{\color{red!80!black}p_i(x|\theta)}{\color{red!80!black}q_i(x|\hat{\varphi})} \; \left(1+\log\frac{\color{red!80!black}p_i(x|\theta)}{\color{red!80!black}g_i(x|\varphi)}\right) \; \log p_i(x|\theta) \right\rangle_{x\sim q_i(x|\hat{\varphi})} \;.
\end{split}
\label{eq:kl_losses}
\end{align}
Comparing the first loss to Eq.\eqref{eq:neyman_losses}, we see that
the log-likelihood is only weighted with a single MC weight, so the
$\chi_N^2$ loss penalizes large discrepancies stronger, specifically,
low values of $q_i$ in regions of high density $p_i$.

%%%%%%%%%%%%%%%%%%%%%%%%%%%%%%%%%%%%%%%%%%%%%%%%%%%
\subsection*{Reverse KL-divergence}

The mode-seeking behavior of the reverse
KL-divergence~\cite{Heinrich:2022xfa}
\begin{align}
  %D_\text{RKL}(p_i|\theta||q_i|\varphi)&\equiv
  D_{\text{RKL},i}
  &=\int_\Phi\d^d x\;g_i(x|\varphi)\log\frac{g_i(x|\varphi)}{p_i(x|\theta)}
%  &=\int_\Phi\d^d x\;q_i(x|\varphi)\log q_i(x|\varphi) - \int_\Phi\d^d x\;q_i(x|\varphi)\log p_i(x|\theta)\;.
  \label{eq:rkl_loss_def}
\end{align}
can be beneficial in the training of the normalizing flow, as it
pushes the flow to assign zero density where $p_i(x|\theta)$ is zero
and focuses on the modes of $p_i(x|\theta)$. Unlike for the
forward KL-divergence, the gradient with respect to $\varphi$ is now more complex,
\begin{align}
\begin{split}
\nabla_\varphi D_{\text{RKL},i}
&=\int_\Phi\d^d x\;\nabla_\varphi g_i(x|\varphi)\log g_i(x|\varphi)
+\int_\Phi\d^d x\; g_i(x|\varphi)\nabla_\varphi \log g_i(x|\varphi) \\
&\phantom{=}-\int_\Phi\d^d x\;\nabla_\varphi g_i(x|\varphi)\log p_i(x|\theta)\\
&=\int_\Phi\d^d x\;g_i(x|\varphi)\left(1+\log\frac{g_i(x|\varphi)}{p_i(x|\theta)}\right)\nabla_\varphi\log g_i(x|\varphi)\\
&= \left\langle \frac{g_i(x|\varphi)}{q_i(x|\hat{\varphi})}\left(1+\log\frac{g_i(x|\varphi)}{p_i(x|\theta)}\right)\nabla_\varphi\log g_i(x|\varphi) \right\rangle_{x\sim q_i(x|\hat{\varphi})} \;.
\end{split}
\end{align}
while the training of the channel weights just requires
\begin{align}
\nabla_\theta D_{\text{RKL},i}
=-\nabla_\theta\int_\Phi\d^d x\;g_i(x|\varphi)\log p_i(x|\theta)
= - \left\langle \frac{g_i(x|\varphi)}{q_i(x|\hat{\varphi})} \; \nabla_\theta \log p_i(x|\theta) \right\rangle_{x\sim q_i(x|\hat{\varphi})} \; .
\end{align}
Consequently, we can write for the total loss function
\begin{align}
\begin{split}
    \loss^\text{int}_\text{RKL}
  &= \sum_i a_i \; \left\langle \frac{\color{red!80!black}g_i(x|\varphi)}{\color{red!80!black}q_i(x|\hat{\varphi})}\left(1+\log\frac{\color{red!80!black}g_i(x|\varphi)}{\color{red!80!black}p_i(x|\theta)}\right)\log g_i(x|\varphi) \right\rangle_{x\sim q_i(x|\hat{\varphi})} \\
  \loss^\text{weights}_\text{RKL}
  &=- \sum_i a_i \; \left\langle \frac{\color{red!80!black}g_i(x|\varphi)}{\color{red!80!black}q_i(x|\hat{\varphi})}\log p_i(x|\theta) \right\rangle_{x\sim q_i(x|\hat{\varphi})} \; .
\end{split}
\label{eq:rkl_losses}
\end{align}
In contrast to the KL divergence, which is mass-distributing, this
loss for the normalizing flow only includes the logarithm of the MC
weight and an additional positive $\log g_i(x|\varphi)$ term,
reflecting the mode-seeking behavior. Furthermore, the RKL loss only
requires $p_i(x|\theta)$ and not its derivative $\nabla_x
p_i(x|\theta)$, when taking the gradients before reparametrization.

%%%%%%%%%%%%%%%%%%%%%%%%%%%%%%%%%%%%%%%%%%%%%%%%%%%
\subsection*{Single-pass gradient computation}
%\label{sec:single-pass}

\begin{figure}[t]
    \centering
    \includegraphics[page=3, width=0.98\textwidth]{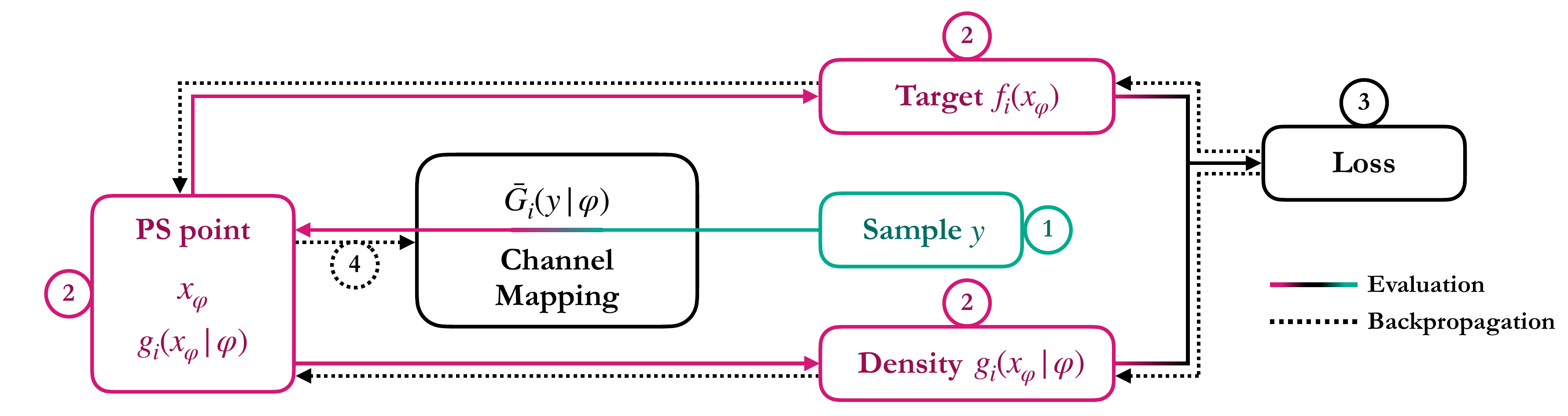}
    \caption{Workflow of the single-pass training of the INN.}
    \label{fig:single_pass_training}
\end{figure}

Finally, when evaluating one of the above-described losses during
online training, we need to be careful. In practice, we want to follow 
the steps
\begin{enumerate}
    \item sample points $y\sim\text{uniform}$;
    \item map $y \to x_\varphi\equiv x(y|\varphi) = \gbar(y|\varphi)$ and evaluate the density $\bar{g}(y|\varphi)=g(x_\varphi|\varphi)^{-1}$;
    \item evaluate the target function $f(x_\varphi) \sim p(x_\varphi)$;
    \item calculate a divergence-based loss between $p(x_\varphi)$ and $g(x_\varphi|\varphi)$;
    \item compute gradients of the loss and optimize the network.
\end{enumerate}
The training workflow is summarized in Fig.~\ref{fig:single_pass_training} and also shows the backpropagation of the gradients coming from the loss function.
For example, the KL-loss is 
\begin{align}
  D_\text{KL} [p(x_\varphi),g(x_\varphi|\varphi)]
  &= \int \d^d y\left.\,p(x_\varphi)\, \log\frac{p(x_\varphi)}{g(x_\varphi|\varphi)}\right\vert_{x_\varphi=\gbar(y|\varphi)}\;.
  \label{eq:kl_latent_space}
\end{align}
For optimization during training, we require its gradient with respect to the network parameters $\varphi$
\begin{align}
    \nabla_\varphi D_\text{KL} [p(x_\varphi),g(x_\varphi|\varphi)]\;,
\end{align}
which would also require us to calculate
\begin{align}
  \frac{\partial p(x_\varphi)}{\partial \varphi} 
  = \frac{\partial p(x_\varphi)}{\partial x_\varphi} \;
  \frac{\partial x_\varphi}{\partial \varphi} \; .
\end{align}
However, the first term is intractable for common event generators, as the amplitude is not differentiable. To circumvent this limitation, we define the loss as a proper function of $x$, such that we do not require the gradient of $p(x)$. This means, we replace Eq.\eqref{eq:kl_latent_space} with
\begin{align}
  D_\text{KL} [p(x),g(x|\varphi)]
  &= \int \d^d x\,p(x)\, \log\frac{p(x)}{g(x|\varphi)}\;.
  \label{eq:kl_phase_space}
\end{align}
In this form, we also need the density $g(x|\varphi)$ as a proper function of $x$ to obtain the  correct gradients. We illustrate this for a two-dimensional toy flow $G$ with one trainable parameter $\varphi$,
\begin{alignat}{3}
  \text{forward $y = G(x|\varphi)$:}
  \qquad &y_1 &&= x_1 + \varphi \qquad y_2 &&= x_2 \cdot \exp x_1 \notag \\
  \text{inverse $x = \gbar(y|\varphi)$:}
  \qquad &x_1 &&= y_1 - \varphi \qquad x_2 &&= y_2 \cdot \exp[-y_1 + \varphi]\;.
\end{alignat}
The corresponding Jacobians are
\begin{align}
  g(x|\varphi) 
  = \left\vert\frac{\partial G(x|\varphi)}{\partial x}\right\vert
  = \exp x_1
  \qquad \text{and} \qquad 
   \bar{g}(y|\varphi)
   &= \left\vert\frac{\partial \gbar(y|\varphi)}{\partial y}\right\vert
   = \exp[-y_1 + \varphi] \; .
\end{align}
While $g(x|\varphi)=\bar{g}(y|\varphi)^{-1}$, $g$ is still a function of $x$ and $\bar{g}$ is a function of $y$. Their gradients with respect to $\varphi$ will therefore be different,
\begin{align}
  \frac{\partial g(x|\varphi)}{\partial \varphi} = 0
  \qquad \text{and} \qquad
  \frac{\partial \bar{g}(y|\varphi)}{\partial \varphi} = \exp[-y_1 + \varphi] = \exp[-x_1(y)]\; ,
  \label{eq:grad_bi}
\end{align}
This means that after the inverse pass $x=\gbar(y|\varphi)$, which has
to be evaluated without gradients to avoid unwanted gradients for $p(x)$,
we perform an additional forward pass $y=G(x|\varphi)$, see
Figs.~\ref{fig:gen_training} and~\ref{fig:sample_training}. This 
forward pass evaluates the Jacobian $g(x|\varphi)$ as a proper function
of $x$. In contrast, the inverse pass would return the Jacobian 
$\bar{g}(y|\varphi)$ as a function of $y$ and yield wrong gradients.

%------------------------------------------------------
%------------------------------------------------------

\clearpage
\bibliography{literature} 

\providecommand{\href}[2]{#2}\begingroup\raggedright\begin{thebibliography}{10}

\bibitem{Campbell:2022qmc}
J.~M. Campbell {\em et al.}, {\it {Event Generators for High-Energy Physics
  Experiments}},  in {\em {2022 Snowmass Summer Study}}.
\newblock 3, 2022.
\newblock \href{http://arxiv.org/abs/2203.11110}{{arXiv:2203.11110 [hep-ph]}}.

\bibitem{Butter:2022rso}
A.~Butter, T.~Plehn, S.~Schumann, {\em et al.}, {\it {Machine Learning and LHC
  Event Generation}},  in {\em {2022 Snowmass Summer Study}}.
\newblock 3, 2022.
\newblock \href{http://arxiv.org/abs/2203.07460}{{arXiv:2203.07460 [hep-ph]}}.

\bibitem{Plehn:2022ftl}
T.~Plehn, A.~Butter, B.~Dillon, and C.~Krause, {\it {Modern Machine Learning
  for LHC Physicists}},
  \href{http://arxiv.org/abs/2211.01421}{{arXiv:2211.01421 [hep-ph]}}.

\bibitem{Bishara:2019iwh}
F.~Bishara and M.~Montull, {\it {(Machine) Learning Amplitudes for Faster Event
  Generation}},  \href{http://arxiv.org/abs/1912.11055}{{arXiv:1912.11055
  [hep-ph]}}.

\bibitem{Badger:2020uow}
S.~Badger and J.~Bullock, {\it {Using neural networks for efficient evaluation
  of high multiplicity scattering amplitudes}},
  \href{http://dx.doi.org/10.1007/JHEP06(2020)114}{JHEP {\bfseries 06} (2020)
  114}, \href{http://arxiv.org/abs/2002.07516}{{arXiv:2002.07516 [hep-ph]}}.

\bibitem{Aylett-Bullock:2021hmo}
J.~Aylett-Bullock, S.~Badger, and R.~Moodie, {\it {Optimising simulations for
  diphoton production at hadron colliders using amplitude neural networks}},
  \href{http://dx.doi.org/10.1007/JHEP08(2021)066}{JHEP {\bfseries 08} (2021)
  066}, \href{http://arxiv.org/abs/2106.09474}{{arXiv:2106.09474 [hep-ph]}}.

\bibitem{Maitre:2021uaa}
D.~Ma\^\i{}tre and H.~Truong, {\it {A factorisation-aware Matrix element
  emulator}},  \href{http://dx.doi.org/10.1007/JHEP11(2021)066}{JHEP {\bfseries
  11} (2021)  066}, \href{http://arxiv.org/abs/2107.06625}{{arXiv:2107.06625
  [hep-ph]}}.

\bibitem{Badger:2022hwf}
S.~Badger, A.~Butter, M.~Luchmann, S.~Pitz, and T.~Plehn, {\it {Loop Amplitudes
  from Precision Networks}},
  \href{http://arxiv.org/abs/2206.14831}{{arXiv:2206.14831 [hep-ph]}}.

\bibitem{Danziger:2021eeg}
K.~Danziger, T.~Jan\ss{}en, S.~Schumann, and F.~Siegert, {\it {Accelerating
  Monte Carlo event generation -- rejection sampling using neural network
  event-weight estimates}},
  \href{http://dx.doi.org/10.21468/SciPostPhys.12.5.164}{SciPost Phys.
  {\bfseries 12} (2022)  164},
  \href{http://arxiv.org/abs/2109.11964}{{arXiv:2109.11964 [hep-ph]}}.

\bibitem{Maitre:2022xle}
D.~Ma\^\i{}tre and R.~Santos-Mateos, {\it {Multi-variable Integration with a
  Neural Network}},  \href{http://arxiv.org/abs/2211.02834}{{arXiv:2211.02834
  [hep-ph]}}.

\bibitem{Klimek:2018mza}
M.~D. Klimek and M.~Perelstein, {\it {Neural Network-Based Approach to Phase
  Space Integration}},
  \href{http://dx.doi.org/10.21468/SciPostPhys.9.4.053}{SciPost Phys.
  {\bfseries 9} (2020)  053},
  \href{http://arxiv.org/abs/1810.11509}{{arXiv:1810.11509 [hep-ph]}}.

\bibitem{Chen:2020nfb}
I.-K. Chen, M.~D. Klimek, and M.~Perelstein, {\it {Improved Neural Network
  Monte Carlo Simulation}},
  \href{http://dx.doi.org/10.21468/SciPostPhys.10.1.023}{SciPost Phys.
  {\bfseries 10} (2021)  023},
  \href{http://arxiv.org/abs/2009.07819}{{arXiv:2009.07819 [hep-ph]}}.

\bibitem{Gao:2020vdv}
C.~Gao, J.~Isaacson, and C.~Krause, {\it {i-flow: High-dimensional Integration
  and Sampling with Normalizing Flows}},
  \href{http://dx.doi.org/10.1088/2632-2153/abab62}{Mach. Learn. Sci. Tech.
  {\bfseries 1} (2020) 4, 045023},
  \href{http://arxiv.org/abs/2001.05486}{{arXiv:2001.05486 [physics.comp-ph]}}.

\bibitem{Bothmann:2020ywa}
E.~Bothmann, T.~Jan{\ss}en, M.~Knobbe, T.~Schmale, and S.~Schumann, {\it
  {Exploring phase space with Neural Importance Sampling}},
  \href{http://dx.doi.org/10.21468/SciPostPhys.8.4.069}{SciPost Phys.
  {\bfseries 8} (2020) 4, 069},
  \href{http://arxiv.org/abs/2001.05478}{{arXiv:2001.05478 [hep-ph]}}.

\bibitem{Gao:2020zvv}
C.~Gao, S.~Höche, J.~Isaacson, C.~Krause, and H.~Schulz, {\it {Event
  Generation with Normalizing Flows}},
  \href{http://dx.doi.org/10.1103/PhysRevD.101.076002}{Phys. Rev. D {\bfseries
  101} (2020) 7, 076002},
  \href{http://arxiv.org/abs/2001.10028}{{arXiv:2001.10028 [hep-ph]}}.

\bibitem{Winterhalder:2021ngy}
R.~Winterhalder, V.~Magerya, E.~Villa, S.~P. Jones, M.~Kerner, A.~Butter,
  G.~Heinrich, and T.~Plehn, {\it {Targeting multi-loop integrals with neural
  networks}},  \href{http://dx.doi.org/10.21468/SciPostPhys.12.4.129}{SciPost
  Phys. {\bfseries 12} (2022) 4, 129},
  \href{http://arxiv.org/abs/2112.09145}{{arXiv:2112.09145 [hep-ph]}}.

\bibitem{nflow1}
D.~J. Rezende and S.~Mohamed, {\it Variational inference with normalizing
  flows},  \href{http://proceedings.mlr.press/v37/rezende15.html}{Proceedings
  of the 32nd International Conference on International Conference on Machine
  Learning {\bfseries 37} (2015)  1530},
  \href{http://arxiv.org/abs/1505.05770}{{arXiv:1505.05770 [stat.ML]}}.

\bibitem{Butter:2022lkf}
A.~Butter, S.~Diefenbacher, G.~Kasieczka, B.~Nachman, T.~Plehn, D.~Shih, and
  R.~Winterhalder, {\it {Ephemeral Learning - Augmenting Triggers with
  Online-Trained Normalizing Flows}},
  \href{http://dx.doi.org/10.21468/SciPostPhys.13.4.087}{SciPost Phys.
  {\bfseries 13} (2022) 4, 087},
  \href{http://arxiv.org/abs/2202.09375}{{arXiv:2202.09375 [hep-ph]}}.

\bibitem{inn}
L.~Ardizzone, J.~Kruse, C.~Rother, and U.~Köthe, {\it Analyzing inverse
  problems with invertible neural networks},
  \href{https://openreview.net/forum?id=rJed6j0cKX}{International Conference on
  Learning Representations (2019)  },
  \href{http://arxiv.org/abs/1808.04730}{{1808.04730 [cs.LG]}}.

\bibitem{cinn}
L.~Ardizzone, C.~Lüth, J.~Kruse, C.~Rother, and U.~Köthe, {\it Guided image
  generation with conditional invertible neural networks},
  \href{http://arxiv.org/abs/1907.02392}{{arXiv:1907.02392 [cs.CV]}}.

\bibitem{Butter:2020tvl}
A.~Butter and T.~Plehn, {\em Generative Networks for LHC Events},
  \href{http://dx.doi.org/10.1142/9789811234033_0007}{in {\em Artificial
  Intelligence for High Energy Physics}},
  \href{http://dx.doi.org/10.1142/9789811234033_0007}{ch.~7, pp.~191--240}.
\newblock World Scientific, 2022.
\newblock \href{http://arxiv.org/abs/2008.08558}{{arXiv:2008.08558 [hep-ph]}}.

\bibitem{Verheyen:2020bjw}
B.~Stienen and R.~Verheyen, {\it {Phase Space Sampling and Inference from
  Weighted Events with Autoregressive Flows}},
  \href{http://dx.doi.org/10.21468/SciPostPhys.10.2.038}{SciPost Phys.
  {\bfseries 10} (2021)  038},
  \href{http://arxiv.org/abs/2011.13445}{{arXiv:2011.13445 [hep-ph]}}.

\bibitem{Bellagente:2021yyh}
M.~Bellagente, M.~Hau\ss{}mann, M.~Luchmann, and T.~Plehn, {\it {Understanding
  Event-Generation Networks via Uncertainties}},
  \href{http://dx.doi.org/10.21468/SciPostPhys.13.1.003}{SciPost Phys.
  {\bfseries 13} (2022)  003},
  \href{http://arxiv.org/abs/2104.04543}{{arXiv:2104.04543 [hep-ph]}}.

\bibitem{Butter:2021csz}
A.~Butter, T.~Heimel, S.~Hummerich, T.~Krebs, T.~Plehn, A.~Rousselot, and
  S.~Vent, {\it {Generative Networks for Precision Enthusiasts}},
  \href{http://arxiv.org/abs/2110.13632}{{arXiv:2110.13632 [hep-ph]}}.

\bibitem{Verheyen:2022tov}
R.~Verheyen, {\it {Event Generation and Density Estimation with Surjective
  Normalizing Flows}},
  \href{http://dx.doi.org/10.21468/SciPostPhys.13.3.047}{SciPost Phys.
  {\bfseries 13} (2022) 3, 047},
  \href{http://arxiv.org/abs/2205.01697}{{arXiv:2205.01697 [hep-ph]}}.

\bibitem{Krause:2021ilc}
C.~Krause and D.~Shih, {\it {CaloFlow: Fast and Accurate Generation of
  Calorimeter Showers with Normalizing Flows}},
  \href{http://arxiv.org/abs/2106.05285}{{arXiv:2106.05285 [physics.ins-det]}}.

\bibitem{Krause:2021wez}
C.~Krause and D.~Shih, {\it {CaloFlow II: Even Faster and Still Accurate
  Generation of Calorimeter Showers with Normalizing Flows}},
  \href{http://arxiv.org/abs/2110.11377}{{arXiv:2110.11377 [physics.ins-det]}}.

\bibitem{Krause:2022jna}
C.~Krause, I.~Pang, and D.~Shih, {\it {CaloFlow for CaloChallenge Dataset 1}},
  \href{http://arxiv.org/abs/2210.14245}{{arXiv:2210.14245 [physics.ins-det]}}.

\bibitem{Cresswell:2022tof}
J.~C. Cresswell, B.~L. Ross, G.~Loaiza-Ganem, H.~Reyes-Gonzalez, M.~Letizia,
  and A.~L. Caterini, {\it {CaloMan: Fast generation of calorimeter showers
  with density estimation on learned manifolds}},  in {\em {36th Conference on
  Neural Information Processing Systems}}.
\newblock 11, 2022.
\newblock \href{http://arxiv.org/abs/2211.15380}{{arXiv:2211.15380 [hep-ph]}}.

\bibitem{Bellagente:2020piv}
M.~Bellagente, A.~Butter, G.~Kasieczka, T.~Plehn, A.~Rousselot,
  R.~Winterhalder, L.~Ardizzone, and U.~K\"othe, {\it {Invertible Networks or
  Partons to Detector and Back Again}},
  \href{http://dx.doi.org/10.21468/SciPostPhys.9.5.074}{SciPost Phys.
  {\bfseries 9} (2020)  074},
  \href{http://arxiv.org/abs/2006.06685}{{arXiv:2006.06685 [hep-ph]}}.

\bibitem{Leigh:2022lpn}
M.~Leigh, J.~A. Raine, and T.~Golling, {\it {$\nu$-Flows: conditional neutrino
  regression}},  \href{http://arxiv.org/abs/2207.00664}{{arXiv:2207.00664
  [hep-ph]}}.

\bibitem{Bieringer:2020tnw}
S.~Bieringer, A.~Butter, T.~Heimel, S.~H\"oche, U.~K\"othe, T.~Plehn, and S.~T.
  Radev, {\it {Measuring QCD Splittings with Invertible Networks}},
  \href{http://dx.doi.org/10.21468/SciPostPhys.10.6.126}{SciPost Phys.
  {\bfseries 10} (2021) 6, 126},
  \href{http://arxiv.org/abs/2012.09873}{{arXiv:2012.09873 [hep-ph]}}.

\bibitem{Bister:2021arb}
T.~Bister, M.~Erdmann, U.~K\"othe, and J.~Schulte, {\it {Inference of
  cosmic-ray source properties by conditional invertible neural networks}},
  \href{http://dx.doi.org/10.1140/epjc/s10052-022-10138-x}{Eur. Phys. J. C
  {\bfseries 82} (2022) 2, 171},
  \href{http://arxiv.org/abs/2110.09493}{{arXiv:2110.09493 [astro-ph.IM]}}.

\bibitem{Butter:2022vkj}
A.~Butter, T.~Heimel, T.~Martini, S.~Peitzsch, and T.~Plehn, {\it {Two
  Invertible Networks for the Matrix Element Method}},
  \href{http://arxiv.org/abs/2210.00019}{{arXiv:2210.00019 [hep-ph]}}.

\bibitem{coupling1}
L.~Dinh, D.~Krueger, and Y.~Bengio, {\it Nice: Non-linear independent
  components estimation},
  \href{http://arxiv.org/abs/1410.8516}{{arXiv:1410.8516 [cs.LG]}}.

\bibitem{coupling2}
L.~Dinh, J.~Sohl-Dickstein, and S.~Bengio, {\it Density estimation using real
  nvp},  \href{http://arxiv.org/abs/1605.08803}{{arXiv:1605.08803 [cs.LG]}}.

\bibitem{Alwall:2014hca}
J.~Alwall, R.~Frederix, S.~Frixione, V.~Hirschi, F.~Maltoni, O.~Mattelaer,
  H.~S. Shao, T.~Stelzer, P.~Torrielli, and M.~Zaro, {\it {The automated
  computation of tree-level and next-to-leading order differential cross
  sections, and their matching to parton shower simulations}},
  \href{http://dx.doi.org/10.1007/JHEP07(2014)079}{JHEP {\bfseries 07} (2014)
  079}, \href{http://arxiv.org/abs/1405.0301}{{arXiv:1405.0301 [hep-ph]}}.

\bibitem{Sherpa:2019gpd}
Sherpa, E.~Bothmann {\em et al.}, {\it {Event Generation with Sherpa 2.2}},
  \href{http://dx.doi.org/10.21468/SciPostPhys.7.3.034}{SciPost Phys.
  {\bfseries 7} (2019) 3, 034},
  \href{http://arxiv.org/abs/1905.09127}{{arXiv:1905.09127 [hep-ph]}}.

\bibitem{Kilian:2007gr}
W.~Kilian, T.~Ohl, and J.~Reuter, {\it {WHIZARD: Simulating Multi-Particle
  Processes at LHC and ILC}},
  \href{http://dx.doi.org/10.1140/epjc/s10052-011-1742-y}{Eur. Phys. J. C
  {\bfseries 71} (2011)  1742},
  \href{http://arxiv.org/abs/0708.4233}{{arXiv:0708.4233 [hep-ph]}}.

\bibitem{KLEISS1994141}
R.~Kleiss and R.~Pittau, {\it Weight optimization in multichannel {Monte
  Carlo}},  \href{http://dx.doi.org/10.1016/0010-4655(94)90043-4}{Computer
  Physics Communications {\bfseries 83} (1994) 2, 141}.

\bibitem{Weinzierl:2000wd}
S.~Weinzierl, {\it {Introduction to Monte Carlo methods}},
  \href{http://arxiv.org/abs/hep-ph/0006269}{{arXiv:hep-ph/0006269}}.

\bibitem{vegas1}
G.~P. Lepage, {\it {A New Algorithm for Adaptive Multidimensional
  Integration}},  \href{http://dx.doi.org/10.1016/0021-9991(78)90004-9}{J.
  Comput. Phys. {\bfseries 27} (1978)  192}.

\bibitem{vegas2}
G.~P. Lepage, {\it {VEGAS - an adaptive multi-dimensional integration
  program}},  \href{"https://cds.cern.ch/record/123074"}{Cornell Preprint
  {\bfseries 447} (1980)  }.

\bibitem{Ohl:1998jn}
T.~Ohl, {\it {Vegas revisited: Adaptive Monte Carlo integration beyond
  factorization}},
  \href{http://dx.doi.org/10.1016/S0010-4655(99)00209-X}{Comput. Phys. Commun.
  {\bfseries 120} (1999)  13},
  \href{http://arxiv.org/abs/hep-ph/9806432}{{arXiv:hep-ph/9806432}}.

\bibitem{Brass:2018xbv}
S.~Brass, W.~Kilian, and J.~Reuter, {\it {Parallel Adaptive Monte Carlo
  Integration with the Event Generator WHIZARD}},
  \href{http://dx.doi.org/10.1140/epjc/s10052-019-6840-2}{Eur. Phys. J. C
  {\bfseries 79} (2019) 4, 344},
  \href{http://arxiv.org/abs/1811.09711}{{arXiv:1811.09711 [hep-ph]}}.

\bibitem{Lepage:2020tgj}
G.~P. Lepage, {\it {Adaptive multidimensional integration: VEGAS enhanced}},
  \href{http://dx.doi.org/10.1016/j.jcp.2021.110386}{J. Comput. Phys.
  {\bfseries 439} (2021)  110386},
  \href{http://arxiv.org/abs/2009.05112}{{arXiv:2009.05112 [physics.comp-ph]}}.

\bibitem{Maltoni:2002qb}
F.~Maltoni and T.~Stelzer, {\it {MadEvent: Automatic event generation with
  MadGraph}},  \href{http://dx.doi.org/10.1088/1126-6708/2003/02/027}{JHEP
  {\bfseries 02} (2003)  027},
  \href{http://arxiv.org/abs/hep-ph/0208156}{{arXiv:hep-ph/0208156}}.

\bibitem{Mattelaer:2021xdr}
O.~Mattelaer and K.~Ostrolenk, {\it {Speeding up MadGraph5\_aMC@NLO}},
  \href{http://dx.doi.org/10.1140/epjc/s10052-021-09204-7}{Eur. Phys. J. C
  {\bfseries 81} (2021) 5, 435},
  \href{http://arxiv.org/abs/2102.00773}{{arXiv:2102.00773 [hep-ph]}}.

\bibitem{Press:1989vk}
W.~H. Press and G.~R. Farrar, {\it Recursive stratified sampling for
  multidimensional monte carlo integration},
  \href{http://dx.doi.org/10.1063/1.4822899}{Computers in Physics {\bfseries 4}
  (1990) 2, 190}.

\bibitem{glow}
D.~P. Kingma and P.~Dhariwal, {\it Glow: Generative flow with invertible 1x1
  convolutions},  \href{http://arxiv.org/abs/1807.03039}{{arXiv:1807.03039
  [stat.ML]}}.

\bibitem{EulerAngles}
L.~Euler, {\it {Formulae generales pro translatione quacunque corporum
  rigidorum}},
  \href{http://math.dartmouth.edu/~euler/docs/originals/E478.pdf}{Novi
  Commentarii academiae scientiarum Petropolitanae {\bfseries 20} (1776)  189}.

\bibitem{doi:10.1063/1.1666011}
D.~K. Hoffman, R.~C. Raffenetti, and K.~Ruedenberg, {\it Generalization of
  euler angles to n‐dimensional orthogonal matrices},
  \href{http://dx.doi.org/10.1063/1.1666011}{Journal of Mathematical Physics
  {\bfseries 13} (1972) 4, 528}.

\bibitem{Plehn:2015dqa}
T.~Plehn, {\it {Lectures on LHC Physics}},
  \href{http://dx.doi.org/10.1007/978-3-642-24040-9}{Lect. Notes Phys.
  {\bfseries 844} (2012)  1},
  \href{http://arxiv.org/abs/0910.4182}{{arXiv:0910.4182 [hep-ph]}}.

\bibitem{NNPDF:2021njg}
NNPDF, R.~D. Ball {\em et al.}, {\it {The path to proton structure at 1\%
  accuracy}},  \href{http://dx.doi.org/10.1140/epjc/s10052-022-10328-7}{Eur.
  Phys. J. C {\bfseries 82} (2022) 5, 428},
  \href{http://arxiv.org/abs/2109.02653}{{arXiv:2109.02653 [hep-ph]}}.

\bibitem{Buckley:2014ana}
A.~Buckley, J.~Ferrando, S.~Lloyd, K.~Nordstr\"om, B.~Page, M.~R\"ufenacht,
  M.~Sch\"onherr, and G.~Watt, {\it {LHAPDF6: parton density access in the LHC
  precision era}},
  \href{http://dx.doi.org/10.1140/epjc/s10052-015-3318-8}{Eur. Phys. J. C
  {\bfseries 75} (2015)  132},
  \href{http://arxiv.org/abs/1412.7420}{{arXiv:1412.7420 [hep-ph]}}.

\bibitem{lusifer}
S.~Dittmaier and M.~Roth, {\it {LUSIFER: A LUcid approach to six FERmion
  production}},  \href{http://dx.doi.org/10.1016/S0550-3213(02)00640-5}{Nucl.
  Phys. B {\bfseries 642} (2002)  307},
  \href{http://arxiv.org/abs/hep-ph/0206070}{{arXiv:hep-ph/0206070}}.

\bibitem{durkan2019neural}
C.~Durkan, A.~Bekasov, I.~Murray, and G.~Papamakarios, {\it Neural spline
  flows},
  \href{https://papers.nips.cc/paper/2019/hash/7ac71d433f282034e088473244df8c02-Abstract.html}{Advances
  in Neural Information Processing Systems {\bfseries 32} (2019)  7511},
  \href{http://arxiv.org/abs/1906.04032}{{arXiv:1906.04032 [stat.ML]}}.

\bibitem{Heinrich:2022xfa}
L.~Heinrich and M.~Kagan, {\it {Differentiable Matrix Elements with MadJax}},
  in {\em {20th International Workshop on Advanced Computing and Analysis
  Techniques in Physics Research}: {AI Decoded - Towards Sustainable, Diverse,
  Performant and Effective Scientific Computing}}.
\newblock 2, 2022.
\newblock \href{http://arxiv.org/abs/2203.00057}{{arXiv:2203.00057 [hep-ph]}}.

\end{thebibliography}\endgroup
\end{document}